\documentclass[final]{ws-ijmpe}

\usepackage[super,sort,numbers]{natbib}
\usepackage{color}

\begin{document}

\markboth{Sergey Postnikov, Madappa Prakash}{Thermal and transport properties of a non-relativistic quantum gas
interacting through a delta-shell potential}

\catchline{}{}{}{}{}

\title{THERMAL AND TRANSPORT PROPERTIES OF A NON-RELATIVISTIC QUANTUM GAS INTERACTING THROUGH A DELTA-SHELL POTENTIAL}

\author{\footnotesize SERGEY POSTNIKOV}


\address{Nuclear Theory Center, Indiana University, \\
Bloomington, Indiana 47405, USA \\
spostnik@indiana.edu}

\author{MADAPPA PRAKASH}

\address{Department of Physics and Astronomy, Ohio University,\\
Athens, OH 45701-2979,
USA\\
prakash@harsha.phy.ohiou.edu}

\maketitle

\begin{history}
\received{(received date)}
\revised{(revised date)}
\end{history}

\begin{abstract}
This work extends the seminal work of Gottfried on the two-body quantum physics of particles interacting through a delta-shell potential to many-body physics by studying a system of non-relativistic particles when the thermal De-Broglie wavelength of a particle is smaller than the range of the potential and the density is such that average distance between particles is smaller than the range. The ability of the delta-shell potential to reproduce some basic properties of the deuteron are examined. Relations for moments of bound states are derived. The virial expansion is used to calculate the first quantum correction to the ideal gas pressure in the form of the second virial coefficient. Additionally, all thermodynamical functions are calculated up to the first order quantum corrections. For small departures from equilibrium, the net flows of mass, energy and momentum, characterized  by the coefficients of diffusion, thermal conductivity and shear viscosity, respectively, are calculated. Properties of the gas are examined for various values of physical parameters including the case of infinite scattering length when the unitary limit is achieved.
\end{abstract}

\newpage

\tableofcontents

\newpage

\section{Introduction}
\label{sec1}
Recently, studies of transport properties, particularly viscosities of
interacting particles in a many-body system, have attracted much
attention as observed phenomena in diverse fields of physics (such as
atomic physics and relativistic heavy ions physics) are greatly
influenced by their effects~\cite{ST09}. In atomic physics, the relaxation
of atomic clusters from an ordered state have been
observed~\cite{Kinast05}. In relativistic heavy-ion physics, viscosities
influence the observed elliptic and higher order collective flows versus transverse momentum of
final state hadrons~\cite{HMS09}. In both of these fields, the challenge of
determining the transport properties at varying temperatures and
densities has been vigorously pursued~\cite{ST09}.

In this work, we  study a simple system in which the high
temperature thermal and transport properties of a dilute gas are significantly affected by
the nature of two-body interactions. The two-body interaction chosen
for this study - the delta-shell interaction - allows us to study the
roles of long scattering lengths, finite-range corrections, and
resonances on the thermal and transport properties of a quantum gas
through an extension of the classic work of Gottfried~\cite{Gottfried66,Gottfried04} 
to the many-body context.   
Our pedagogical study reveals several universal
characteristics. Our results stress the need to perform
analyses beyond what we offer here, particularly with regard to
contributions from higher than two-body interactions.

Gottfried's treatment~\cite{Gottfried66,Gottfried04} of the
quantum mechanics of two particles interacting through a delta-shell
potential 
\begin{equation}
\label{Vrdel}
V(r)= - v \, \delta(r-R), 	
\end{equation}
where $v$ is the strength and $R$
is the range, forms the basis of this work in which the thermal and transport
properties of a dilute gas of non-relativistic particles are
calculated for densities $n$ for which the average inter-particle
distance $d \sim n^{-1/3} \gg R$.  The delta-shell potential is
particularly intriguing as the scattering length $a_{sl}$
and the effective range $r_0$ can be tuned at will,
including the interesting case of $a_{sl} \rightarrow \infty$ (the unitary limit) that is
accessible in current atomic physics experiments (see, for e.g., 
Ref. \cite{Kinast05}).
The delta-shell potential also allows us to understand the transport
characteristics of nuclear systems in which the neutron-proton and
neutron-neutron scattering lengths ($a_{sl} \simeq -23.8$ fm and
$-18.5$ fm, respectively~\cite{slengths}) are much longer than the few
fm ranges of strong interactions.  In addition, the
delta-shell potential admits resonances.

\begin{figure*}[t]
\centerline{\includegraphics[width=10cm]{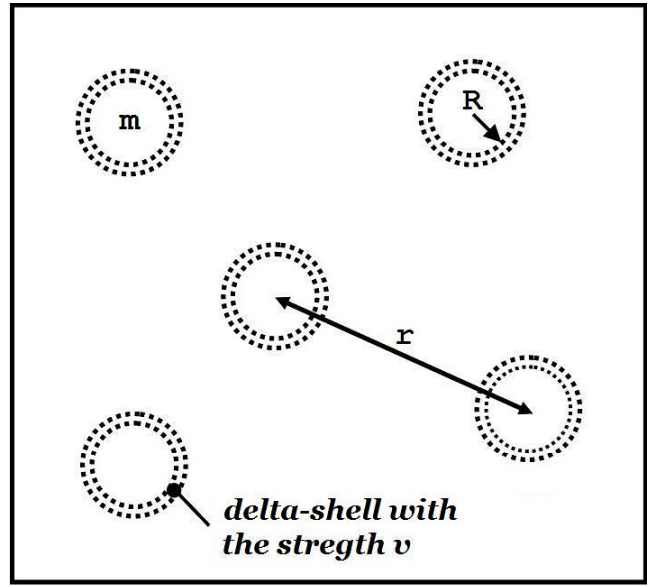}}
\caption{Illustration of a system of particles interacting through a
delta-shell potential $V(r)=-v \, \delta(r-R)$, where $v$ is the
strength and $R$ is the range of the potential. The distance between a pair of particles is denoted by $r$.}
\label{deltaboxx}
\end{figure*}

The schematic diagram in Fig. \ref{deltaboxx} illustrates a gas of particles interacting through the delta-shell potential and the relevant length scales. 

\newpage

This work consists of three main parts: (i) The quantum mechanics of the 2-body problem (sections \ref{sec2} through \ref{sec11}), (ii) Thermal and statistical physics of the delta shell gas (sections  \ref{p2sec1} - \ref{p2sec6}) and (iii) Transport  properties of the delta shell gas (\ref{p3sec1} - \ref{p3sec4}). 

In part (i), we begin by summarizing the text book results of Ref.~\cite{Gottfried66,Gottfried04} and complement them with additional material related to 
the scattering state and bound state solutions of the Schr\"odinger equation for two particles interacting through the delta-shell potential (section \ref{sec2}). These solutions are analytical and allow us to calculate phase shifts for all partial waves (section \ref{sec3}), the density of scattering states (section \ref{sec4}), energies of bound states (section \ref{sec5}), the scattering length and the effective range. 
The conditions for the unitary limit and for resonances to occur are identified. In section \ref{sec6}, 
the ability of the delta-shell potential to reproduce some basic properties of the loosely bound deuteron (binding energy is $-2.23$ MeV) is examined. With the exception of the shape parameter (sensitive only to
moderately high-energy scattering), the calculated
results (e.g., the deuteron root-mean-square radius) agree very well with
experiment (and for good reasons). The disagreement with the shape
parameter is also understandable.
In section \ref{sec7}, a differential equation to determine the S-wave scattering length and the effective range is established. The momentum-space wave function and the form factor for the S-wave are presented in section \ref{sec8}. 
Next, consequences from the Feynman-Hellman theorem \cite{Feynman39} \cite{Hellmann37} for wave functions with a kink (as is the case for the
delta-shell potential) are examined in section \ref{sec9}. 
The normalization constants for bound states are determined in section \ref{sec10}.
Insights gained through an application of 
the virial theorem, which takes a special form for the delta-shell potential is discussed in section \ref{sec11}. 
Relations between the various
moments of the bound-state wave functions are derived section \ref{sec12}. These
relations are analogous to those derived by Kramers\cite{Kramers57} and Pasternak\cite{Pasternak37} for the hydrogen
atom, but for wave functions that exhibit a kink. We show how special care must be
taken to derive these  relations in the presence of a kink.

In sections \ref{p2sec1} and \ref{p2sec2} of part (ii), we calculate  
the first quantum correction to the ideal-gas pressure in the form of
the second virial coefficient \cite{Huang87}. The physical conditions for which the second virial coefficient gives an adequate description is examined in section \ref{p2sec3}. 
The case in which the scattering length becomes infinite (the unitary limit)
is examined in some detail in sections \ref{p2sec4} and \ref{p2sec5}.
In addition, all thermodynamical functions are
calculated including corrections from the second virial
coefficient (section \ref{p2sec6}). 

The transport properties of a delta-shell gas are calculated in part (iii).
Small departures from equilibrium result in net flows of mass, energy
and momentum which are characterized by the
coefficients of diffusion, thermal conductivity and shear viscosity,
respectively.  These coefficients are calculated in first 
and second order of deviations 
(from the equilibrium distribution function)  using the
Chapman-Enskog method \cite{Chapman70} in section \ref{p3sec1}.  
The transport
properties are examined in various physical situations, including the
case of an infinite scattering length and when resonances occur in sections \ref{p3sec2} and \ref{asychap}. In section  \ref{p3sec4}, 
The ratio of shear viscosity to entropy density is calculated and compared to the
recently the proposed universal minimum value of $\simeq 1/(4\pi)$ \cite{Kovtun2005, Salasnich2011}.

Section  \ref{concl} contains a summary with concluding remarks.


\clearpage
	

\section{Quantum mechanics of the two-body problem}
\label{sec2}

We begin by adopting solutions and expressions derived in the book by Gottfried \cite{Gottfried66}.
The Hamiltonian for two particles interacting via the delta-shell potential has the form
\begin{equation}
  \hat H = -\hbar^2 \frac{\Delta}{2\mu}-v \, \delta(r-R),
\label{H}
\end{equation}
where $\mu$ denotes the reduced mass, 
\begin{equation}
\Delta=\frac{1}{r^2}\frac{\partial}{\partial r}r^2\frac{\partial}{\partial r}-\frac{\hat L^2}{r^2}
\end{equation}
is the Laplacian in spherical coordinates ($\hat L$ is the orbital momentum operator), $\hbar$ is the Plank constant, $r$ is the separation distance, $v$ and $R$ are the strength and range parameters of the potential, respectively. Introducing the dimensionless position variable $\rho=k r$ ($k$ is the wave number), the radial part of the stationary Schr\"odinger equation
\begin{equation}
\hat H \, \psi(r) \, Y_{lm}(\theta,\phi)=E \, \psi(r) \, Y_{lm}(\theta,\phi)
\label{shr_eqn}
\end{equation}
($Y_{lm}(\theta,\phi)$ are the spherical harmonics which are eigenfunctions of the operator $\hat L^2$ with eigenvalues $l(l+1)$) takes the form
\begin{equation}
  u''(\rho)+\left(1-\frac{l(l+1)}{\rho^2}\right)u(\rho)= -\frac{\Lambda}{k} \, \delta(\rho- k R) u(\rho),
\label{radial_eqn_u}
\end{equation}
where
\begin{equation}
 u(\rho)=\frac{\rho}{k} \, \psi(\rho)\,, \quad 
 \Lambda=v\frac{2\mu}{\hbar^2}\,, \quad {\rm and} \quad
 E=\frac{\hbar^2 k^2}{2\mu}\,,
 \end{equation} 
and $l$ is the usual orbital quantum number. Above, primes denote appropriate spatial derivatives. The general solution of Eq. (\ref{radial_eqn_u}) is given by a combination of Bessel functions of the first kind, $J_{l+\frac{1}{2}}(\rho)$, and the second kind, $N_{l+\frac{1}{2}}(\rho)$ ~\cite{AbramowitzANDStegun}:
\begin{equation}
  u(\rho)=c_1 \sqrt{\rho} J_{l+\frac{1}{2}}(\rho)+c_2 \sqrt{\rho} N_{l+\frac{1}{2}}(\rho),
\label{general_sln_u}
\end{equation}
or in terms of the spherical Bessel functions $j_l(\rho)$ and $n_l(\rho)$:
\begin{equation}
  \psi(\rho)\equiv\frac{u(\rho)}{\rho}=A_l j_l(\rho)+B_l n_l(\rho).
\label{general_sln_psi}
\end{equation}
The $\delta$-function part of Eq. (\ref{radial_eqn_u}) requires that at $r=R$,
\begin{equation}
  \psi'(k R + 0)-\psi'(k R - 0) = -\frac{\Lambda}{k} \, \psi(k R).
\label{cond_dpsi_a}
\end{equation}
The physical boundary condition that the radial wave function be continuous at $r=R$ implies that
 \begin{equation}
  \psi(k R - 0)=\psi(k R + 0).
\label{cond_psi_a}
\end{equation}
The condition that the wave function is well behaved at the origin requires
 \begin{equation}
  u(0)=0.
\label{cond_u_0}
\end{equation}
For a bound state, the boundary condition
 \begin{equation}
  \psi(\rho \to \infty) \to 0
\label{cond_psi_inf}
\end{equation}
assures square integrability.

\section {Phase shifts for scattering states}
\label{sec3}

Positive energy ($E>0$, hence the wave number $k$ is real) solutions correspond to scattering states. As the potential is confined to a finite volume, the calculation of the phase shift $\delta_l$ is straightforward. From the asymptotic behavior ($\rho\to\infty$) of the general solution in Eq. (\ref{general_sln_psi}), one has
 \begin{equation}
  \psi(\rho \to \infty) \to A_l\frac{\sin(\rho-\frac{\pi l}{2})}{\rho}- B_l\frac{\cos(\rho-\frac{\pi l}{2})}{\rho}=\sqrt{A_l^2+B_l^2} \, \frac{\sin(\rho-\frac{\pi l}{2}+\delta_l)}{\rho},
\label{psi_inf}
\end{equation}
whence
\begin{equation}
  \tan(\delta_l)=- \, \frac{B_l}{A_l}.
\label{tan_dl}
\end{equation}
Utilizing Eq. (\ref{general_sln_psi}), the ratio on the right-hand side of the above equation is found by satisfying the boundary conditions in Eq. (\ref{cond_dpsi_a}), Eq. (\ref{cond_psi_a}) and Eq. (\ref{cond_u_0}). The result is \cite{Gottfried66}
 \begin{equation}
  \tan(\delta_l)=\frac{g \, x \, j_l^2(x)}{1+g \, x \, j_l(x) \, n_l(x)},
\label{tan_dl_gx}
\end{equation}
where
\begin{equation}
\label{gdef}
g\equiv\Lambda \, R = \frac{2\mu v}{\hbar^2}R,
\end{equation}
and
\begin{equation}
\label{xdef}
x=k R
\end{equation}
are dimensionless parameters.\\
In deriving Eq. (\ref{tan_dl_gx}), use of the Wronskian relation
\begin{equation}
  j_l(x) \, n'_l(x)-j'_l(x) \, n_l(x)=\frac{1}{x^2},
\label{bessel_eq}
\end{equation}
has been made.

The partial-wave cross section $\sigma_l(k)$ and the total cross section $\sigma(k)$ are given by \cite{Gottfried66}
\begin{equation}
\sigma_l(k)\equiv \frac{4\pi}{k^2}(2l+1)\sin^2(\delta_l)=4 \pi R^2 (2l+1)\frac{\sin^2(\delta_l)}{x^2},
\label{sigma_l}
\end{equation}
\begin{equation}
\sigma(k)=\sum_{l=0}^{\infty}\sigma_l(k).
\label{sigma_tot}
\end{equation}
Using Eq. (\ref{tan_dl_gx}) and trigonometric relations, we find that
\begin{equation}
\sin^2(\delta_l)=\left[ 1+\left(\frac{1+g \, x \, j_l(x) \, n_l(x)}{g \, x \, j_l^2(x)}\right)^2\right]^{-1}.
\label{sin2_d0}
\end{equation}

\begin{figure*}[t]
\centerline{\includegraphics[width=12cm]{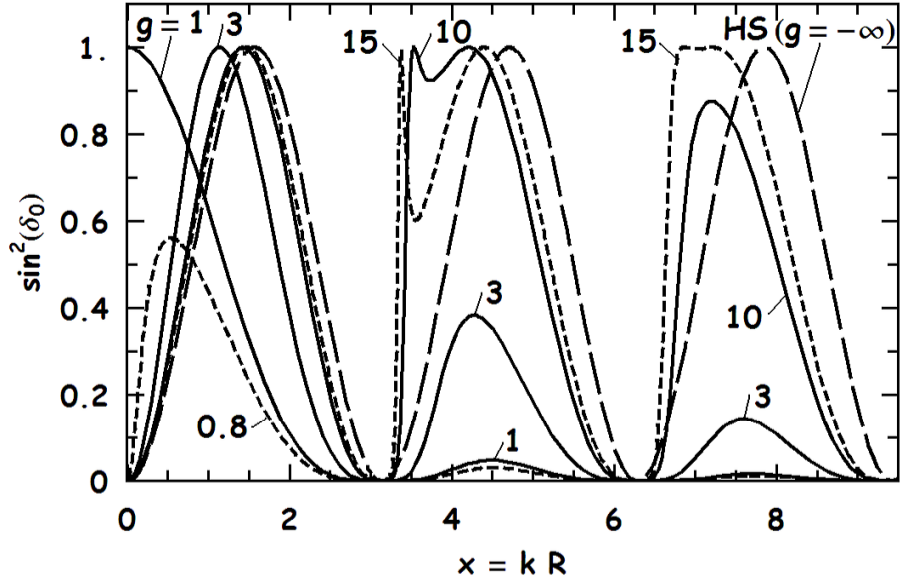}}
\vskip -20pt
 \caption{A comparison of $\sin^2(\delta_0)$ for the delta-shell (Eq.~(\ref{sin2_d0})) and hard-sphere(HS) potentials. The abscissa shows the energy variable $x=k R$. Results are for $g=0.8, 1, 3, 10$ and $15$.}
  \label{sind}
\end{figure*}

A plot of $\sin^2(\delta_0)$ for various values of $g$ is shown in Fig. \ref{sind} for the delta-shell and the hard-sphere potentials. The hard-sphere result is the limiting case of Eq. (\ref{tan_dl_gx}) when $g \to -\infty$, whence $\tan(\delta_l)=\frac{j_l(x)}{n_l(x)}$ and therefore
\begin{equation}
 \sin^2(\delta_l)=\left[1+\left(\frac{n_l(x)}{j_l(x)}\right)^2\right]^{-1}.
\label{sin2_dHS}
\end{equation}

In the case of the delta-shell potential,  
sharp resonances occur as the wave function can propagate inside the shell as well as bounce off of it.  Resonances weaken with increasing energy as a more energetic particle is less influenced by the delta-shell  potential.  For contrast, results for the hard-sphere potential  (long-dashed lines) are also shown.

\begin{figure*}[tb]
\centerline{\includegraphics[width=11cm]{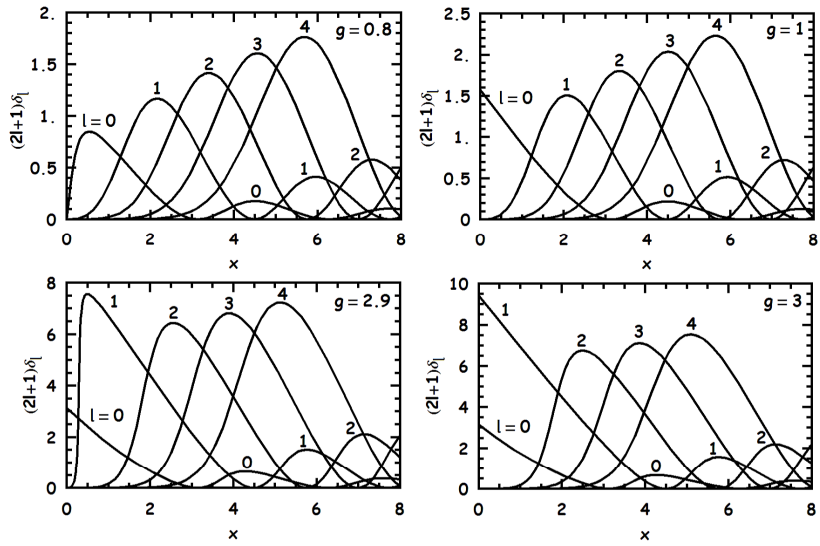}}
\vskip -20pt
 \caption{The quantity $(2 l+1)\delta_l$ as function of $x=k R$ for the indicated values of the orbital angular momentum quantum number $l$.  Values of $g$ are shown in the inset.}
  \label{ldl}
\end{figure*}

In order to highlight the relative strengths of the various partial waves, the quantity $(2 l+1)\delta_l$ as function of $x=k R$ is shown in Fig. \ref{ldl} for values of $l=0, 1, ..., 6$ for several values of $g$. 

\begin{figure*}[tb]
\centerline{\includegraphics[width=11cm]{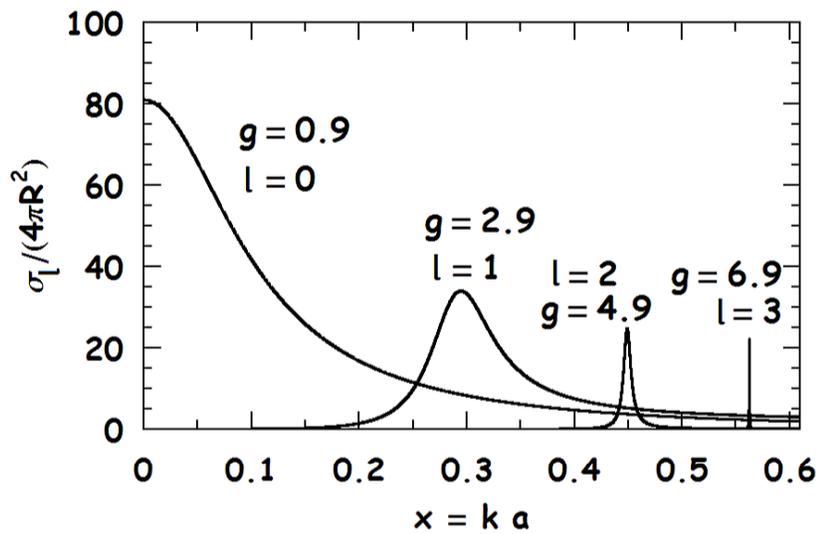}}
\vskip -20pt
 \caption{Partial wave cross sections for  $g\approx2l+1$ for which resonant features occur.}
  \label{res_sl}
\end{figure*}

Fig. \ref{res_sl} shows the partial wave cross sections $\sigma_l$ for  $g\approx 2 l+1$ for which resonant
features are clearly seen; the resonances become narrow with increasing $l$.
%

\clearpage

\section{The density of states and the phase shifts}
\label{sec4}

If particles are confined to a large spherical box with radius $L\to\infty$, the wave function vanishes at the walls of the box \cite{Hirschfelder67}. Imposing this condition on Eq. (\ref{psi_inf}), one has
\begin{equation}
\begin{array}{ll}
 k L - \frac{l\pi}{2}+\delta_l(k)=\pi m \, &\text{(interacting system)},\\
 k L - \frac{l\pi}{2}=\pi m \, &\text{(noninteracting system)},
\end{array}
\label{walls}
\end{equation}
where $m$ is an integer.
\par The number of levels $\Delta m$ per range $\Delta k$ of wave number is given by
\begin{equation}
\begin{array}{ll}
 \Delta m=\frac{L}{\pi}\Delta k+\frac{1}{\pi}\frac{\Delta \delta_l(k)}{\Delta k}\Delta k \\
\\
\Delta m^{(0)}=\frac{L}{\pi}\Delta k^{(0)},
\end{array}
\label{deltam}
\end{equation}
where the superscript $(0)$ refers to the non-interacting system.
The corresponding density of states is $d_l(k)=(2 l+1)\frac{\Delta m}{\Delta k}$ and $d_l^{(0)}(k)=(2 l+1)\frac{\Delta m^{(0)}}{\Delta k^{(0)}}$. Setting $\Delta k\to0$ (box of infinite size, $L\to\infty$), the momentum derivative of the phase shifts can be expressed as
\begin{equation}
  d_l(k)-d_l^{(0)}(k)=\frac{2l+1}{\pi}\frac{\partial \delta_l(k)}{\partial k}.
\label{shift_deriv}
\end{equation}
The derivative of the phase shift can also be expressed as
\begin{equation}
 \frac{\partial \delta_l(x)}{\partial x}=\frac{1}{1+\tan^2(\delta_l)}
\frac{\partial \tan(\delta_l)}{\partial x},
\label{shift_dx}
\end{equation}
where $x=k R$.
For the delta-shell potential
\begin{equation}
 \frac{\partial \delta_l(x)}{\partial x}=\frac{g j_l(x)((2l+1-g)j_l(x)-2xj_{l+1}(x))}{g^2 x^2 j_l^4(x)+(g x j_l(x) n_l(x)+1)^2},
\label{shift_dx_tan}
\end{equation}
where $\tan(\delta_l)$ from Eq. (\ref{tan_dl_gx}) has been used. By setting $g\to -\infty$ in Eq. (\ref{shift_dx_tan}), we recover the result for the case of hard spheres:
\begin{equation}
\frac{\partial \delta_l(x)}{\partial x}=-\frac{1}{x^2} \, \frac{1}{ j_l^2(x)+ n_l^2(x)}.
\label{shift_dx_hs}
\end{equation}
To establish the low-energy behavior, we expand the phase shifts in a power series of $x$ around zero momentum:
  
\begin{equation}
\frac{\partial \delta_l(x)}{\partial x}\approx
\begin{cases}
\frac{x^{2l}}{[(2l-1)!!]^2}\frac{g}{[(2l+1)-g]} & \text{for $g\neq2l+1$,}\\
-\frac{2}{3} & \text{for $g=1;$ \, $l=0$,}\\
-\frac{x^{2l-2}}{2}\frac{(2l-1)^2(2l+3)(2l+1)}{[(2l+1)!!]^2} & \text{for $g=2l+1;$ \, $l>0$,}
\end{cases}
\label{shift_dx_low}
\end{equation}
or
\begin{equation}
\tan(\delta_l)\approx
\begin{cases}
\frac{x^{2l+1}}{[(2l+1)!!]^2}\frac{g(2l+1)}{[(2l+1)-g]} & \text{for $g\neq2l+1$,}\\
-\frac{x^{2l-1}}{2}\frac{(2l-1)(2l+1)(2l+3)}{[(2l+1)!!]^2} & \text{for $g=2l+1$.}
\end{cases}
\label{tanshift_dx_low}
\end{equation}

We see here that when $g$ approaches $2l+1$, the usual $k^{2l+1}$ behavior
for the phase shift does not work.
\begin{figure*}[tb]
\centerline{\includegraphics[width=12cm]{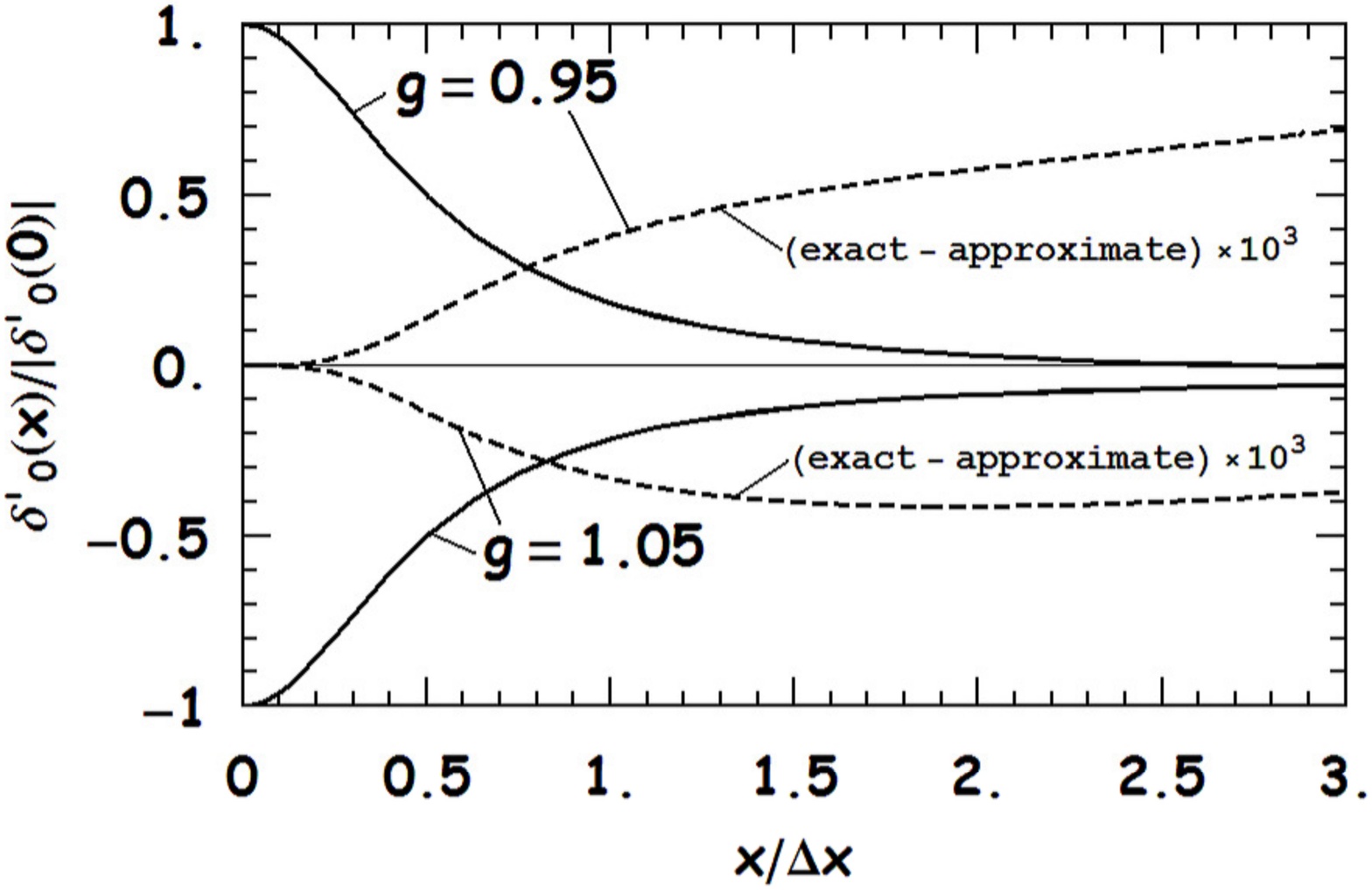}}
\vskip -20pt
\caption{An illustration of the S-wave phase shift derivative for low momenta near a resonance.}
  \label{fllf}
\end{figure*}
\begin{figure*}[tb]
\centerline{\includegraphics[width=12cm]{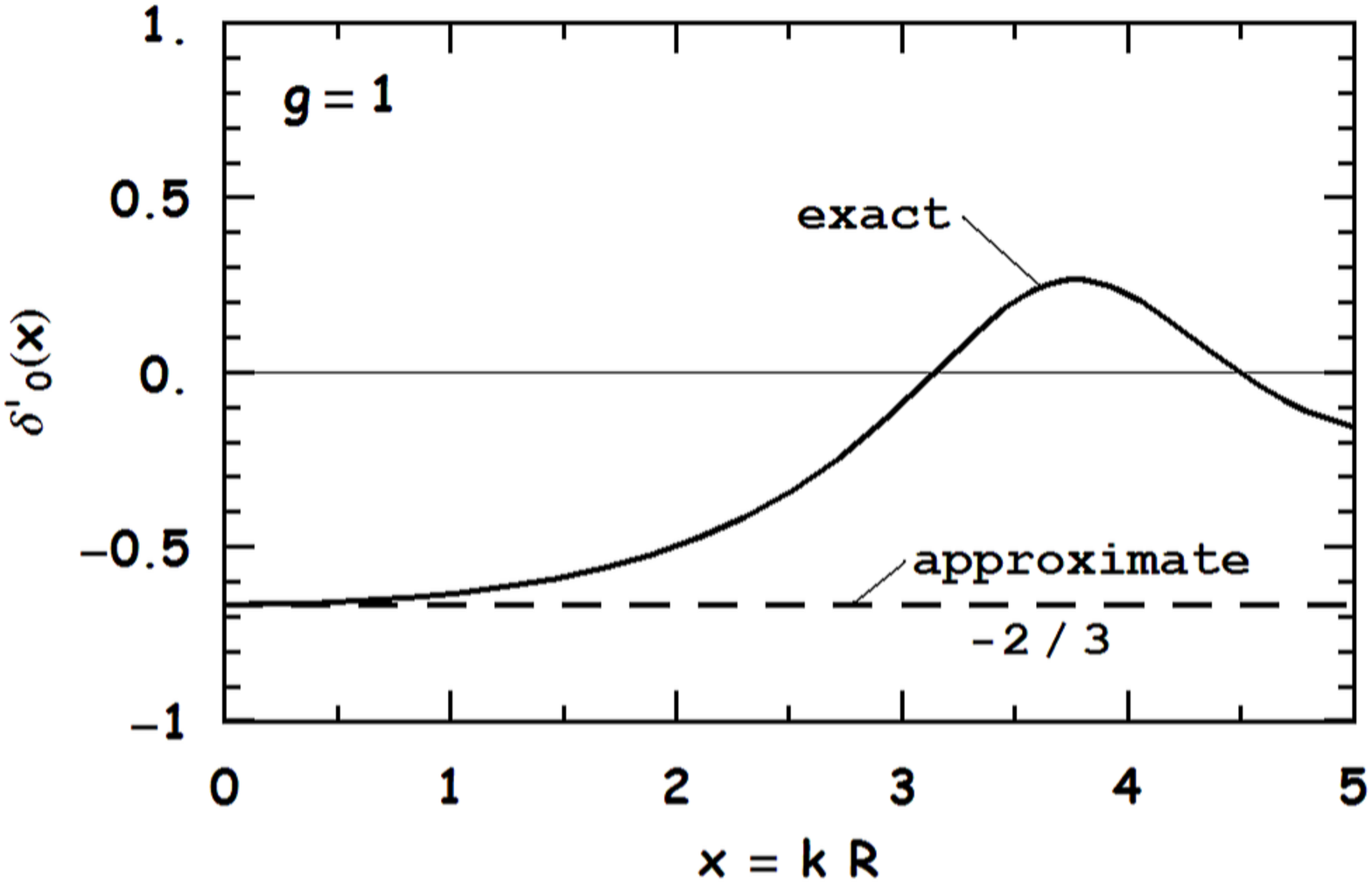}}
\vskip -20pt
\caption{An illustration of the S-wave phase shift derivative for low momenta at a resonance.}
  \label{fllf2}
\end{figure*}
\begin{figure*}[tb]
\centerline{\includegraphics[width=12cm]{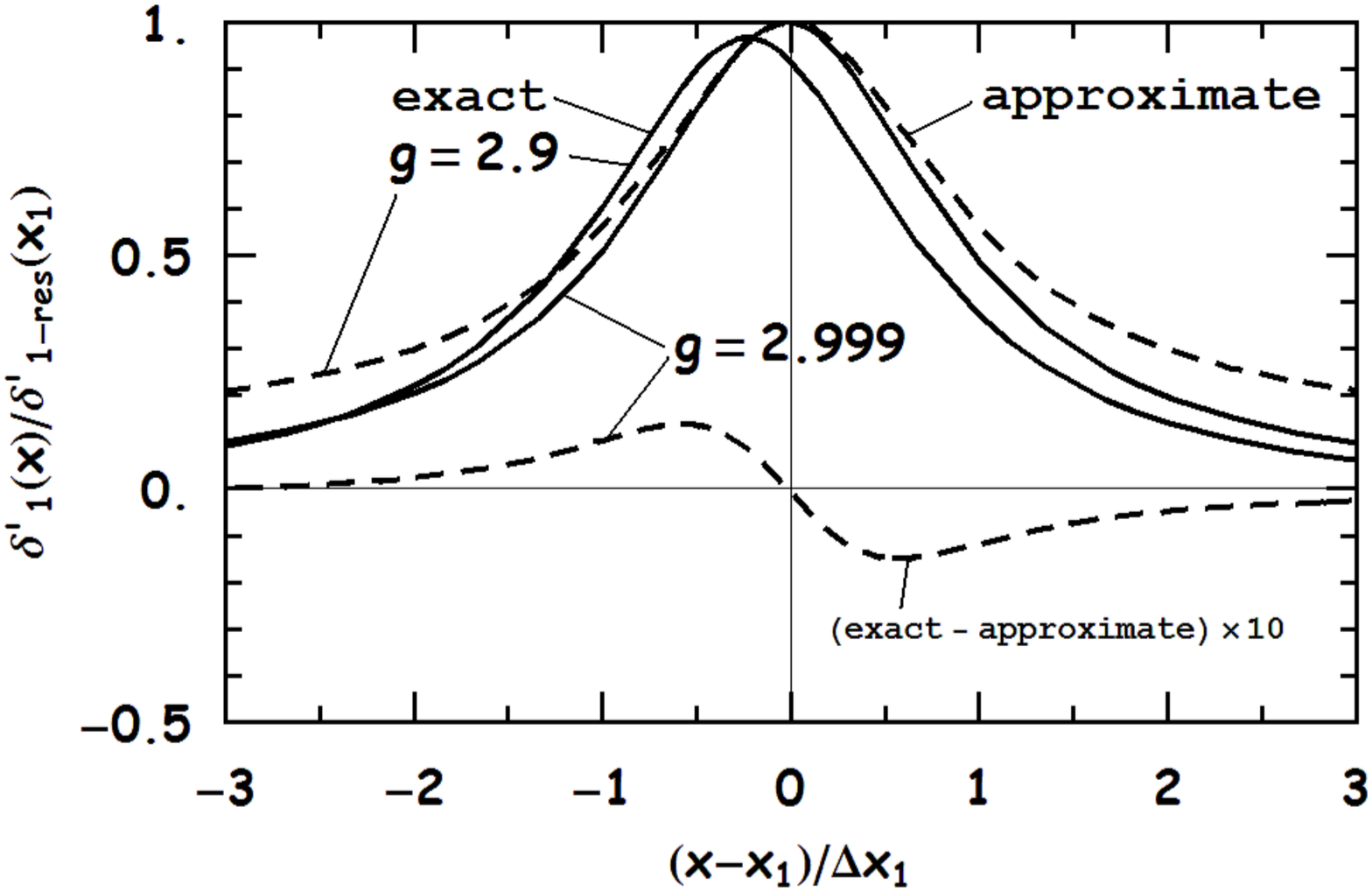}}
\vskip -20pt
\caption{An illustration of the P-wave phase shift derivative near a resonance.}
  \label{flg3}
\end{figure*}
Let us investigate the resonance behavior closely. We start with the S-wave first. If we expand the numerator and the denominator of Eq. (\ref{shift_dx_tan}) in a power series of $x$, we obtain
\begin{equation}
\delta'_{0-res}(x)\equiv\frac{3(g-1)g-(g-3)g x^2}{-3(g-1)^2+(g-4)g x^2} \, .
\label{ddeltaap}
\end{equation}
We can infer the peak value from this equation to be
\begin{equation}
\delta'_{0-res}(0)\equiv\frac{g}{1-g},
\label{ddeltaappk}
\end{equation}
and the width at half the peak
\begin{equation}
\Delta x\equiv \frac{2\sqrt{3(1-g)^2}}{\sqrt{6-g(4-g)}} \, .
\label{ddeltaapw}
\end{equation}
From the above relations and from Fig. \ref{fllf} and Fig. \ref{fllf2}, it is clear that as $g\to1$ the function becomes more like a Dirac-$\delta$ function ($\delta'_{0-res}(0)\to\infty$ and 
$\Delta x\to 0$, but $\delta'_{0-res}(0)\Delta x$ is a finite number),
so with the normalization given by the integral over Eq. (\ref{ddeltaap}), we have

\begin{equation}
\delta'_{0-res}(g\approx1)\approx \text{sign}(1-g)\pi\frac{3^{3/2}}{(4-g)^{3/2}\sqrt{g}}\delta(x)-\frac{(3-g)}{(4-g)},
\label{ddeltadirac}
\end{equation}
where $\text{sign}(0)\delta(x) \equiv0 $ and $1=\int_{-\infty}^{\infty}\delta(x)dx=2\int_{0}^{\infty}\delta(x)dx$ due to symmetry.

Now we consider the P-wave for which $l=1$. As we see from Fig. \ref{flg3}, when $g\approx3$ the position $x_1$ of a sharp resonance for which $\sin^2(\delta_1)=1$ from Eq. (\ref{sin2_d0}) is small and can be found from
\begin{equation}
\frac{1+g \, x \, j_1(x) \, n_1(x)}{g \, x \, j_1^2(x)}=0.
\label{x1_eq}
\end{equation}
By expanding in a Taylor series, the result is
\begin{equation}
x_1^2(g\approx3) \approx \left(\frac{3-g}{g-1}\right)\frac{5}{3}.
\label{x1_gl}
\end{equation}

Expanding the numerator and the denominator of Eq. (\ref{shift_dx_tan}) in a power series of $\epsilon=x-x_1$ from Eq. (\ref{x1_gl}), we obtain
\begin{equation}
\delta'_{1-res}(x)\equiv-\frac{360(5(g-3)-51 \epsilon^2)}{625 (g-3)^2+5184 \epsilon^2} \, ,
\label{ddelta1ap}
\end{equation}
with the peak value and the width given by
\begin{equation}
\delta'_{1-res}(x_1)\equiv\frac{72}{25(3-g)},
\label{ddelta1appk}
\end{equation}
\begin{equation}
\Delta x_1 \equiv \frac{25(3-g)}{72} \, .
\label{ddelta1apw}
\end{equation}
Again the resonance structure can be approximated by a Dirac-$\delta$ function similar to Eq. (\ref{ddeltadirac}):

\begin{equation}
\delta'_{1-res}(g\approx3)\approx H(3-g)\, \pi \left(1-\frac{2125}{1728}(3-g)\right) \, \delta\left(x-\sqrt{\frac{5(3-g)}{2 g}}\right)+\frac{85}{24},
\label{ddelta1dirac}
\end{equation}
where $H(\epsilon)$ is the unit step function: $H(\epsilon)=1$ for $\epsilon>0$ and $0$ otherwise.

\section {Energies of bound states}
\label{sec5}

If the energy is negative ($E<0$ implies that the wave number $k$ is pure imaginary), the system
is bound. The radial wave function in this case is found by requiring Eq.~(\ref{general_sln_psi}) to satisfy the conditions in Eq. (\ref{cond_psi_a}), Eq. (\ref{cond_u_0}) and Eq. (\ref{cond_psi_inf}). The result is
\begin{equation}
  \psi_l(r<R)=N_\psi \, j_l(k r) h_l(k R),
\label{psi_in}
\end{equation}
\begin{equation}
  \psi_l(r>R)=N_\psi \, j_l(k R) h_l(k r),
\label{psi_out}
\end{equation}
where $h_l=j_l+i \, n_l$ is the spherical Hankel function and $N_\psi$ is the normalization factor.
Bound-state energies are given by poles of the scattering amplitude in the complex energy plane.
Applying the boundary condition in Eq. (\ref{cond_dpsi_a}) yields an equation for the bound state energies\cite{Gottfried66}
\begin{equation}
  i \, x \, j_l(x) \, h_l(x)=\frac{1}{g},
\label{b_eq_x}
\end{equation}
where $x=k R$. In obtaining the above relation, the equality in Eq. (\ref{bessel_eq}) has been used. As $x$ is pure imaginary, it is convenient to introduce a real valued variable through $x=i y$. The right hand side of Eq. (\ref{b_eq_x}) then becomes a real function $f_l(y)$, which satisfies
\begin{equation}
  f_l(y)\equiv-y \, j_l(iy) \, h_l(iy).
\label{fl_def}
\end{equation}
For $l=0$, one has
\begin{equation}
  j_0(iy)=\frac{\sinh(y)}{y}, \,\,  h_0(iy)=-\frac{e^{-y}}{y}.
\label{j0_h0}
\end{equation}
For $l>0$, the recurrence formula
\begin{equation}
  b_{l+1}(iy)=i\left(-\frac{l}{y}b_l(iy)+\frac{db_l(iy)}{dy}\right),
\label{rec_fml}
\end{equation}
is useful, where $b$ is either $j$ or $h$. Eq. (\ref{rec_fml}) enables expressions of $f_l$ for
successive values of $l$ to be generated:
\begin{eqnarray}
\begin{array}{llll}
  f_0(y)=\sinh(y)\frac{e^{-y}}{y},\\
  f_1(y)=\left(\cosh(y)-\frac{\sinh(y)}{y}\right)(1+\frac{1}{y})\frac{e^{-y}}{y},\\
  f_2(y)=\left((\frac{3}{y^2}+1)\sinh(y)-\frac{3}{y}\cosh(y)\right)\left(1+\frac{3}{y}+\frac{3}{y^2}\right)\frac{e^{-y}}{y}.
\end{array}
\label{series_fl}
\end{eqnarray}

Eq. (\ref{b_eq_x}) is of the form
\begin{equation}
  f_l(y)=\frac{1}{g}.
\label{eqn_fl_g}
\end{equation}%
\begin{figure*}[tb]
\centerline{\includegraphics[width=12cm]{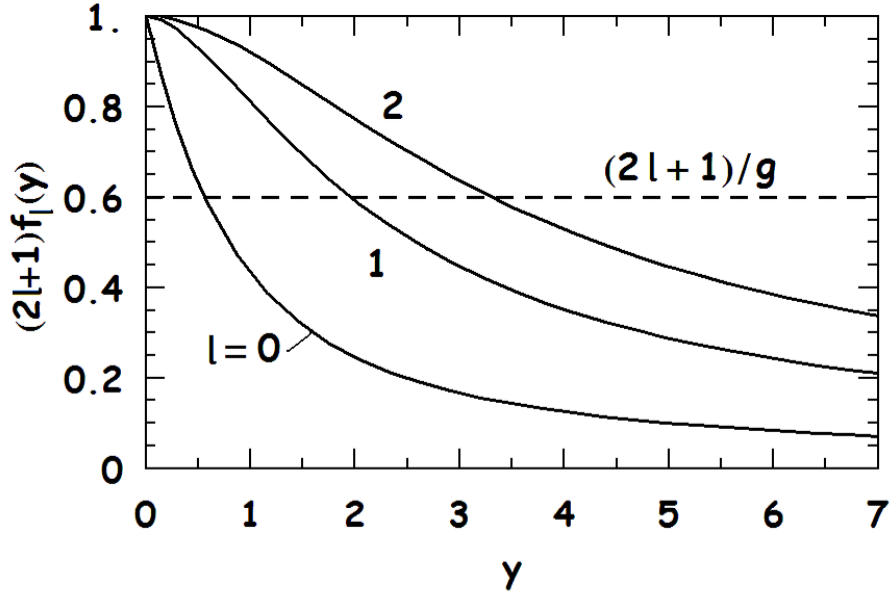}}
\vskip -20pt
\caption{An illustration of the determination of bound states using Eq. (\ref{eqn_fl_g}).}
  \label{ffl}
\end{figure*}
Written as in Eq. (\ref{eqn_fl_g}), the bound state energies are easily determined by a graphical procedure. Figure \ref{ffl} shows the main characteristic of the function $f_l$; it gradually decreases as $y$ increases. The value of $f_l$ at $y=0$ and the asymptotic form for $y\to \infty$ are found by appropriate expansions of the Bessel and Hankel functions\cite{Gottfried66}:
\begin{equation}
  f_l(0)=\frac{1}{2l+1},
\label{fl_0}
\end{equation}
\begin{equation}
  f_l(y\to \infty)\to\frac{1}{2 y}.
\label{fl_inf}
\end{equation}
The intersection of $f_l(y)$ with the constant line $\frac{1}{g}$ determines the bound state energy for a given value of $l$. The
magnitude of $g$ sets the total number of bound states. Explicitly,
\par $g<1$: no bound states at all,
\par $g\ge1$: there are several bound states from $l=0$ up to $l_{max}=[\frac{g-1}{2}]$. 
\par The notation $[q]$ means the integer part of $q$.

\begin{figure*}[tb]
\centerline{\includegraphics[width=12cm]{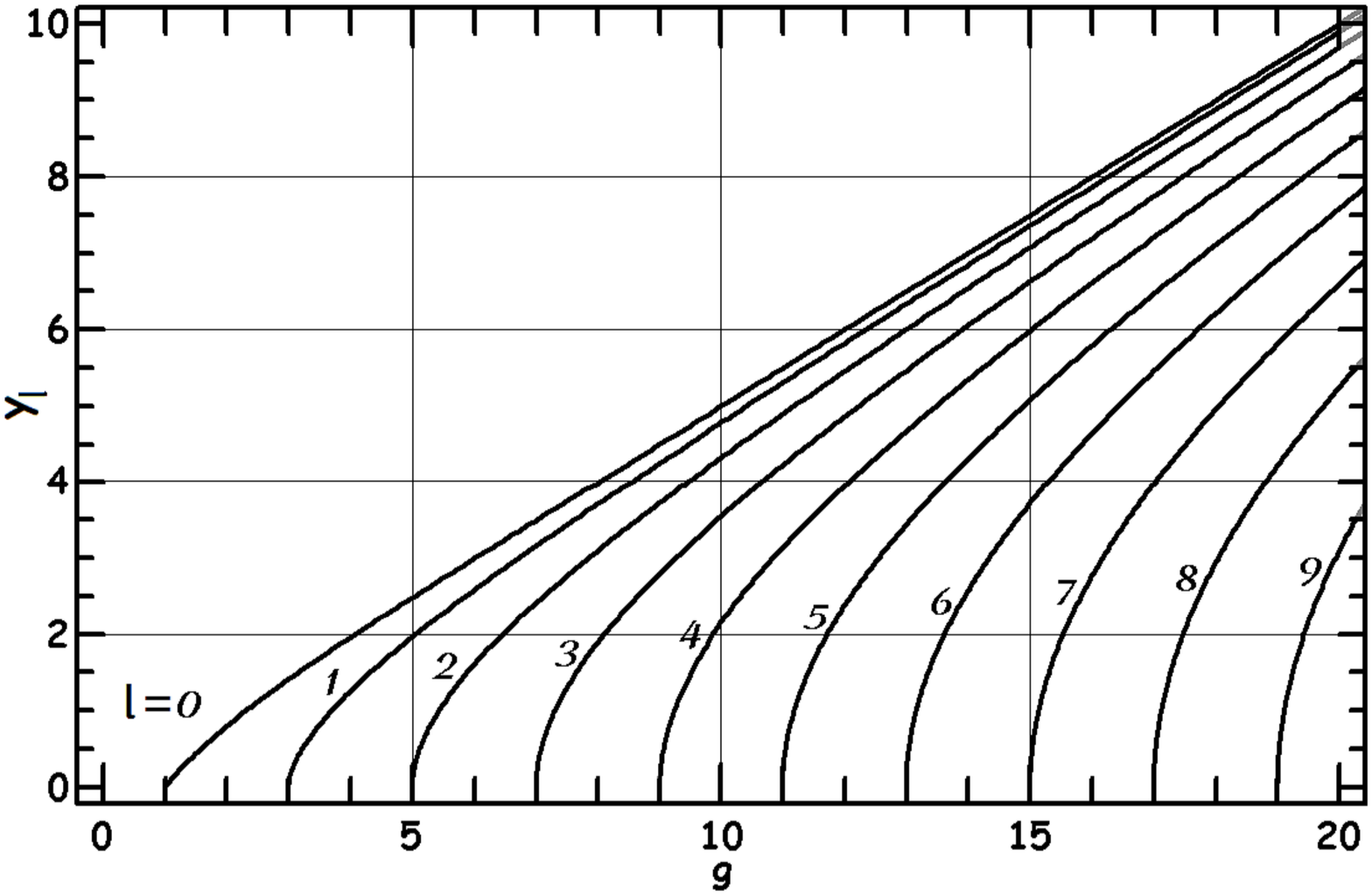}}
 \vskip -20pt
  \caption{Solutions $y_l(g)$ of the equation $f_l(y)=\frac{1}{g}$ for different  partial waves $l$ versus the strength parameter $g$.}
  \label{y_l}
\end{figure*}

The binding solutions $y_l$ for low values of $l=0,1,...$ are plotted in Fig. \ref{y_l} as a function of the strength parameter $g$. The bound
  state energies are given by $-\frac{\hbar^2 {y_l}^2}{2\mu R^2}$.
If $g\approx2l+1$, the energies are close to a resonance value.  When the energy is small,  the solution in Eq. (\ref{eqn_fl_g}) can be well approximated by a power series in $y$; the result is
\begin{equation}
	y_l^2(g\approx 2l+1)\approx
	\begin{cases}
	\left(\frac{3 g-\sqrt{3(8-5 g)g}}{4 g}\right)^2\approx\left(\frac{g-1}{g}\right)^2 & \text{for $l=0$,}\\
	\left(\frac{g-(2l+1)}{g}\right)\frac{(2l-1)(2l+3)}{2} & \text{for $l>0$.}
	\end{cases}
	\label{appyb}
\end{equation}
Also, for the case $g \gg 2l+1$, we have from Eq. (\ref{fl_inf})
\begin{equation}
	y_l(g \gg 2l+1)\approx \frac{g}{2}.
	\label{ybgg}
\end{equation}

\section{The scattering length, effective range and the delta-shell model of the deuteron}
\label{sec6}

The properties of the deuteron, the simplest nucleus consisting of a
neutron and a proton, are well measured in experiments. Some basic
properties include the scattering length $a_{sl}$, the effective range
$r_0$ and the shape parameter $P$ which are defined through an
expansion of $k\cot(\delta_0)$ as \cite{Sitenko75}
\begin{equation}
  k \, \cot(\delta_0)=-\frac{1}{a_{sl}}+r_0\frac{k^2}{2}-P \, r_0^3 k^4+O(k^6).
\label{asl_r0_P_def}
\end{equation}
For $l>0$ partial waves, Newton\cite{Newton66} generalizes the above to read as

\begin{equation}
  k^{2l+1} \, \cot(\delta_l)=-\frac{1}{a^{(l)}_{sl}}+r^{(l)}_0\frac{k^2}{2}+O(k^4).
\label{asll_r0_P_def}
\end{equation}

Expanding Eq. (\ref{tan_dl_gx}) for the delta-shell in a similar manner yields
\begin{equation}
k \, \cot(\delta_0)=\frac{1-g}{R \, g}+R\left(1+\frac{1}{g}\right)\frac{k^2}{3}+R^3\frac{(3+g)}{g}\frac{k^4}{45}+o(k^6),
\label{ctan}
\end{equation}
from which we infer:
\begin{equation}
  \text{scattering length: }\frac{a_{sl}}{R}=\frac{g}{g-1},
\label{asl}
\end{equation}
\begin{equation}
  \text{effective range: }\frac{r_0}{R}=\frac{2}{3}\left(1+\frac{1}{g}\right),
\label{r_0}
\end{equation}
\begin{equation}
  \text{shape parameter: }P=-\frac{3}{40}\frac{g^2(3+g)}{(1+g)^3}.
\label{P}
\end{equation}
Generalizing for $l\geq0$, the $l^{th}$-wave scattering length parameter $a^{(l)}_{sl}$ and $l^{th}$-wave range parameter $r^{(l)}_0$ are
\begin{equation}
  \frac{a^{(l)}_{sl}}{R^{2l+1}}=\frac{(2l+1)}{\left((2l+1)!!\right)^2}\left(\frac{g}{g-(2l+1)}\right),
\label{asll}
\end{equation}
\begin{equation}
  \frac{r^{(l)}_0}{R^{1-2l}}=\frac{2 \left((2l+1)!!\right)^2}{(2l+3)(2l-1)}\left(\frac{2l-1}{g}-1\right).
\label{rl_0}
\end{equation}

\begin{figure*}[tb]
\centerline{\includegraphics[width=12cm]{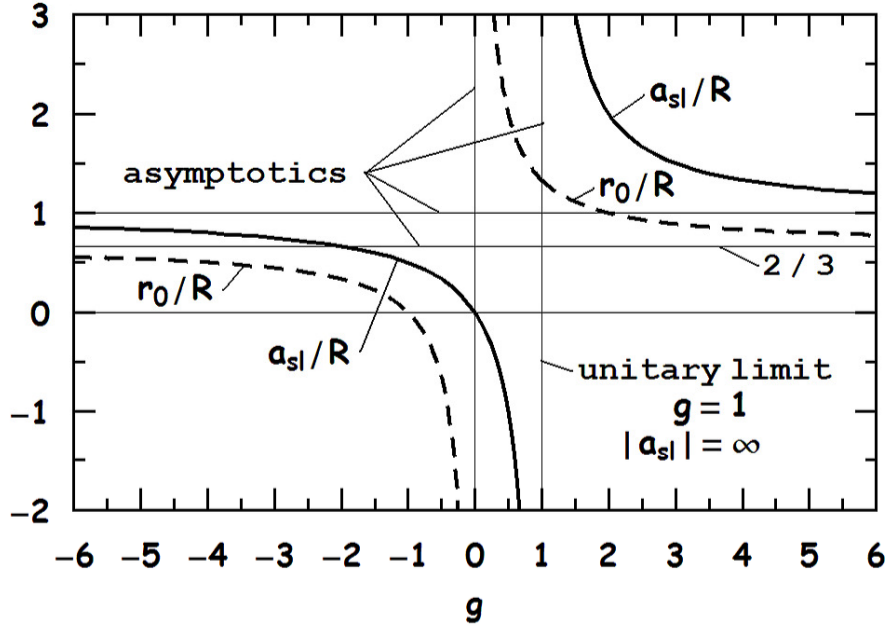}}
\vskip -20pt
  \caption{S-wave ($l=0$) scattering length and effective range as functions of the
  strength parameter $g$.}
  \label{asl_}
\end{figure*}

These results reflect the leading order behaviors of the partial wave cross sections and have multiple dimensions of length.
Fig. \ref{asl_} shows the scattering length and effective range as
functions of the strength parameter $g$.  It is worthwhile to note
that one can make the scattering length in Eq. (\ref{asll}) 
very large by choosing the value of $g$ very close to $2l+1$.

\begin{figure*}[tb]
\centerline{\includegraphics[width=12cm]{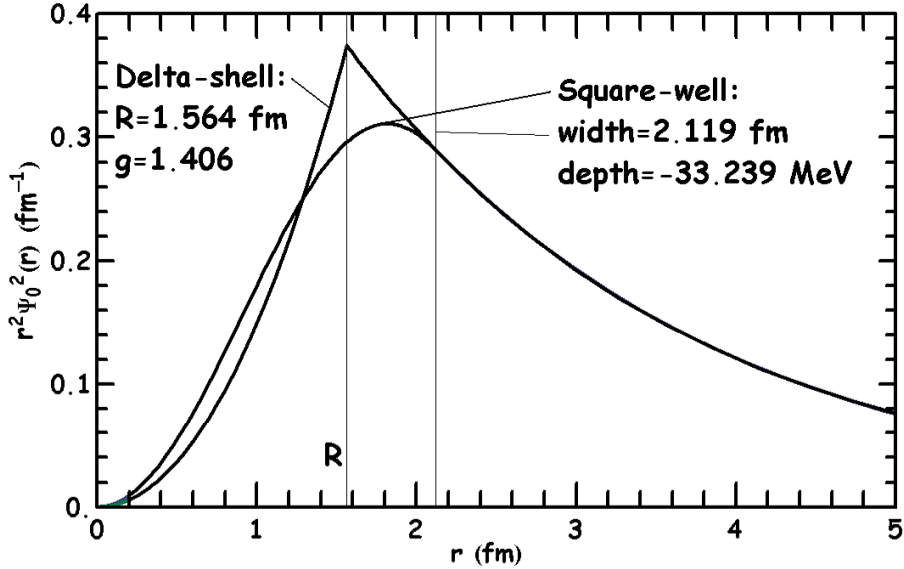}}
\vskip -20pt
 \caption{The deuteron probability density distributions for the delta-shell and square well models.}
  \label{psi_dsh_sqw}
\end{figure*}

In the triplet configuration, the deuteron is dominated by the S-state
and has a scattering length\cite{Epelbaum2004} of $5.42$ fm. One
can adjust the parameters $R$ and $g$ to get the binding energy to the
measured value\cite{Epelbaum2004} of $2.2246$ MeV. This fitting
results in the model parameters $R=1.564$ fm and $g=1.406$ or
$\Lambda=0.899$ fm$^{-1}$. With these numbers, expressions (\ref{r_0})
and (\ref{P}) give the effective range $r_0=1.79$ fm which is very
close to the experimental value\cite{Epelbaum2004} $1.76$ fm, but
the shape parameter $P=-0.047$ differs substantially from the experimental
value\cite{Epelbaum2004} of $-0.007$. Not unexpectedly, the delta
shell potential, being spherically symmetric and simplistic, differs from
the real potential probed at higher energies \cite{Moszkowski93}. The bound-state wave function can be
used to calculate the root mean square radius of the deuteron:
\begin{equation}
  r_{rms}\equiv\frac{1}{2}\sqrt{\frac{\int d^3 r \,r^2\, |\psi(r)|^2}{\int d^3 r \, |\psi(r)|^2}}.
\label{r_rms}
\end{equation}
The result is $1.95$ fm, which is close to the experimental
value\cite{Epelbaum2004} $1.97$ fm. The probability densities for
the delta-shell and square-well models of the deuteron are shown in
Fig. \ref{psi_dsh_sqw}. They have very similar distributions in the
outer region as expected from low-energy scattering in which the core
of the potential is obscured \cite{Sitenko75}.

\section{Differential equation for S-wave \\
scattering length and effective range}
\label{sec7}

Here, we derive a set of differential equations
for the numerical calculation of the S-wave scattering length and 
effective range following the method outlined by Fl\"{u}gge \cite{Flugge99}.
We first cast the potential as
\begin{equation}
\label{formV}
    V(r)=V_0 \, \phi(\xi),
\end{equation}
where $V_0$ has dimensions of energy and $\phi(\xi)$ is a dimensionless function of $\xi=\frac{r}{R}$,
with $R$ being a characteristic length scale of the potential.
    The S-wave radial Schr\"{o}dinger equation then reads
    \begin{equation}
    \label{difShsl}
        u''(\xi)+\left( g \, \phi(\xi)+x^2 \right) u(\xi)=0,
    \end{equation}
where $g=-\frac{2\mu V_0}{\hbar^2}R^2$, $k^2=\frac{2\mu E}{\hbar^2}$ and $x=k R$.
Let us expand the solution as a series in $x$:
\begin{equation}
    \label{seriesxsol}
    u(\xi)=u_0(\xi)+u_1(\xi)x+u_2(\xi)x^2+... \, .
\end{equation}
Substitution of Eq. (\ref{seriesxsol}) into Eq. (\ref{difShsl}) yields
\begin{equation}
\label{iterdifeqS}
\left\{
\begin{array}{ll}
            u_0''(\xi)+g \, \phi(\xi)u_0(\xi)=0,\\
            u_1''(\xi)+g \, \phi(\xi)u_1(\xi)=0,\\
            u_2''(\xi)+g \, \phi(\xi)u_2(\xi)+u_0(\xi)=0.
\end{array}
\right.
\end{equation}
The condition that the wave function is well behaved at the origin requires
\begin{equation}
	\label{u00bc}
	u(0)=0\,; 
\end{equation}
hence, from Eq. (\ref{seriesxsol}), it  follows that 
\begin{equation}
	\label{u0ibc}
	u_i(0)=0,
\end{equation}
where $i=0,1,2,...$.
Using the freedom of a particular normalization, we can choose
\begin{equation}
\label{iterICS} 
  u_{0}'(0)=1.
\end{equation}
%
Expansion of the asymptotic solution as $x\xi=k r \to \infty$ as a series in $\xi$ gives
\begin{equation}
\label{asymseries}
\begin{split}
    u(x\xi\to\infty)\to &A \sin(\delta_0+x\xi)=\\
    &A \sin(\delta_0)\left( 1+x \, \cot(\delta_0) \, \xi-\frac{x^2}{2}\xi^2-x \, \cot(\delta_0) \, \frac{x^2 \xi^3}{6}+... \right)=\\
    &C\left(1-\frac{\xi}{\alpha}\right)+C\left(\frac{\varrho}{2}\xi-\frac{\xi^2}{2}+\frac{\xi^3}{6 \alpha}\right)x^2+... \, ,
\end{split}
\end{equation}
and from Eq. (\ref{asll_r0_P_def}) we have
\begin{equation}
\label{xcot}
x\,\cot(\delta_0)=-\frac{1}{\alpha}+\frac{\varrho}{2}x^2+... \, ,
\end{equation}
where $\alpha=\frac{a_{sl}}{R}$, $\varrho=\frac{r_0}{R}$ and $C=A \sin(\delta_0)$ is a normalization constant.
Equation (\ref{iterdifeqS}) and Eq. (\ref{asymseries}) imply
\begin{equation}
\label{u0u2eta0eta2}
\begin{split}
    &u_0(x\xi\to\infty)\to C\left(1-\frac{\xi}{\alpha}\right),\\
    &u_1(x\xi\to\infty)\to 0,\\
    &u_2(x\xi\to\infty)\to C\left(\frac{\varrho}{2}\xi-\frac{\xi^2}{2}+\frac{\xi^3}{6 \alpha}\right).\\
\end{split}
\end{equation}
Therefore
\begin{equation}
\label{alphaxi}
    \alpha(\xi)\equiv \xi-\frac{u_0(\xi)}{u_0'(\xi)} \stackrel{\xi\to\infty}{\rightarrow} \alpha,
\end{equation}
\begin{equation}
\label{rhoaxi}
    \varrho(\xi)\equiv \xi-\frac{u_{20}(\xi)}{u_{20}'(\xi)} \stackrel{\xi\to\infty}{\rightarrow} \varrho,
\end{equation}
where $u_{20}(\xi)\equiv u_2'(\xi)+u_0(\xi)\frac{\xi}{2}$.
These formulas are useful for the numerical evaluation of the scattering length and the effective range through
integration of the set in Eq. (\ref{iterdifeqS}) with the initial conditions (\ref{u0ibc}) and (\ref{iterICS}).
Furthermore, in the case of the scattering length we can derive an alternative equation
by combining Eq. (\ref{alphaxi}), Eq. (\ref{iterdifeqS}) and Eq. (\ref{iterICS}):
\begin{equation}
\label{difSL}
\left\{
\begin{array}{ll}
    \alpha(0)=0,\\
    \alpha'(\xi)+(\xi-\alpha(\xi))^2 g \, \phi(\xi)=0\,.
\end{array}
\right.
\end{equation}
This can be integrated numerically for any spherically symmetric potential (for example, the Yukawa potential $\phi(\xi)=e^{-\xi}/\xi$),
but in the case of the delta-shell ($\phi(\xi)=\delta(\xi-1)$ and $V_0=-\frac{v}{R}$) an analytical solution can be obtained.
Our task here is to apply the method above to alternatively derive the scattering length for a delta-shell potential.
To begin with, the delta-shell function can be represented as
\begin{equation}
\label{deltasq}
    \delta(\xi-1)\to
    \begin{cases}
        n & |\xi-1|<\frac{1}{2n},\\
        0 & |\xi-1|\geq\frac{1}{2n},
    \end{cases}
\end{equation}
with the limit $n\to\infty$, taken at the end. With this representation the solution of Eq. (\ref{difSL}) for $\xi>1$ is
\begin{equation}
\label{solnasl}
    \alpha(\xi>1+\frac{1}{2n})=\frac{\left(4 n +g(4 n^2-1)\right)\sin(\sqrt{g/n})-4\sqrt{g n}\cos(\sqrt{g/n})}{2g n (2n-1)\sin(\sqrt{g/n})-4 n\sqrt{g n}\cos(\sqrt{g/n})},
\end{equation}
where we have used the continuity conditions
\begin{equation}
\label{asl1}
\alpha \left. \left(1-\frac{1}{2n}\right)\right|_{-0}=\alpha \left.\left(1-\frac{1}{2n}\right)\right|_{+0},
\end{equation}
\begin{equation}
\label{asl2}
\alpha \left. \left(1+\frac{1}{2n}\right)\right|_{-0}=\alpha \left.\left(1+\frac{1}{2n}\right)\right|_{+0},
\end{equation}
Finally, taking the limit as $n\to\infty$ we have
\begin{equation}
\label{liminfsl}
    \frac{a_{sl}}{R}=\lim_{n\to\infty}\alpha(\xi>1+\frac{1}{2n})=\frac{g}{g-1}
\end{equation}
assuming $g\neq1$. This coincides with the result in Eq. (\ref{asl}).

\section{Momentum-space wave-function \\
and form-factor for the S-wave}
\label{sec8}

The S-state wave-function with the bound state energy $E=-\frac{\hbar^2 \kappa^2}{2 \mu}$ has the form

\begin{equation}
\Psi_0(r<R)= N_0 e^{-\kappa R}\frac{\sinh(\kappa r)}{r},
\label{psi0in}
\end{equation}
\begin{equation}
\Psi_0(r>R)= N_0 \sinh(\kappa R)\frac{e^{-\kappa r}}{r},
\label{psi0out}
\end{equation}
where the normalization factor
\begin{equation}
N_0 = 2 \frac{e^{\kappa R}}{\sqrt{4 \pi}}\sqrt{\frac{\kappa}{e^{2 \kappa R}-2 \kappa R -1}}\,.
\label{N0}
\end{equation}
Using Eq. (\ref{r_rms}), the $rms$ radius turns out to be
\begin{equation}
r_{rms}=\frac{R}{2}\sqrt{1+\frac{1}{2 \kappa^2 R^2}+\frac{4 \kappa R}{3(e^{2 \kappa R}-2 \kappa R -1)}}.
\label{rms0}
\end{equation}

The wave-function in momentum space is given by
\begin{equation}
\Psi_0(k)=\frac{1}{(2 \pi)^{\frac{3}{2}}} \int d^3 r \, e^{i \vec{k} \vec{r}} \Psi_0(r).
\label{psi0k}
\end{equation}
Performing the angular integration yields

\begin{equation}
\Psi_0(k)=\frac{2}{k \sqrt{2 \pi}} \int_0^{\infty} d r \, r \sin(k r) \Psi_0(r)\,.
\label{psik0r}
\end{equation}
Utilizing Eq. (\ref{psi0in}) and Eq. (\ref{psi0out}), we obtain
\begin{equation}
\Psi_0(k)=\frac{\kappa e^{\kappa R} \sin(k R)}{\pi k (k^2+\kappa^2)}
\sqrt{\frac{2 \kappa}{e^{2 \kappa R}-2 \kappa R -1}},
\label{psik0c}
\end{equation}

\begin{figure*}[tb]
\centerline{\includegraphics[width=12cm]{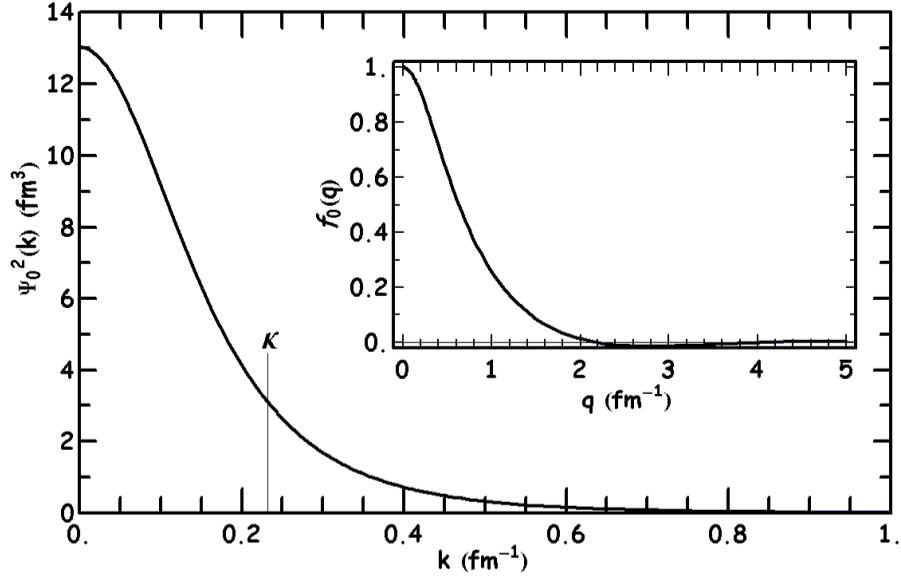}}
\vskip -20pt
 \caption{The delta-shell model momentum space wave-function and form-factor for the deuteron as a function of momentum $k$ and momentum transfer $q$. The second derivative of the curve at $q=0$ gives $r_{rms}=1.95$ fm.}
\label{psik}
\end{figure*}

The s-state elastic form-factor
\begin{equation}
f_0(q)=\int d^3 r \, e^{i \vec{q} \vec{r}} |\Psi_0(r)|^2=\frac{4 \pi}{q} \int_0^{\infty} d r \, r \sin(q r) |\Psi_0(r)|^2,
\label{f0q}
\end{equation}
where $q$ is the momentum transfer in the scattering process,
 can be calculated numerically (as for example by the Filon method \cite{Abramowitz70}). Fig. \ref{psik} shows the momentum space wave-function and the form factor for the delta-shell model of the deuteron.
Note that the relation between the slope of the form-factor for $q\to0$ and the $rms$ radius
\begin{equation}
r_{rms}=\frac{1}{2}\sqrt{-3 \frac{\partial^2 f_0}{\partial q^2}\bigr{|_{q=0}}}
\label{f0sl}
\end{equation}
can be used as a check of the numerical procedure used to calculate Eq. (\ref{f0q}).

\section{The Feynman-Hellmann theorem applied to the delta-shell model}
\label{sec9}

The Feynman-Hellmann\cite{Feynman39,Hellmann37} theorem allows us to study the dependence of the bound
state energy $E$ on the parameters $v$ and $R$ of the delta-shell potential.
The theorem asserts that
\begin{equation}
\label{dHdl}
    \frac{\partial E}{\partial \lambda}=\left<\frac{\partial H}{\partial \lambda}\right>_{\Psi_{lm}},
\end{equation}
where the angular brackets denote an expectation value in the basis of wavefunctions $\Psi$, and $\lambda$ is any parameter in the Hamiltonian $H$.

In contrast to the commonly studied cases of $V(r)\propto r^q (q>-2)$,
the singular behavior of the delta-shell interaction $V(r)=-v \, \delta(r-R)$ requires
special treatment.

Choosing $R$ as the parameter,
\begin{equation}
\label{dHda}
\begin{split}
    \frac{\partial E}{\partial R}=\left<\frac{\partial V(r)}{\partial R}\right>_{\Psi_{lm}}
    &=-v \, \left<\frac{\partial}{\partial R}\delta(r-R)\right>_{\Psi_{lm}=\psi(r) Y^l_m(\Omega)}
    =v \, \left<\frac{\partial}{\partial r}\delta(r-R)\right>_{\psi(r)}\\
    &=v \kappa \, \left<\frac{\partial}{\partial \rho}\delta(\rho-y)\right>_{u(\rho=\kappa r)}
    \equiv v \kappa \, \left<\delta'(\rho-y)\right>,
\end{split}
\end{equation}
where
\begin{equation}
\label{expect_psi}
    \left<F(r)\right>_\psi\equiv\int_0^\infty \psi F(r) \psi \, r^2 dr,
\end{equation}
and
\begin{equation}
\label{expect_u}
   \left<F(\rho)\right>=\left<F(\rho)\right>_u\equiv\int_0^\infty u F(\rho) \, u \, d\rho.
\end{equation}

To facilitate easy manipulations, a Gaussian representation of the delta function is helpful:
\begin{equation}
\label{deltan}
    \delta_n(r-R)=\frac{1}{2\sqrt{\pi n}}e^{-(r-R)^2/n}\xrightarrow{n\to\infty}\delta(r-R),
\end{equation}
using which one can easily check the validity of  Eq. (\ref{dHda}).

To start with, consider $l=0$ (s-wave) for which the Schr\"odinger Eq. (\ref{radial_eqn_u})
becomes
\begin{equation}
\label{eqn_u_0}
u_{0n}''(\rho)=\left( 1-\frac{\Lambda}{\kappa} \, \delta_n(\rho-y) \right) u_{0n}(\rho),
\end{equation}
where $u_{0n}(\rho)$ are continuous. Note first the identity
\begin{equation}
\label{int_du_ddu}
    \int_0^\infty u_{0n}' u_{0n}'' d\rho =\int_0^\infty d\left(\frac{[u_{0n}']^2}{2}\right)
=-\frac{[u_{0n}'(0)]^2}{2}.
\end{equation}
Alternatively, multiplying Eq. (\ref{eqn_u_0}) by $u'_{0n}$ and integrating both sides of Eq. (\ref{eqn_u_0}) we obtain
\begin{equation}
\label{int_du_eqn}
\begin{split}
    \int_0^\infty u_{0n}' u_{0n}'' d\rho & =\int_0^\infty \left(1-\frac{\Lambda}{\kappa} \, \delta_n(\rho-y)\right)d\left(\frac{[u_{0n}]^2}{2}\right)\\
& =-\frac{[u_{0n}(0)]^2}{2}+\frac{\Lambda}{2 \kappa} \int_0^\infty u_{0n} \delta'_n(\rho-y) u_{0n} \, d\rho=
\frac{\Lambda}{2 \kappa}\left< \delta'_n(\rho-y) \right>_{0n},
\end{split}
\end{equation}
where the condition in Eq. (\ref{cond_u_0}) has been used. In the limit $n\to\infty$,
Eq. (\ref{int_du_ddu}) and Eq. (\ref{int_du_eqn}) imply that
\begin{equation}
\label{ddelta}
    \left< \delta'(\rho-x) \right>_0= - \kappa \frac{[u_{0}'(0)]^2}{\Lambda}.
\end{equation}

For the case of $l>0$, the Schr\"odinger Eq. (\ref{radial_eqn_u})
takes the form
\begin{equation}
\label{eqn_u_l}
u_{ln}''(\rho)=\left(\frac{l(l+1)}{\rho^2}+1-\frac{\Lambda}{\kappa} \, \delta_n(\rho-y)\right)u_{ln}(\rho).
\end{equation}
In this case, Eq. (\ref{int_du_eqn}) becomes
\begin{equation}
\label{int_dul_eqn}
\begin{split}
    \int_0^\infty u_{ln}' u_{ln}'' d\rho & =\int_0^\infty \left(\frac{l(l+1)}{\rho^2}+1-\frac{\Lambda}{\kappa} \, \delta_n(\rho-y)\right)d\left(\frac{[u_{ln}]^2}{2}\right)\\
& =-l(l+1)\frac{[u_{ln}(0)]^2}{2\rho^2}+l(l+1)\int_0^\infty u_{ln} \frac{1}{\rho^3}
u_{ln} \, d\rho +
\frac{\Lambda}{2 \kappa}\left< \delta'_n(\rho-y) \right>_{ln}\\
& =l(l+1)\left<\rho^{-3} \right>_{ln}+\frac{\Lambda}{2 \kappa}\left< \delta'_n(\rho-y) \right>_{ln},
\end{split}
\end{equation}
where $\left<\rho^{-3} \right>$ is the third inverse moment. The first term in the
second line vanishes because $u_{ln}(\rho\to0)\propto\rho^{l+1}$(this is the regular solution given by Eq. (\ref{eqn_u_l}) in the limit $\rho\to0$). Thus for $l>0$, the left hand side of Eq. (\ref{int_dul_eqn}), using Eq. (\ref{int_du_ddu}), becomes zero as $u'_{ln}(\rho\to0)\propto\rho^{l}$.
Taking the limit $n\to\infty$, we obtain
\begin{equation}
\label{ddeltal}
    \left< \delta'(\rho-y) \right>_l=-2 \kappa \frac{l(l+1)}{\Lambda}\left<\rho^{-3} \right>_{l}.
\end{equation}

Eq. (\ref{ddelta}) and Eq. (\ref{ddeltal}) together with Eq. (\ref{dHda}) yield
\begin{equation}
\label{dEda}
 \frac{\partial E}{\partial R}=2 E \cdot
 \begin{cases}
     \frac{[u_{0}'(0)]^2}{2} & \text{$l=0$,}\\
     l(l+1)\left<\rho^{-3} \right>_{l} & \text{$l_{max}\geq l>0$,}
 \end{cases}
\end{equation}
where $l_{max}=\left[\frac{g-1}{2}\right]$ is the largest angular momentum allowed in order for a bound state to form
when the parameter $g$ is fixed (see \ref{sec5}).

Choosing $v$ as the parameter, the theorem in Eq. (\ref{dHdl}) gives
\begin{equation}
\label{dHdv}
    \frac{\partial E}{\partial v}=\left<\frac{\partial V(r)}{\partial v}\right>_{\Psi_{lm}=\Psi_l(r) Y^l_m(\Omega)}=-\left<\delta(r-R)\right>_{\Psi_l(r)}=-R^2 \Psi^2_l(R).
\end{equation}
The expressions derived using the Feynman-Hellmann theorem will be utilized to advantage in subsequent sections.

\section{Wave function normalization for bound states}
\label{sec10}

The Feynmann-Helmann theorem Eq. (\ref{dHdl}) can be used to normalize the $l\neq0$ partial waves.
If the parameter $g$ is held fixed while varying the parameter $R$, then the radial wave function of the bound state takes the form
\begin{equation}
\label{norm_psi}
    \Psi_l(r)=\frac{N_l(y)}{R^{3/2}}\cdot
    \begin{cases}
    h_l(i y) \, j_l(i y \frac{r}{R}) & ~{\rm for}~~r<R,\\
    j_l(i y) \, h_l(i y \frac{r}{R}) & ~{\rm for}~~r>R,
    \end{cases}
\end{equation}
where $y$ is the solution of Eq. (\ref{eqn_fl_g}); note that now $y$ is fixed too.
In this case, the partial derivative of the wave function Eq. (\ref{norm_psi}) with respect
to the parameter $R$ (keeping $g$ fixed) is given by
\begin{equation}
\label{dpsida}
    \left(\frac{\partial\Psi_l}{\partial R}\right)_g = -\frac{3}{2 R}\Psi_l+\frac{N_l(y)}{R^{3/2}}\cdot
    \begin{cases}
    h_l(i y) \, (-\frac{l}{R}j_l(i y \frac{r}{R})+\frac{i y r}{R^2}j_{l+1}(i y \frac{r}{R}))\,; &~ r<R\\
    j_l(i y) \, (-\frac{l}{R}h_l(i y \frac{r}{R})+\frac{i y r}{R^2}h_{l+1}(i y \frac{r}{R}))\,; & ~r>R \,,
    \end{cases}
\end{equation}
where the recurrence formula in Eq. (\ref{rec_fml}) was used. In a compact form
\begin{equation}
\label{dpsida2}
    \left(\frac{\partial\Psi_l}{\partial R}\right)_g = -\frac{3+2 l}{2 R}\Psi_l+y r\frac{N_l(y)}{R^{7/2}}\cdot
    \begin{cases}
    i h_l(i y) \, j_{l+1}(i y \frac{r}{R}) & ~{\rm for}~~r<R\,\\
    i j_l(i y) \, h_{l+1}(i y \frac{r}{R}) & ~{\rm for}~~r>R\,.
    \end{cases}
\end{equation}
The next step is to use the equality
\begin{eqnarray}
\label{eqty_pa}
    \left(\frac{\partial}{\partial R}\left<\Psi_l\left|\delta(r-R)\right|\Psi_l\right>\right)_g & =
\left< \left( \frac{\partial\Psi_l}{\partial R} \right)_g \right|\delta(r-R)\left| \Psi_l \right>+
\left< \Psi_l \right|\delta(r-R)\left| \left( \frac{\partial\Psi_l}{\partial R} \right)_g \right>\nonumber \\
& + \left< \Psi_l \right| \left( \frac{\partial \delta(r-R)}{\partial R} \right)_g \left| \Psi_l \right> \,,
\end{eqnarray}
which follows from the chain rule for derivatives. First, we calculate the left hand side of Eq. (\ref{eqty_pa}):
\begin{eqnarray}
\label{eqty_LHS}
\left(\frac{\partial (R^2 \Psi_l^2(r=R))}{\partial R}\right)_g = -R \Psi_l^2(r=R),
\end{eqnarray}
since $R^2 \Psi_l^2(r=R)\propto \frac{1}{R}$ as follows from Eq. (\ref{norm_psi}). For the calculation of the first
two terms on the right hand side of Eq. (\ref{eqty_pa}), we make use of Eq. (\ref{norm_psi}) and Eq. (\ref{dpsida})
to obtain
\begin{eqnarray}
\label{psideltadpsi}
\left\langle \Psi_l \right| \delta(r-R) \left| \left( \frac{\partial\Psi_l}{\partial R} \right)_g \right\rangle=
-\frac{3+2 l}{2} \, R \Psi^2_l(r=R)+ \nonumber \\
y \frac{N^2_l(y)}{R^{2}} h_l(i y) j_l(i y) \cdot 
    \begin{cases}
    i h_l(i y) \, j_{l+1}(i y)\,; & ~r=R-0\,,\\
    i j_l(i y) \, h_{l+1}(i y)\; & ~r=R+0\,.
    \end{cases}
\end{eqnarray}
Note that the difference between the expressions inside the curly bracket is equal to $\frac{1}{y^2}$
from the relations in Eq. (\ref{bessel_eq}) and Eq. (\ref{rec_fml}). The discontinuity at $r=R$ must be treated
with some care. The average of the left and right limiting values as $r\to R \pm 0$ is
\begin{eqnarray}
\label{psideltadpsim}
\left\langle  \Psi_l \right|\delta(r-R)\left| \left( \frac{\partial\Psi_l}{\partial R} \right)_g \right\rangle=
-\frac{3+2 l}{2} \, R \Psi^2_l(r=R)+ \nonumber \\ y \frac{N^2_l(y)}{R^{2}} h_l(i y) j_l(i y)
 \left(i h_l(i y) \, j_{l+1}(i y) - \frac{1}{2 y^2}\right),
\end{eqnarray}
or equivalently,
\begin{eqnarray}
\label{psideltadpsim2}
\left\langle  \Psi_l \right|\delta(r-R)\left| \left( \frac{\partial\Psi_l}{\partial R} \right)_g \right\rangle=
-\frac{3+2 l}{2} \, R \Psi^2_l(r=R)+ \nonumber \\ y \frac{N^2_l(y)}{R^{2}} h_l(i y) j_l(i y)
 \left(i j_l(i y) \, h_{l+1}(i y) + \frac{1}{2 y^2}\right).
\end{eqnarray}
Finally, we substitute Eq. (\ref{psideltadpsim}) and Eq. (\ref{eqty_LHS}) into Eq. (\ref{eqty_pa})
to get
\begin{equation}
\label{psiddelpsi}
\begin{split}
	\left\langle  \Psi_l \right| \left( \frac{\partial \delta(r-R)}{\partial R} \right)_g \left| \Psi_l \right\rangle=
2(l+1)\, R \Psi^2_l(r=R)-&\\
\frac{N^2_l(y)}{y R^{2}} h_l(i y) j_l(i y)&\left(2 i y^2 h_l(i y) \, j_{l+1}(i y) - 1\right).
\end{split}
\end{equation}
Now the Feynman-Hellmann theorem can be applied in the form
\begin{equation}
\label{dHdaN}
\begin{split}
    \left(\frac{\partial E(R)}{\partial R}\right)_g=\left<\left( \frac{\partial V(r)}{\partial R}\right)_g \right>=
-v(R) \left<\left( \frac{\partial \delta(r-R)}{\partial R}\right)_g \right>
-\frac{\partial v(R)}{\partial R} \left< \delta(r-R) \right>,
\end{split}
\end{equation}
where $v(R)=\frac{\hbar^2 g}{2 \mu R}$ from Eq. (\ref{gdef}), and $E(R)=-\frac{\hbar^2 y^2}{2 \mu R^2}$
by the definition of $y$. Evaluating the derivatives above and the expectation value of the delta function, we find
\begin{equation}
\label{FHN2}
    -2 \frac{y^2}{g R^2}=\left<\left( \frac{\partial \delta(r-R)}{\partial R}\right)_g \right>-R \Psi^2_l(r=R),
\end{equation}
where in the intermediate step we divided the whole expression by $v$ and used the definition of $g$ in Eq. (\ref{gdef}).
Inserting Eq. (\ref{psideltadpsi}) into Eq. (\ref{FHN2}) and using the explicit form of the wave function in Eq. (\ref{norm_psi})
 gives the equation for normalization $N^2_l(y)$ as
\begin{equation}
\label{FHN22}
    2 \frac{y^2}{g}=(2l+1)N_l^2(y) h_l^2(i y) j_l^2(i y)-
\frac{N^2_l(y)}{y} h_l(i y) j_l(i y) \left(2 i y^2 h_l(i y) \, j_{l+1}(i y) - 1 \right),
\end{equation}
where $R$ cancels out as expected, as $N_l(y)$ is independent of $R$ for $g$ held fixed.
Recalling that $-y j_l(i y) h_l(i y)=\frac{1}{g}$ as the equation for the bound states (\ref{eqn_fl_g}),
we can solve the above equation for the normalization constant $N_l(y)$ with the result
\begin{equation}
\label{Nl2}
   N_l^2(y)=\frac{2 y^4}{1+(2l+1)y \, j_l(i y)h_l(i y)-2 y^2 \, i j_{l+1}(i y)h_l(i y)}.
\end{equation}
Eq. (\ref{fl_def}) allows to rewrite the above result in the shorter form
\begin{equation}
\label{Nl2s}
   N_l^2(y)=\frac{2 g y^4}{g-(2l+1)-2 g y^2 \, i j_{l+1}(i y)h_l(i y)}.
\end{equation}
A quick way to calculate the normalization constant $N_l(y)$ is to use Eq. (\ref{dHdv}).
The partial derivative of the energy with respect to the parameter $v$ is easily calculated from Eq. (\ref{eqn_fl_g}):
\begin{equation}
\label{dEdv}
    \frac{\partial E}{\partial v}=-2\frac{E}{y}\frac{\partial y}{\partial v}
    =2\frac{E}{v}\left(\frac{j_l(iy) h_l(iy)}{j_l(iy) h_l(iy)+y\left(\frac{d j_l(iy)}{d y} h_l(iy)+j_l(iy) \frac {d h_l(iy)}{d y}\right)}\right).
\end{equation}
Application of Eq. (\ref{rec_fml}) and Eq. (\ref{norm_psi}) leads to the same expression as Eq. (\ref{Nl2}).
Also, this result coincides with the expression obtained by a straightforward integration of
the square of the wave function Eq. (\ref{norm_psi}) using Lommel's integral \cite{Arfken2000}
\begin{equation}
\label{Lommels}
    (\alpha^2-\beta^2)\int_{x_0}^{x_1} x B_n(\alpha x) B_n(\beta x) dx = x \left.\left[\beta B_n(\alpha x) B_{n-1}(\beta x)-
    \alpha B_{n-1}(\alpha x) B_{n}(\beta x)\right] \right|_{x_0}^{x_1}\,,
\end{equation}
where $B_n$ is the Bessel function of either the first or the second kind.

\section{Virial theorem}
\label{sec11}
\begin{figure*}[tb]
\centerline{\includegraphics[width=12cm]{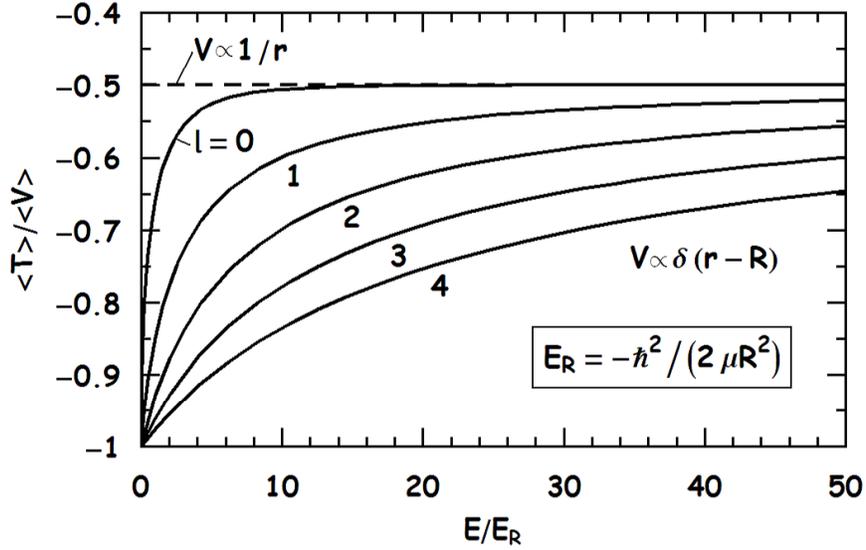}}
\vskip -20pt
\caption{Ratio of the average kinteic and  potential energies (from the   
virial theorem) for the delta-shell potential.  The horizontal dashed line shows the result for $V(r) \propto \frac 1r$.} 
  \label{virialtheorem}
\end{figure*}
The virial theorem relates the expectation value of the kinetic energy $\langle T \rangle$ with the expectation value of the potential energy $\langle V \rangle$ for a system in a bound state. For the well known power law potentials $V_n(r)\propto r^n$, the virial theorem gives 
\begin{equation}
\label{VTpower}
    \frac{\langle T \rangle}{\langle V_n \rangle}=\frac{n}{2}.
\end{equation}
In the following, we establish the virial relation for the delta-shell potential. First, from the Hamiltonian $H=T+V$
it is follows that
\begin{equation}
\label{HVT}
   E=\langle T \rangle + \langle V \rangle,
\end{equation}
where $E=-\frac{\hbar^2 y^2}{2 \mu R^2}$ is a bound state energy. Next, we have
\begin{equation}
\label{V}
   \langle V(r) \rangle= -v \langle \delta(r-R) \rangle_{\Psi_l(r)} =\frac{-v N_l^2(y)}{R \, g(y)^2 \, y^2}=\frac{N_l^2(y)}{g(y) \, y^4}E,
\end{equation}
from Eq. (\ref{Nl2s}), Eq. (\ref{eqn_fl_g}), Eq. (\ref{gdef}) and Eq. (\ref{norm_psi}). The kinetic term is found from Eq. (\ref{HVT})
\begin{equation}
\label{T}
  \langle T \rangle = E- \langle V \rangle=\left(1-\frac{N_l^2(y)}{g(y) \, y^4}\right)E.
\end{equation}
Hence
\begin{equation}
\label{VTdelta}
  \frac{\langle T \rangle}{\langle V \rangle}=\left(\frac{g(y)\, y^4}{N_l^2(y)}-1\right),
\end{equation}
where $g(y)=1/f_l(y)$ is a bound state energy from Eq. (\ref{eqn_fl_g}) and $f_l(y)$ is defined in Eq. (\ref{fl_def}).
Explicitly

\begin{equation}
\label{exp_virth}
    \frac{\langle T \rangle}{\langle V \rangle}=-\frac{1}{2}\left( 2l+3+\frac{1}{y j_l(iy) h_l(iy)}-2 iy \frac{j_{l+1}(iy)}{j_l(iy)} \right).
\end{equation}

The result is energy and angular momentum dependent as shown in Fig. \ref{virialtheorem}.
For deeply bound states ($y\to \infty$), the result asymptotically behaves as in the case of the Coulomb potential $\frac{1}{r}$ approaching the value of $-1/2$.

\section{Relationship between moments:\\
Kramers-Pasternak-like relation}
\label{sec12}

Kramers \cite{Kramers57} and Pasternak \cite{Pasternak37}
independently noted that the various moments $\langle
\rho^{p}\rangle\equiv\int_0^\infty u(\rho) \rho^p u(\rho)~ d\rho$ of
the bound state wave functions of the hydrogen atom (for which the Bohr
radius  provides the natural length scale) were related according
to
\begin{equation}
\label{KramersH}
    (p+1)y \langle \rho^{p} \rangle-(2 p+1)\langle \rho^{p-1} \rangle+
\frac{p}{4}y \left((2l+1)^2-p^2\right)\langle \rho^{p-2} \rangle=0,
\end{equation}
where $p\geq-(2l+1)$.

Here we present Kramers-Pasternak-like relations that connect the various
moments of the wave function for the delta-shell potential. It must be
noted that, unlike for the hydrogen atom, the wave functions exhibit a
kink at the shell size $R$ of the potential, and some care must be
taken to enforce the jump conditions.

Let us consider the following expectation value in the basis of wavefunctions $u(\rho)$
\begin{equation}
\label{dpexp}
\begin{split}
    \langle \rho^{s}\delta'(\rho-y) \rangle=\int_0^\infty u(\rho)
    \rho^{s} \frac{\partial \delta(\rho-y)}{\partial \rho} u(\rho) d\rho=\\
    \int_0^\infty \frac{\partial}{\partial \rho}\left(u^2(\rho)
    \rho^{s}\right) \delta(\rho-y) d\rho=\\
    2 u(y)u'(y) y^s + s \, y^{s-1} \, u^2(y)=y^s \langle \delta'(\rho-y) \rangle+s \, y^{s-1} \, u^2(y),
\end{split}
\end{equation}
where $s$ is an integer. Therefore we obtain
\begin{equation}
\label{n0case}
    \langle \rho^{s} \delta' \rangle=y^{s-1}(s \, u^2(y)+y \langle \delta' \rangle).
\end{equation}
Next we define a moment of order $q$ as
\begin{equation}
\label{momentsdef}
    \langle \rho^{q}\rangle\equiv\int_0^\infty u(\rho) \rho^q u(\rho) d\rho,
\end{equation}
with the requirement $q\geq-(2l+2)$ assuring the convergence of the integral (\ref{momentsdef}).

Now we return to the radial Schr\"{o}dinger equation (\ref{eqn_u_l}).
Integration of Eq. (\ref{momentsdef}) by parts yields the useful formula
\begin{equation}
\label{usefulmoment}
    \int_0^\infty u_n(\rho) \rho^{q+1} u_n'(\rho) d\rho=-\frac{(q+1)}{2}\langle \rho^{q}\rangle_n.
\end{equation}
Multiplying Eq. (\ref{eqn_u_l}) by $u_n(\rho) \rho^{p-1}$ and integrating over $\rho$, we get
\begin{equation}
\label{LHSmoments}
    \int_0^\infty u_n(\rho) \rho^{p-1} u_n^{\prime\prime}(\rho) d\rho=l(l+1)\langle \rho^{p-3} \rangle_n+
    \langle \rho^{p-1} \rangle_n-\frac{g}{y}\langle \rho^{p-1} \delta_n(\rho-y) \rangle_n,
\end{equation}
where $p>-2l$. On the other hand, we can integrate by parts to get
\begin{equation}
\begin{split}
\label{RHSmoments1}
    \int_0^\infty u_n \rho^{p-1} d u_n^{\prime}=&-(p-1)\int_0^\infty u_n \rho^{p-2} u_n^{\prime}d\rho
    -\int_0^\infty (u_n^{\prime})^2 \rho^{p-1}d\rho\\
    =&-\int_0^\infty (u_n^{\prime})^2 \rho^{p-1}d\rho+\frac{1}{2}(p-1)(p-2)\langle \rho^{p-3} \rangle_n,
\end{split}
\end{equation}
where we have used Eq. (\ref{usefulmoment}) in the last step. Proceeding further with integration by parts
\begin{equation}
\begin{split}
\label{RHSmom2}
    \frac{1}{p}\int_0^\infty (u_n^{\prime})^2 d\rho^{p}=-\frac{2}{p}\int_0^\infty u_n^{\prime}\rho^{p}u_n^{\prime\prime}d\rho&\\
    =-\frac{2l(l+1)}{p}\int_0^\infty u_n \rho^{p-2} u_n^{\prime}d\rho-\frac{2}{p}\int_0^\infty u_n \rho^{p} u_n^{\prime}d\rho
    +\frac{g}{p y}\int_0^\infty  \rho^{p}\delta_n(\rho-y) d u_n^2&\\
    =l(l+1)\frac{p-2}{p}\langle \rho^{p-3} \rangle_n+\langle \rho^{p-1}\rangle_n-\frac{g}{y}\langle \rho^{p-1} \delta_n(\rho-y) \rangle_n,
    -\frac{g}{p y}\langle \rho^{p} \delta'_n(\rho-y) \rangle_n&.
\end{split}
\end{equation}
Hence
\begin{equation}
\begin{split}
\label{RHSmom3}
    \int_0^\infty u_n \rho^{p-1} d u_n^{\prime}=\frac{1}{2}(p-1)(p-2)\langle \rho^{p-3} \rangle_n&\\
    -l(l+1)\frac{p-2}{p}\langle \rho^{p-3} \rangle_n
    -\langle \rho^{p-1}\rangle_n&
    +\frac{g}{y}\langle \rho^{p-1} \delta_n(\rho-y) \rangle_n
    +\frac{g}{p y}\langle \rho^{p} \delta'_n(\rho-y) \rangle_n.
\end{split}
\end{equation}
Setting Eq. (\ref{RHSmom3}) equal to Eq. (\ref{LHSmoments}), we have
\begin{equation}
\begin{split}
\label{RHSeqLHS}
    \langle \rho^{p} \delta'_n(\rho-y) \rangle_n=\frac{y(p-1)}{g}\left( 2l(l+1)-\frac{p}{2}(p-2)\right)\langle \rho^{p-3} \rangle_n \\
    +\frac{2 y p}{g}\langle \rho^{p-1} \rangle_n-2 p \langle \rho^{p-1} \delta_n(\rho-y) \rangle_n\,.
\end{split}
\end{equation}
In the limit $n\to\infty$,
\begin{equation}
\label{ddeltarhop}
    \langle \rho^{p} \delta'(\rho-y) \rangle=\frac{y(p-1)}{g}\left( 2l(l+1)-\frac{p}{2}(p-2)\right)\langle \rho^{p-3} \rangle
    +\frac{2 y p}{g}\langle \rho^{p-1} \rangle-2 p y^{p-1} u^2(y),
\end{equation}
where $p>-2l$.
In particular,
\begin{equation}
\label{p1}
    \langle \rho \delta'(\rho-y) \rangle=\frac{2 y^2}{g R}-2 u^2(y),
\end{equation}
\begin{equation}
\label{p2}
    \langle \rho^{2} \delta'(\rho-y) \rangle=l(l+1)\frac{2 y}{g}\langle \rho^{-1} \rangle
    +\frac{4 y}{g}\langle \rho \rangle-4 y u^2(y),
\end{equation}
\begin{equation}
\label{p3}
    \langle \rho^{3} \delta'(\rho-y) \rangle=\frac{y^2}{g \, R}\left(4 l(l+1)-3\right)+\frac{6 y}{g}\langle \rho^2 \rangle-6 y^2 u^2(y),
\end{equation}
where the normalization $\langle 1 \rangle=y/R$ is used.
Taking Eq. (\ref{p1}), we can rewrite Eq. (\ref{n0case}) as
\begin{equation}
\label{nscase}
    \langle \rho^{s} \delta' \rangle=y^{s-1}\left((s-3) \, u^2(y)+\frac{2 y^2}{g \, R}\right).
\end{equation}
Eq. (\ref{ddeltarhop}) allows us to derive Kramers-Pasternak-like relations for the moments of the delta-shell potential as
\begin{equation}
\label{KrPas}
\begin{split} 
   (s-1)\left(4 l(l+1)-s(s-2)\right)\langle \rho^{s-3} \rangle+4 s \langle \rho^{s-1} \rangle=\\
   2 y^{s-2}\left(3 g(s-1) \, u^2(y)+\frac{2 y^2}{R}\right),
\end{split}
\end{equation}
where $s>-2l$. The case $s=3$ gives
\begin{equation}
\label{u2Kr}
u^2(y)=\frac{2l(l+1)-y^2-3/2}{3 g \, R}+\frac{\langle \rho^{2} \rangle}{g y},	
\end{equation}
and therefore the final expression is
\begin{equation}
\label{KrPas2}
\begin{split} 
   (s-1)\left(4 l(l+1)-s(s-2)\right)R\langle \rho^{s-3} \rangle+4 s R\langle \rho^{s-1} \rangle-6(s-1)y^{s-3}R\langle \rho^{2} \rangle=\\
   2 y^{s-2}\left((s-1)\left(2l(l+1)-y^2-3/2\right)+2 y^2\right) \,.
\end{split}
\end{equation}
The above expression, which is the Kramers-Pasternak-like relation for the delta-shell potential, is the main result of this section.

Also we recall from Eq. (\ref{ddelta}) and Eq. (\ref{ddeltal}) the relation
\begin{equation}
\label{p0}
\langle \delta'(\rho-y) \rangle=-\frac{y}{g}\cdot
 \begin{cases}
     [u_{0}'(0)]^2 & \text{$l=0$,}\\
     2l(l+1)\langle\rho^{-3}\rangle & \text{$l_{max}\geq l>0$.}
 \end{cases}
\end{equation}

If we take partial derivatives of both sides of
Eq. (\ref{fl_def}) with respect to the parameter $R$, we can calculate
the expression for $\left<\rho^{-3} \right>_{l}$ explicitly. The result is
\begin{equation}
\label{rhom3}
    \left<\rho^{-3} \right>_{l}=\frac{1}{R \, l(l+1)}\left(\frac{N_l^2(y)}{2 g y^4}-1\right),
\end{equation}
where $l_{max}\geq l>0$.

Substitution of these relations in Eq. (\ref{n0case}) provides one of the moment relations.\\
In the case of $l=0$, we have

\begin{equation}
\label{l0mom}
    \langle \rho \rangle_0=\frac{y^2}{R}+\frac{y^2}{4}[u_{0}'(0)]^2=\frac{y^2}{R}+\frac{R^2}{4}\Psi_0^2(0)=
    \frac{y^2}{R}\left(\frac{e^{2y}-y-1}{e^{2y}-2y-1}\right).
\end{equation}
For $l>0$,
\begin{equation}
\label{lmom}
    2\langle \rho \rangle_l+l(l+1)\langle \rho^{-1} \rangle_l-y^2 l(l+1)\langle \rho^{-3} \rangle_l=\frac{2 y^2}{R}.
\end{equation}
We note that since the mean value $\langle ... \rangle\propto R^{-1}$ and $\Psi^2\propto R^{-3}$ (see Eq. (\ref{norm_psi})), the above relations scale
with the parameter $R$ and can be calculated knowing only $y$ and $l$.

\clearpage

\section*{Statistical physics of a dilute non-relativistic delta-shell gas}


In a dilute gas at sufficiently high temperature, the equilibrium thermal properties such as energy, pressure, specific heat and entropy are adequately described by  the familiar ideal gas laws for non-interacting particles. Under similar physical conditions of dilute density and high temperature, potential interactions between the particles in the system bring about corrections to the ideal gas behaviors for all of the state variables. In this section, we calculate the first quantum corrections due to the presence of interactions, specifically through the delta-shell potential, highlighting at the same time the differences that arise from Fermi and Bose statistics.      As far as we are aware, such corrections for the delta-shell gas have not been considered before.

\section {The first quantum correction to the ideal-gas law}
\label{p2sec1}

To start with, we outline the formalism following the text book by Huang\cite{Huang87}.
For a system of $N$ identical particles, the partition function $Q_N$ has the form
\begin{equation}
  Q_N(V,T)=Tr \, e^{-\beta \hat H}=\int d^{3N}r \sum_{\alpha}\Psi_\alpha^*(1,...,N) \, e^{-\beta \mathcal H} \, \Psi_\alpha(1,...,N),
  \label{QN}
\end{equation}
where $\Psi_\alpha$ is a complete set of orthonormal wave functions, $\vec{r}_j$ is the coordinate of particle $j=1,...,N$, $\beta=\frac{1}{k_B T}$ ($k_B$ is the Boltzmann constant),  $V$ and $T$ stand for the volume and temperature of the gas, respectively.
\par If the separation $|\vec{r_i}-\vec{r_j}|$ between any pair of particles is larger than both the thermal de-Broglie wavelength $\lambda=\frac{h}{\sqrt{2\pi \, m \, k_B T}}$ and the range of interaction $r_0$, then the cluster expansion can be applied. Explicitly, the partition function is given by
\begin{equation}
  Q_N(V,T)=\sum_{\{m_n\}}\prod_{n=1}^{N}\frac{1}{m_n !}\left(\frac{V}{\lambda^3}b_n(V,T)\right)^{m_n},
  \label{QN_cluster}
\end{equation}
where the set $\{m_n\}$ satisfies $\sum_{n=1}^{N}n\,m_n=N$ and $b_n(V,T)$ is the $n^{th}$ cluster integral
\begin{equation}
  b_n(V,T)\equiv\frac{1}{n!\lambda^{3n-3}V}\int d^3 r_1 ... d^3 r_n \, U_l(1,...,n).
  \label{cluster_int}
\end{equation}
If we define the Slater sum
\begin{equation}
  W_M(1,...,M)\equiv M! \lambda^{3 M} \sum_{\alpha}\Psi_\alpha^*(1,...,M) \, e^{-\beta \mathcal H} \, \Psi_\alpha(1,...,M),
  \label{W_M}
\end{equation}
then a sequence of cluster functions $U_n(1,...,n)$ can be generated:
 \begin{eqnarray}
 \begin{array}{llll}
   W_1(1) = U_1(1) \\
   W_2(1,2) = U_1(1) U_1(2)+U_2(1,2) \\
   W_3(1,2,3) = U_1(1) U_1(2) U_1(3)+U_1(1) U_2(2,3)+U_1(2) U_2(3,1) \\
    +U_1(3) U_2(1,2) + U_3(1,2,3) \\
   ...
   \end{array}
   \label{Udef}
  \end{eqnarray}
Introducing the fugacity $z=e^{\beta \mu}$, where $\mu$ is the chemical potential,
the grand canonical partition function 
$\mathcal{L}(z,V,T)=\sum_{N=0}^{\infty}z^N Q_N(V,T)$ in the cluster  expansion has the compact form
\begin{equation}
\log(\mathcal{L}(z,V,T))=\frac{V}{\lambda^3}\sum_{n=1}^{\infty}b_n z^n,
 \label{grand}
\end{equation}
and is used to obtain the equation of state for the gas:
\begin{equation}
\frac{P V}{k_B T}=\log(\mathcal{L}(z,V,T)),
 \label{state_PV}
\end{equation}
\begin{equation}
    \langle N \rangle = z \frac{\partial}{\partial z}\log(\mathcal{L}(z,V,T)).
 \label{state_N}
\end{equation}
The equation of state in parametric form is expressed in terms of the cluster integrals
\begin{equation}
\frac{P}{k_B T}=\frac{1}{\lambda^3}\sum_{n=1}^{\infty} b_n z^n,
 \label{state_Pz}
\end{equation}
\begin{equation}
\frac{1}{v}=\frac{1}{\lambda^3}\sum_{n=1}^{\infty}n \, b_n z^n,
 \label{state_vz}
\end{equation}
where $v=V/\langle N \rangle$ is the volume per particle.
Now, one sets the volume to infinity, $V\to \infty$, to define a new quantity
\begin{equation}
\bar b_n(T)=\lim_{V \to \infty}b_n(T,V),
 \label{bbar}
\end{equation}
whence Eq. (\ref{state_Pz}) and Eq. (\ref{state_vz}) are unchanged, but $b_n(T,V)$ is replaced by $\bar b_n(T)$. The virial expansion of the equation of state takes the form
\begin{equation}
\frac{P v}{k_B T}=\sum_{n=1}^{\infty}a_n(T)\left(\frac{\lambda^3}{v}\right)^{n-1},
 \label{virial_expn}
\end{equation}
where $a_n$ are the virial coefficients. As seen from Eq. (\ref{state_Pz}) and Eq. (\ref{state_vz}), the virial coefficients can be expressed in terms of the cluster integrals $\bar b_n$ as
\begin{eqnarray}
\begin{array}{lllll}
  a_1 = \bar b_1 = 1 \, \qquad &\to& \text{ideal-gas law}\\
  a_2 = - \bar b_2\, \qquad &\to&  \text{the first quantum correction}\\
  a_3 = 4 \bar b_2^2 - 2 \bar b_3 \,  \qquad &\to&  \text{ the second quantum correction}\\
  a_4 = -20 \bar b_2^3+18 \bar b_2 \bar b_3- 3\bar b_4 \\
  ...
\end{array}
\label{ans}
\end{eqnarray}
The first correction to the ideal-gas equation of state is given by the second virial coefficient which entails the calculation of the cluster-two integral $b_2$. Recall that for noninteracting ideal quantum gases the cluster integrals are \cite{Huang87}
\begin{equation}
 \bar b_n^{(0)} = \left\{
\begin{array}{ll}
n^{-\frac{5}{2}} \qquad & \text{for an ideal Bose gas,}\\
(-1)^{n+1} n^{-\frac{5}{2}} \qquad & \text{for an ideal Fermi gas.}
\end{array} \right.
 \label{bn0}
\end{equation}

\section {Cluster-two integral and the first correction}
\label{p2sec2}

For two interacting particles with the center of mass coordinate $\vec{R}$ and separation $\vec{r}$, the cluster-two integral can be calculated as \cite{Huang87}
\begin{equation}
\begin{array}{lll}
\bar b_2-\bar b_2^{(0)}&=&\frac{1}{2 \lambda^3 V}\int d^3R \, d^3r[W_2(1,2)-W_2^{(0)}(1,2)]\\
&=&2\sqrt{2}\int d^3r\sum_n[|\psi_n(\vec{r})|^2\,e^{-\beta E_n}-|\psi_n^{(0)}(\vec{r})|^2\,e^{-\beta E_n^{(0)}}]\\
&=& 2\sqrt{2}\sum_n(e^{-\beta E_n}-e^{-\beta E_n^{(0)}}),
\end{array}
 \label{b2_b20}
\end{equation}
where the unit normalization of the two-body wave function
\begin{equation}
\int d^3r|\psi_n(\vec{r})|^2=\int d^3r |\psi_n^{(0)}(\vec{r})|^2=1
 \label{norm}
\end{equation}
is used.
\par In general, the energy spectrum consists of discrete (for bound) and continuum (for scattering) states. Let $d(k)dk$ be the number of states with the wave number lying between $k$ and $dk$. Equation (\ref{b2_b20}) takes the form
\begin{equation}
\bar b_2-\bar b_2^{(0)}=2^{3/2}\left [\sum_{Bound}e^{-\beta E_{Bound}}+\int_0^\infty dk\left(d(k)-d^{(0)}(k)\right) e^{-\beta\frac{\hbar^2 k^2}{2\mu}}\right]
 \label{b2_b20c}
\end{equation}
as first shown by Beth and Uhlenbeck\cite{Beth37}.  The expression for the
density of states was found before in Eq. (\ref{shift_deriv}) and can
be used here to obtain the working expression for the second virial
coefficient (\ref{ans}):
\begin{equation}
a_2=-\bar b_2^{(0)}-2^{3/2}{\sum_{l}}'(2l+1)\left [e^{-\beta E_l}-\frac{I_l(T)}{\pi}\right],
 \label{b2_b20I}
\end{equation}
where $2l+1$ is the multiplicity of the energy level $E_l$ and the prime on the summation sign indicates the use of even $l$
for Bosons and odd $l$ for Fermions. 

For the delta-shell potential, one must be careful
when $g=2l+1$ for which $E_{Bound(l)}=0$. As these zero-energy bound states
are not normalizable, they are excluded from the discrete sum, but are accounted for by the continuous part of the energy spectrum at $k=0$. 
It is also related to a proper definition of the phase shift 
as is done in formulating Levinson's theorem \cite{Levinson1974}, which connects the zero-energy phase shift $\delta_l(0)$
with the number of discrete bound states \cite{Newton66}. 

Some comments about the physics elucidated by the theorem are instructive. Consider zero-energy scattering by a 
potential, and, at first, ignore subtleties associated with the zero-energy bound states. (The subtleties arise for  the zero angular momentum $l = 0$ state only). Denote the phase shift for a particle incident at an energy $E$ with 
an angular momentum  $l \hbar$ by $\delta_l(E)$. It is given an absolute meaning, that is, with no ambiguity with 
regard to multiples of $\pi$, by setting $\delta_l( E\to \infty )\to 0$. The theorem states that
\begin{equation}
\label{LevinsonN}
\delta_l(E=0)=\pi N_l,
\end{equation}
where $N_l$ is the number of bound states with angular momentum $l \hbar$.

In the case of the delta-shell potential, the theorem gives
\begin{equation}
\label{levinson0}
\delta_0(0)=\pi \cdot
    \begin{cases}
    0 & \text{for } g<1,\\
    1/2 & \text{for } g=1,\\
    1 & \text{for  } g>1,
    \end{cases}
\end{equation}
and
\begin{equation}
\label{levinsonl}
\delta_{l>0}(0)=\pi \cdot
    \begin{cases}
    0 & \text{for } g<2l+1,\\
    1 & \text{for } g\geq2l+1.
    \end{cases}
\end{equation}

The temperature
dependent partial wave integral $I_l(T)$ is given by
\begin{equation}
I_l(T)\equiv\int_{0}^\infty dx \left [-\frac{\partial \delta_l(x)}{\partial x}\right ] e^{-\gamma(T) \, x^2},
 \label{Il_def}
\end{equation}
where $\gamma(T)=\frac{1}{2\pi}\left(\frac{\lambda(T)}{R}\right)^2$. 
For the virial approximation to hold, $\lambda > R$ and $\frac{\partial \delta_l}{\partial x} \propto x^{2l}$. Far from the resonances which occur for  $g=2l+1$, these considerations imply
\begin{equation}
I_l \propto \left(\frac{R}{\lambda}\right)^{2l+1}.
 \label{Il_prop}
\end{equation}
Therefore, the smaller the partial wave, the larger the contribution
to the first correction to the ideal gas equation of state.
\begin{figure*}[tb]
\centerline{\includegraphics[width=12cm]{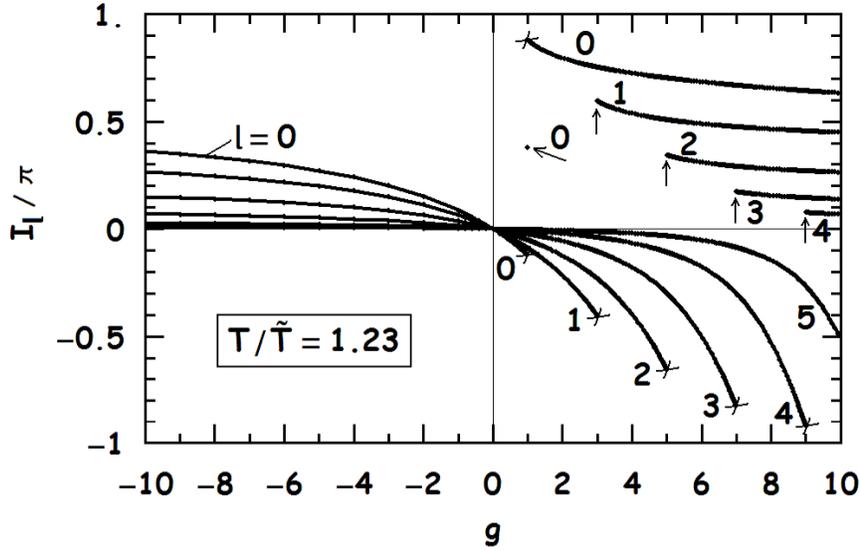}}
\vskip -20pt
\caption{The partial wave integral $I_l(T\simeq\tilde{T})$,
where $\tilde{T}$ is defined by $\lambda(\tilde{T})=R$, as a function
of the parameter $g$ for $l=0,1,...,5$. For $l>0$,and at resonance points where
$g=2l+1$, the function jumps discontinuously through a value of $\pi$.
For $l=0$, there are two discontinuous jumps each of magnitude $\frac{\pi}{2}$.
Crosses indicate the limiting values which the function never takes.}
\label{IlTx}
\end{figure*}
Figure \ref{IlTx} shows the partial wave integral
$I_l(T\simeq\tilde{T})$, where $\tilde{T}$ is defined by
$\lambda(\tilde{T})=R$ (whence $k_B\tilde{T}=\frac{1}{2\pi}\frac{(h c)^2}{R^2 m
c^2}$), as a function of the parameter $g$.
For $l>0$, the function $I_l(T\simeq\tilde{T})$ exhibits a
discontinuous jump at the resonance point $g=2l+1$. 
Let us closely examine these discontinuities for the cases of $S-$ and $P$-waves. Splitting the integral, we have
\begin{equation}
\begin{split}
I_0(g\approx1)\approx & -\int_{3\Delta x}^{\infty}dx  \delta'_0(x) e^{-\gamma(T) \, x^2}-\int_0^{3\Delta x}dx \delta'_{0-res}(x) e^{-\gamma(T) \, x^2}\approx \\
& I_0(g=1)+\text{sign}(g-1)\left(\frac{\pi}{2}-\sqrt{\pi \gamma(T)}|g-1|\right),
\end{split}
 \label{I0_jmp}
\end{equation}
and
\begin{equation}
\begin{split}
I_1(g\approx3)\approx & -\int_{x_1+3\Delta x_1}^{\infty}dx  \delta'_1(x) e^{-\gamma(T) \, x^2}-\int_0^{x_1+3\Delta x_1}dx \delta'_{1-res}(x) e^{-\gamma(T) \, x^2}\approx \\
& I_1(g=3)-U(3-g)\, \pi \left(1-\frac{2125}{1728}(3-g)\right) \, e^{-\gamma(T) \, \frac{5(3-g)}{2 g}},
\end{split}
 \label{I1_jmp}
\end{equation}
where $\Delta x$, $\Delta x_{1}$ and $x_1$ all $\to 0$ and we have used Eq. (\ref{ddeltaap}) and Eq. (\ref{ddelta1dirac}). The magnitude of
the jump is $\pi$ (only for $l=0$, there are two jumps of $\pi/2$ each), which is temperature independent.  This jump
feature is due to the fact that the derivative of the phase
shift $\frac{\partial \delta_l(x)}{\partial x}$ tends to become
delta-function-like in the proximity of the resonance values of $g$
with a normalization factor of $\pi$.  For $l>0$, the
delta-function--like part is off-set from the origin and appears only
if the limit is approached by $g$ from the left hand side. So the value of
the function at the exact resonance point belongs to the right hand
side piece of $I_l(T)$. If the temperature is taken to be much lower
than $\tilde{T}$, the integrals tend to zero except in the proximity
of the resonance points. Here, the ``$\pi$''-jump persists, but the
region in which the function significantly differs from zero narrows.

For $l=0$, there are two discontinuous jumps of magnitude
$\frac{\pi}{2}$ each.  In this case, the delta-function-like peak stays
centered around the origin (that gives a factor of $\frac{1}{2}$) and
flips sign depending on the manner in which (left or right) the limit ($g\to1$)
is taken. However, there is no delta function at the exact resonance value.
These arguments explain the discontinuous pattern for $I_0(T)$.

The above features can  also be understood by interpreting the derivative of the phase shift as
the density of levels. The system receives a non-vanishing contribution from zero energy levels
for the resonance configuration as seen from the low energy expansions of Eq. (\ref{shift_dx_low}). This is also the case for almost-bound (loosely bound) states near the resonance values of $g$.

\section{Region of validity for the second virial coefficient}
\label{p2sec3}

The two main physical conditions in our model gas are low density ($n
R^3<1$) and temperature such that $\lambda(T)>R$.  These conditions set
the limits for the range of the potential parameter, namely the
strength $g$, for the second virial approximation to be valid.
\begin{figure*}[tb]
\centerline{\includegraphics[width=12cm]{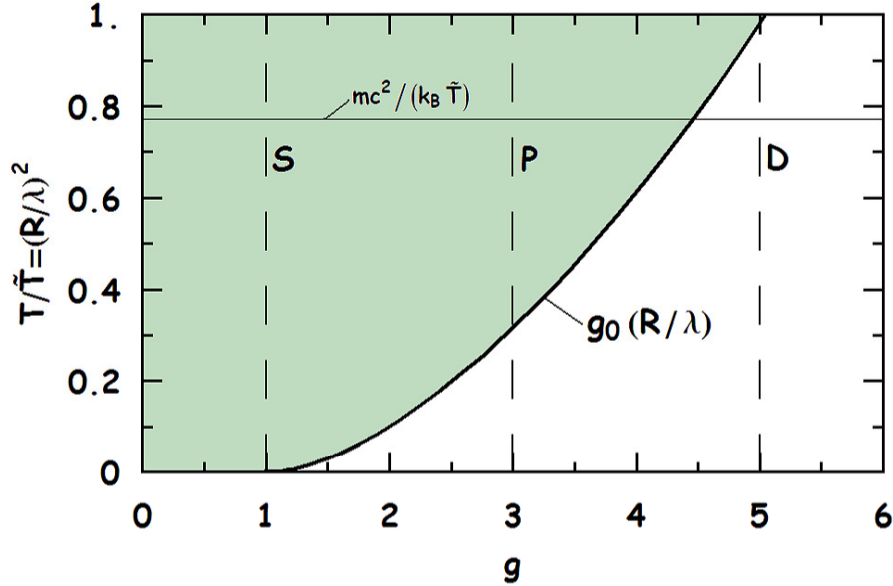}}
\vskip -20 pt
\caption{The safe region in the parameter space of $T$ and $g$ for the
second virial coefficient. The left boundary line is given by
$g_0\left(\frac{R}{\lambda}\right)\equiv
\frac{1}{f_0\left(y=\sqrt{2\pi}\left(\frac{R}{\lambda}\right)\right)}$.
At the vertical dashed lines indicated by letters of the corresponding
angular momentum values, one has resonance conditions $g=2l+1$.
The horizontal dashed line sets the upper boundary from
the non-relativity condition for the gas.}
\label{region}
\end{figure*}
As shown in Fig. \ref{region}, for the upper
temperature boundary $\left(\frac{R}{\lambda}\right)^2\leq1$ or if the
mass of the particles is sufficiently small, then
$\left(\frac{R}{\lambda}\right)^2\ll\frac{m c^2}{k_B \tilde{T}}$ (the temperature scale $k_B\tilde{T}= 2 \pi (\hbar c)^2 / (R^2 m
c^2)$) for the gas to be non-relativistic.  For the second virial
coefficient, two-body bound states contribute through the term
$e^{-\beta E_{Bound}}$.  To keep that contribution low, we have to
stay in the region in which $g\leq g_0\left(\frac{R}{\lambda}\right)$;
that is on the left of the curve
$g_0\left(\frac{R}{\lambda}\right)\equiv
\frac{1}{f_0\left(y=\sqrt{2\pi}\left(\frac{R}{\lambda}\right)\right)}$, 
where the thermal energy equals the energy of the first bound
state. Altogether, one has to stay in the grey region indicated not to
exceed the relativity limit.  The vertical dashed lines labeled by
letters of the corresponding angular momentum values show the
resonance values $g=2l+1$.  The regions around these values require
special care in the evaluation of integrals $I_l(T)$,
$I_l^{(2)}(T)$ and $I_l^{(4)}(T)$ (see Eq. (\ref{I2l}) \& Eq. (\ref{I4l})).

\section{The second virial coefficient in terms of the scattering length and the range parameter}
\label{p2sec4}

In this section, we derive an expression for the second virial coefficient in terms
of the scattering length and the range parameter. Using the expansion in Eq. (\ref{asl_r0_P_def}), we can obtain a series
in powers of the momentum for the function in the square brackets in the kernel of the integral in Eq. (\ref{Il_def}). From Eq. (\ref{asl_r0_P_def}),
\begin{equation}
\label{dtgserx}
\begin{split}
    \frac{\partial \delta_l(x)}{\partial x}=&\frac{1}{1+\tan^2(\delta_l)}\frac{\partial \tan(\delta_l)}{\partial x}=\\
   & -(2l+1)\frac{a^{(l)}_{sl}}{R^{2l+1}}x^{2l}+\delta_{0,l}\left(\frac{a^{(l)}_{sl}}{R}\right)^3 x^{2l+2}
    -\frac{(2l+3)}{2}\frac{(a^{(l)}_{sl})^2 r_0^{(l)}}{R^{2l+3}} x^{2l+2}\\
    &+O(x^{2l+4}),
\end{split}
\end{equation}
where $\delta_{0,l}$ is the Kronecker delta.
Now the integral $I_l$ in Eq. (\ref{Il_def}) can be calculated term by term giving
\begin{equation}
\label{IllowT}
    I_l(T)=(2l+1)!!\frac{\pi^{l+1}}{\sqrt{2}}\frac{a^{(l)}_{sl}}{\lambda^{2l+1}}
-\delta_{0,l}\frac{\pi^2}{\sqrt{2}}\left(\frac{a^{(l)}_{sl}}{\lambda}\right)^3
+\frac{(2l+3)!!}{2}\frac{\pi^{l+2}}{\sqrt{2}}\frac{(a^{(l)}_{sl})^2r_0^{(l)}}{\lambda^{2l+3}}+O(\lambda^{-2l-5}).
\end{equation}
This result captures the manner in which the scattering length and the effective range influence the second virial coefficient.
Note that the temperature dependence resides in the De-Broglie wavelength $\lambda$.  Applied to the delta-shell potential 
\begin{equation}
\label{IllowTd}
\begin{split}
    I_l(T)=&\frac{1}{(2l-1)!!}\frac{\pi^{l+1}}{\sqrt{2}}\frac{g}{(g-(2l+1))}\left(\frac{R}{\lambda}\right)^{2l+1}
+\delta_{0,l}\frac{\pi^2}{\sqrt{2}}\frac{g^3}{(1-g)^3}\left(\frac{R}{\lambda}\right)^3\\
&-\frac{(2l+1)}{2(2l-1)(2l-1)!!}\frac{\pi^{l+2}}{\sqrt{2}}\frac{g(g-(2l-1))}{(g-(2l+1))}\left(\frac{R}{\lambda}\right)^{2l+3}
+O\left(\left(\frac{R}{\lambda}\right)^{2l+5}\right),
\end{split}
\end{equation}
following Eq. (\ref{asl}) and (\ref{r_0}). In order for the above series to be a good approximation, the
temperature must be such that $\lambda \gg R$ and the strength parameter $g$ must not be near the resonance values $2l+1$.

In the case of an infinite scattering length, $a^{(l)}_{sl}\to \infty$ (or $g=2l+1$), that is, in the unitary limit, one can adopt the same steps as above,
 but without the term involving the scattering length. The result is

\begin{equation}
\label{I0LowInf}
     I_{0}(T)=\frac{\pi}{2\sqrt{2}}\frac{r_{0}}{\lambda}+O(\lambda^{-3}),
\end{equation}
and
\begin{equation}
\label{IlLowInf}
     I_{l>0}(T)=-(2l-1)!!\frac{\sqrt{2}\pi^l}{r^{(l)}_{0}\lambda^{2l-1}}+O(\lambda^{-2l-1}).
\end{equation}
For the delta-shell potential, 
\begin{equation}
\label{I0LowInfd}
     I_0(T)=\frac{\pi}{3 \sqrt{2}}\frac{R}{\lambda}+O\left(\left(\frac{R}{\lambda}\right)^3\right),
\end{equation}
and
\begin{equation}
\label{IlLowInfd}
     I_{l>0}(T)=\frac{(2l+3)(2l-1)}{(2l+1)!!}\frac{\pi^l}{\sqrt{2}}\left(\frac{R}{\lambda}\right)^{2l-1}
     +O\left(\left(\frac{R}{\lambda}\right)^{2l+1}\right).
\end{equation}
In the last case the range of the potential $R$ becomes the relevant length scale.


It is worthwhile to note that when the scattering length $a_s^(l) \to \infty$, it ceases to be of relevance in the final results for physical quantities which now depend on the remaining finite quantities such as the delta-shell radius $R$, the effective range $r_0$  and the de-Broglie wavelength $\lambda$. 

\section{Gas of nonzero spin particles}
\label{p2sec5}

If particles have nonzero spin, then in addition to even or odd
partial wave (spatially symmetric and antisymmetric wave
functions) contributions from Bose or Fermi statistics, the
appropriate symmetry separation of states must be made
\cite{Hirschfelder67}. The results for spin $s$ then read as
\begin{equation}
\begin{array}{ll}
a_{2}^{(s)}=\frac{s+1}{2s+1}a_{Bose}^{(0)}+\frac{s}{2s+1}a_{Fermi}^{(0)}&
\text{ for integer } s,\\
\\
a_{2}^{(s)}=\frac{s+1}{2s+1}a_{Fermi}^{(0)}+\frac{s}{2s+1}a_{Bose}^{(0)}
& \text{ for half-integer } s.
\end{array}
 \label{spin}
\end{equation}

In order to illustrate the role played by spin, we consider the cases of spin-zero, spin-half and spin-one particles in what follows. 

\begin{figure*}[tb]
\centerline{\includegraphics[width=11cm]{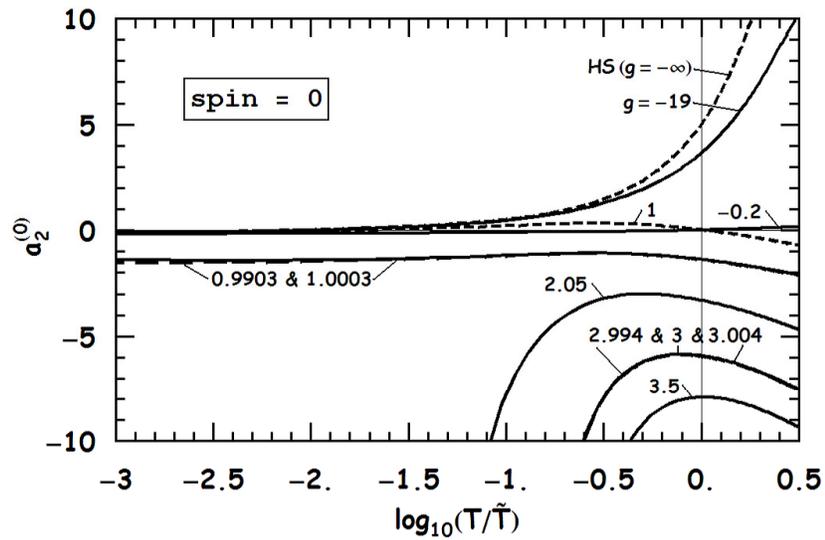}}
\vskip -20pt
\caption{Second virial coefficient as a function of temperature for various values of $g$ for spin-zero particles. Special cases are shown by dashed lines. See text for explanation.}
\label{a2Ts0}
\end{figure*}

Figure \ref{a2Ts0} shows the second virial coefficient for the case of spin-zero particles
as a function of temperature for various values of the interaction parameter $g$.
For comparison, results for hard spheres (HS) are also shown in this figure.

For the repulsive potential ($g<0$), the second virial coefficient is positive and increases with temperature. When the interaction
becomes weak ($g\approx0$), the coefficient tends to zero. For the attractive potential ($g>0$), $a_2$ turns negative.

The curve which corresponds to resonance ($g=1$) in the S-wave channel stands out. It is negative for $T>\tilde{T}$ and positive for $T<\tilde{T}$. However the curve is negative when $g\approx1$, near but not exactly at resonance. 

\begin{figure*}[tb]
\centerline{\includegraphics[width=11cm]{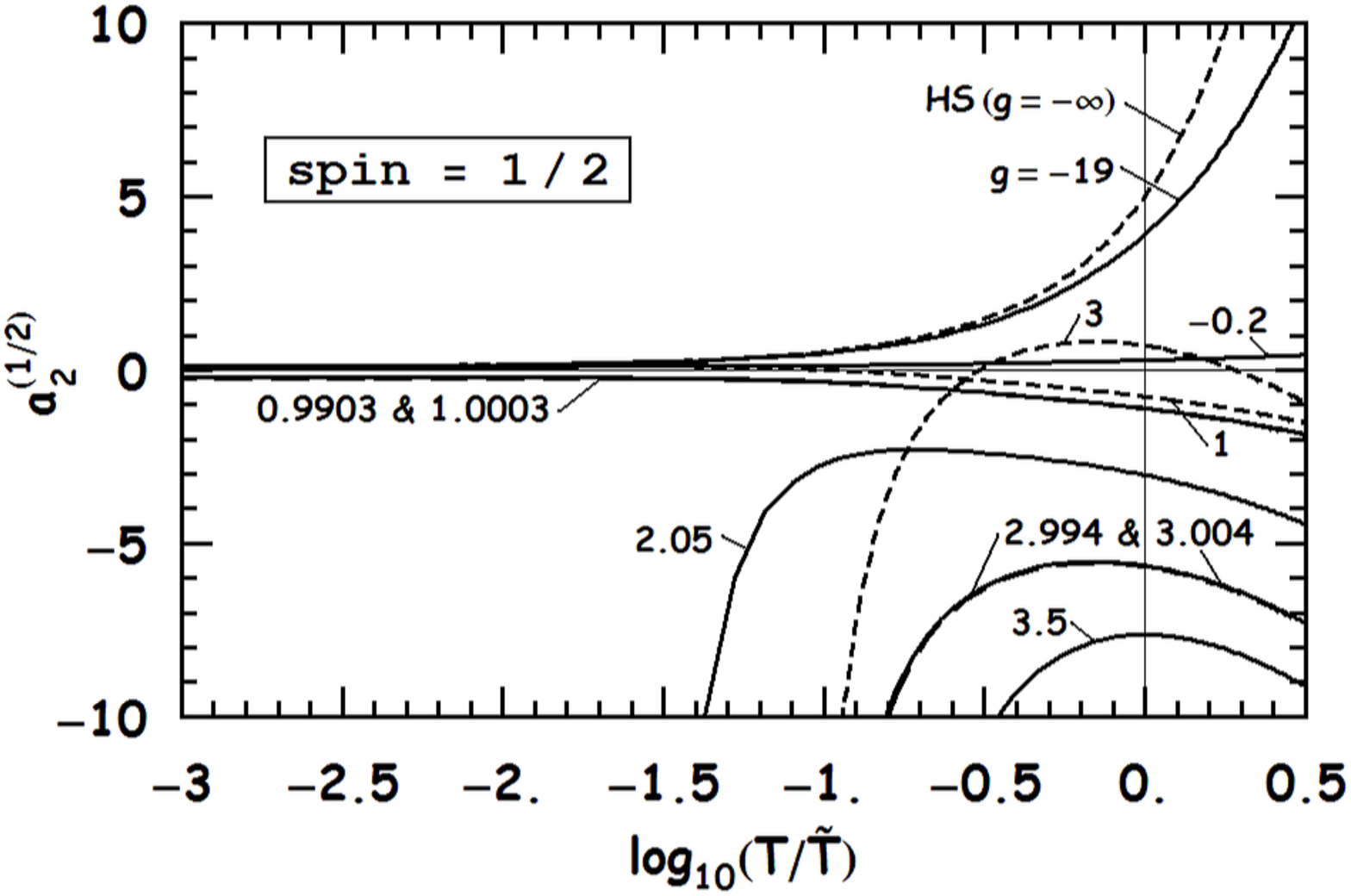}}
\vskip -20pt
\caption{Second virial coefficient as a function of temperature for various values of $g$, for spin- half particles. Special cases are shown by dashed lines. See text for explanation.}
\label{a2Ts1o2}
\end{figure*}

\begin{figure*}[tb]
\centerline{\includegraphics[width=11cm]{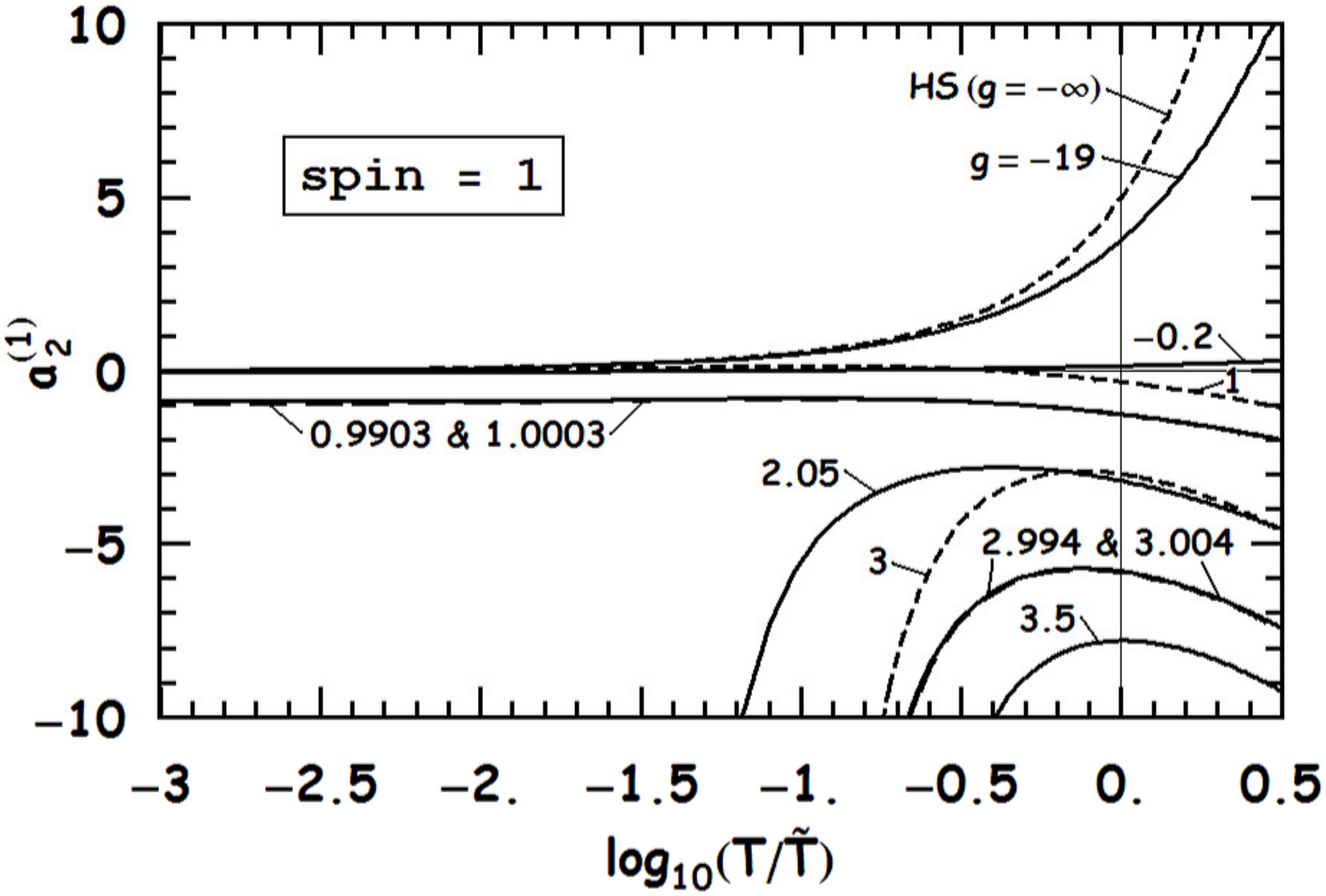}}
\vskip -20pt
\caption{Second virial coefficient as a function of temperature for various values of $g$, for spin-one particles. Special cases are shown by dashed lines. See text for explanation.}
\label{a2Ts1}
\end{figure*}

The curve which corresponds to the P-wave ($g=3$) is absent due to odd $l$ being forbidden in the scattering of a pair of particles with spin zero. 
In the cases of spin-half and spin-one particles, results for which are shown in Fig. \ref{a2Ts1o2} and Fig. \ref{a2Ts1}, all $l$'s contribute. For non-zero spin, effects of the P-wave resonance ($g=3$) stand out and that of the S-wave resonance ($g=1$) changes its value. 

\begin{figure*}[tb]
\centerline{\includegraphics[width=11cm]{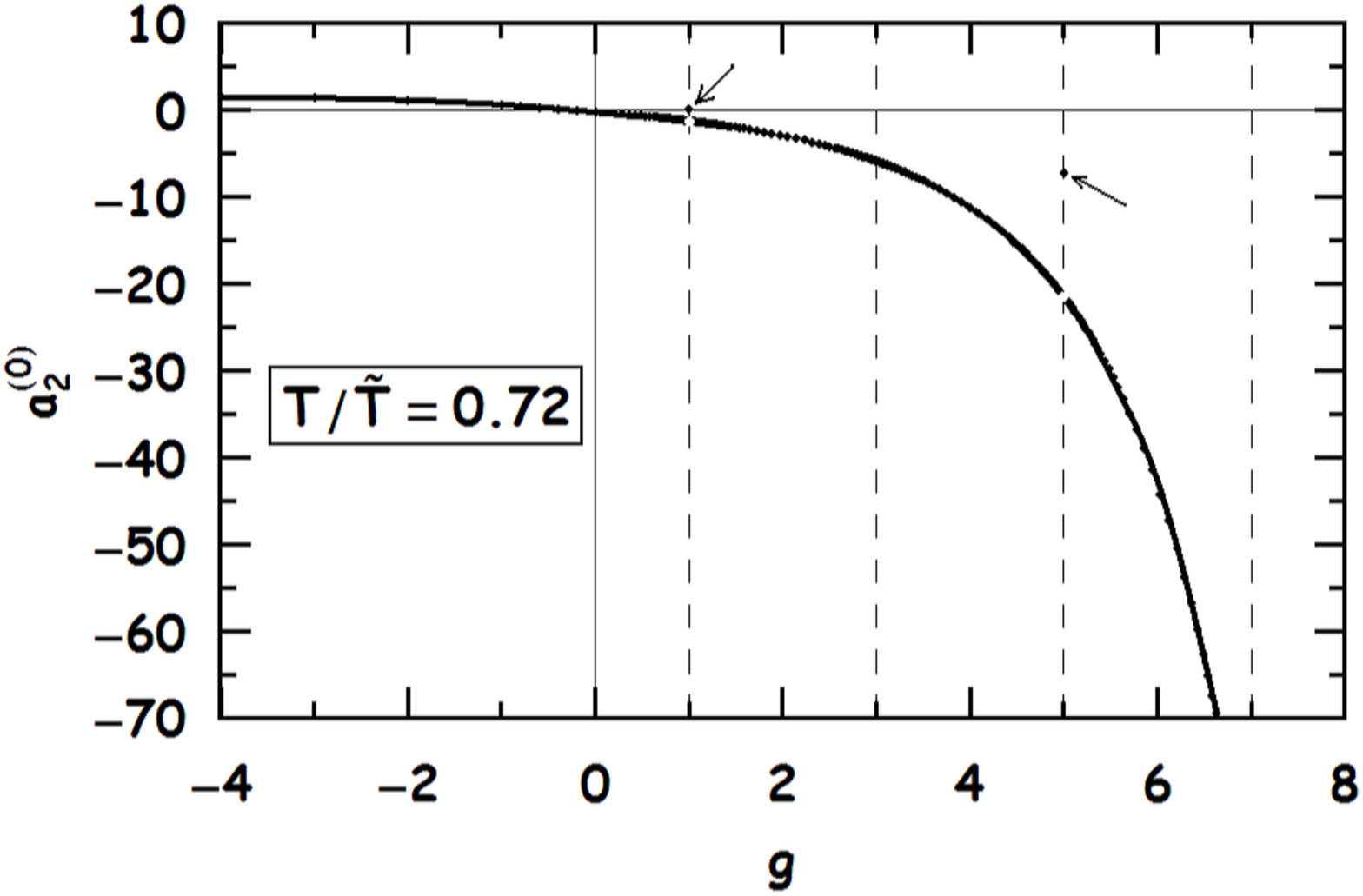}}
\vskip -20pt
\caption{Second virial coefficient as a function of $g$ for temperature $T/\tilde{T}=0.72$, for spin- zero particles. Resonance values are indicated by arrows.}
\label{a2gs0_}
\end{figure*}

\begin{figure*}[tb]
\centerline{\includegraphics[width=11cm]{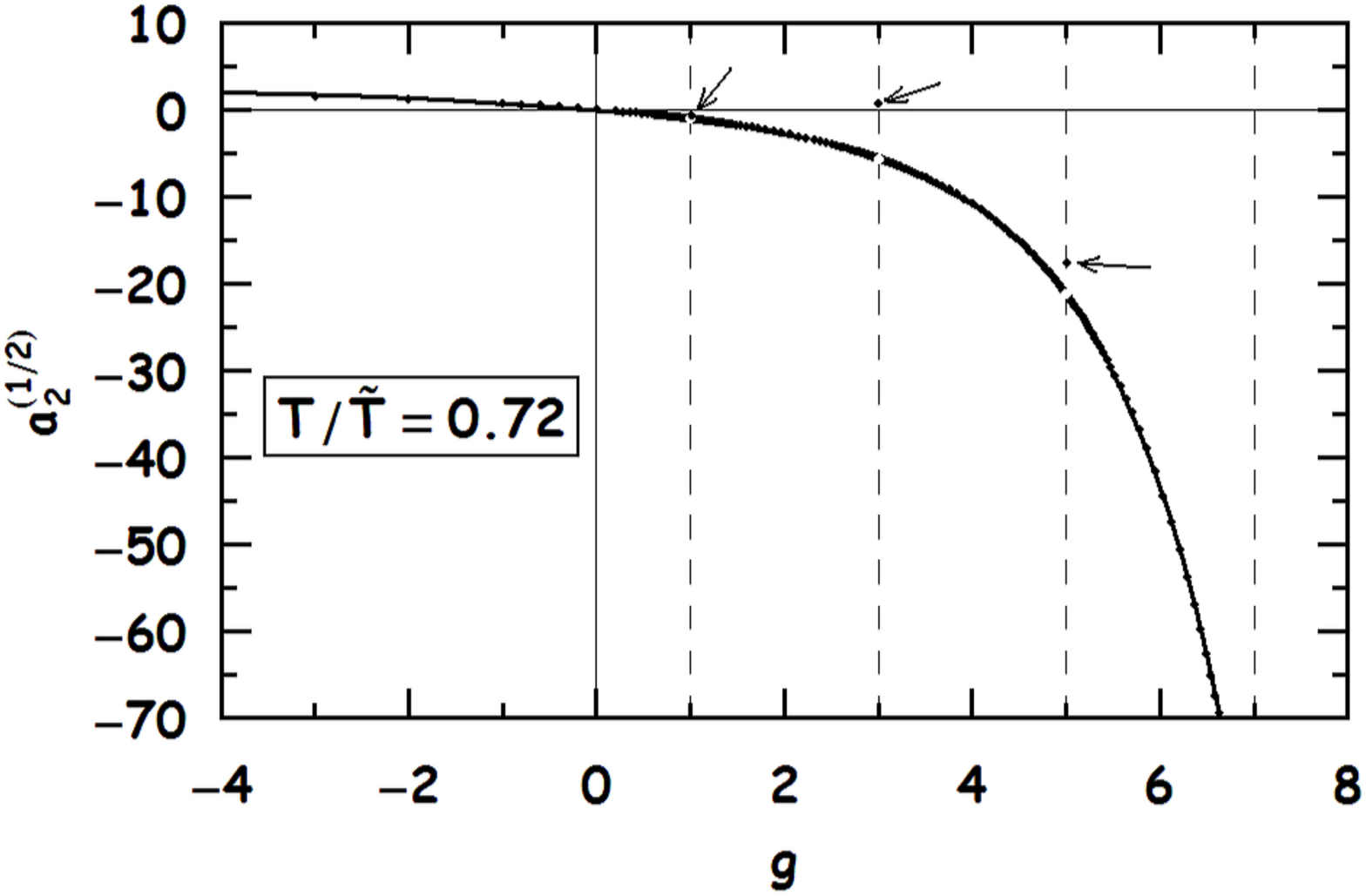}}
\vskip -20pt
\caption{Second virial coefficient as a function of $g$ for temperature $T/\tilde{T}=0.72$, for spin-half particles. Resonance values are indicated  by arrows.}
\label{a2gs1o2_}
\end{figure*}

\begin{figure*}[tb]
\centerline{\includegraphics[width=11cm]{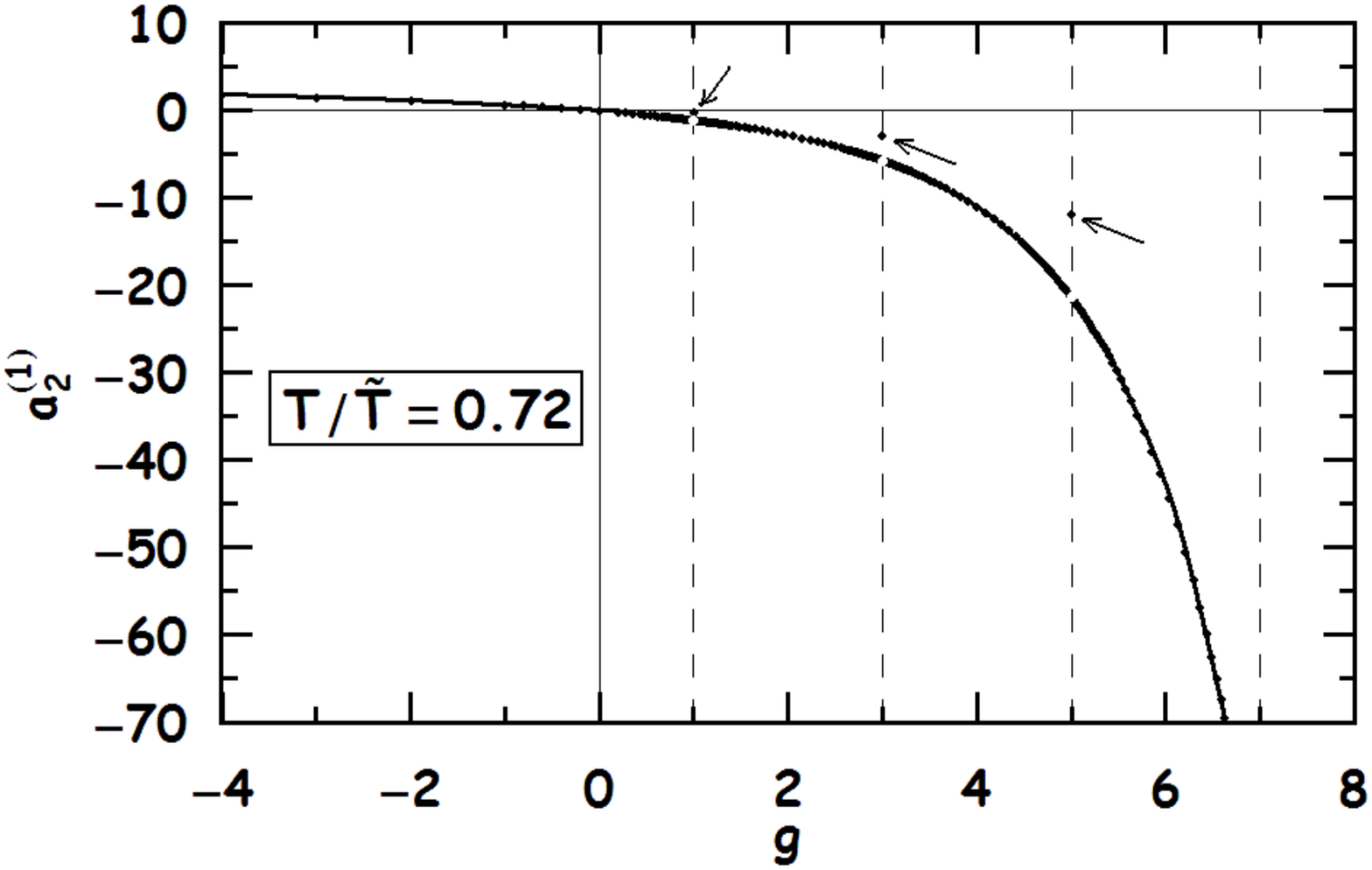}}
\vskip -20pt
\caption{Second virial coefficient as a function of $g$ for temperature $T/\tilde{T}=0.72$, for spin-one particles. Resonance values are indicated by arrows.}
\label{a2gs1_}
\end{figure*}

In Fig. \ref{a2gs0_}, Fig. \ref{a2gs1o2_} and Fig. \ref{a2gs1_}, corresponding to different spin values ($s=0, 1/2,~{\rm and}~ 1$, respectively), the second virial coefficient is shown as a function of the strength parameter $g$ for one value of temperature $T/\tilde{T}=0.72$. Positive values of $a^{s}_2$ for repulsive case ($g<0$), negative values of $a^{s}_2$ for attractive case ($g>0$) and prominent resonance points are shown in these figures.

Negative (positive) values of the second virial coefficient decrease (increase) the pressure of the system with respect to that of the ideal gas. For $T/\tilde{T} \to 0$, the coefficient tends to the corresponding ideal quantum gas values in Eq. (\ref{bn0}). 

\clearpage

\section{Thermodynamical properties in terms of the second virial coefficient}
\label{p2sec6}

We begin with the cluster expansion of the grand partition function in Eq. (\ref{grand}) up to the second order term:
\begin{equation}
    \log(\mathcal{L}(z,V,T))=\frac{V}{\lambda^3}\left(z+b_2(T) z^2+O(z^3)\right)=\frac{V}{\lambda^3}\left(e^{\frac{\mu_c}{k_B T}}+b_2(T) e^{\frac{2\mu_c}{k_B T}}+O(e^{\frac{3\mu_c}{k_B T}})\right),
\label{grand_b2}
\end{equation}
where $z=e^{\frac{\mu_c}{k_B T}}$ is the fugacity.
The pressure is then \cite{Kapusta89}
\begin{equation}
\label{pb2}
\begin{split}
    p&=k_B T \left( \frac{\partial \log(\mathcal{L}(\mu_c,V,T))}{\partial V} \right)_{T,\mu_c}
    =k_B T \left( \frac{\partial \log(\mathcal{L}(z,V,T))}{\partial V} \right)_{T,z}=\\
    &=\frac{k_B T}{\lambda^3}\left(z+b_2(T) z^2+O(z^3)\right).
\end{split}
\end{equation}
The number density
\begin{equation}
\label{nb2}
    n=\frac{T}{V} \left( \frac{\partial \log(\mathcal{L}(\mu_c,V,T))}{\partial \mu_c} \right)_{V,T}=
    \frac{1}{\lambda^3}\left( e^{\frac{\mu_c}{k_B T}}+2 b_2(T) e^{\frac{2\mu_c}{k_B T}}+O(e^{\frac{3\mu_c}{k_B T}}) \right),
\end{equation}
or in equivalent form
\begin{equation}
\label{nzb2}
    n \lambda^3= z+2 b_2(T) z^2+O(z^3).
\end{equation}
Note that $n \lambda^3$ serves as the small parameter for expansion in this method.
Inverting Eq. (\ref{nzb2}) for fugacity, we obtain
\begin{equation}
\label{znb2}
z=n \lambda^3-2 b_2(T) (n \lambda^3)^2+O \left( (n \lambda^3)^3 \right) .
\end{equation}
Inserting this $z$ in Eq. (\ref{pb2}), we have the virial equation of state
\begin{equation}
\label{pvirb2}
 p=n k_B T \left( 1-b_2(T) (n \lambda^3)+O\left((n \lambda^3)^2\right)\right),
\end{equation}
and as $b_2(T)\to-a_2(T)$ from Eq. (\ref{ans}),
\begin{equation}
\label{pa2vir}
    p=n k_B T \left( 1+a_2(T) (n \lambda^3)+O\left((n \lambda^3)^2\right)\right).
\end{equation}
The entropy is given by \cite{Kapusta89}
\begin{equation}
\label{Sb2}
    S=k_B \left( \frac{\partial}{\partial T}T \log\left(\mathcal{L}(\mu_c,V,T)\right) \right)_{V,\mu_c}.
\end{equation}
After inserting Eq. (\ref{grand_b2}) and doing some algebra, we obtain the entropy density
due to interactions as
\begin{equation}
\label{s_a2}
s-\left(\frac{5}{2}n k_B -n k_B \log(n \lambda^3)\right)= n k_B \left(\frac{1}{2}a_2(T)-T\frac{d a_2(T)}{dT}\right)(n \lambda^3)+O\left((n \lambda^3)^2\right),
\end{equation}
where the term in parentheses on the left hand side gives the ideal gas contribution.
In dimensionless form
\begin{equation}
\label{ds_a2}
\delta s\equiv\frac{s-s_0}{\tilde{s}}= \left(\frac{1}{2}a_2(T)-T\frac{d a_2(T)}{dT}\right)\left(\frac{\lambda}{R}\right)^3
+O\left( \left(\frac{\lambda}{R}\right)^6 (n R^3) \right),
\end{equation}
where the ideal gas entropy density
\begin{equation}
\label{s0_a2}
s_0=\frac{5}{2}n k_B -n k_B \log(n \lambda^3),
\end{equation}
and the scaling value
\begin{equation}
\label{st_a2}
\tilde{s}\equiv n k_B (n R^3).
\end{equation}
The quantity $n R^3$ can be thought of as a measure of diluteness, as we assume $n R^3<1$.

\par
\begin{figure*}[tb]
\centerline{\includegraphics[width=11cm]{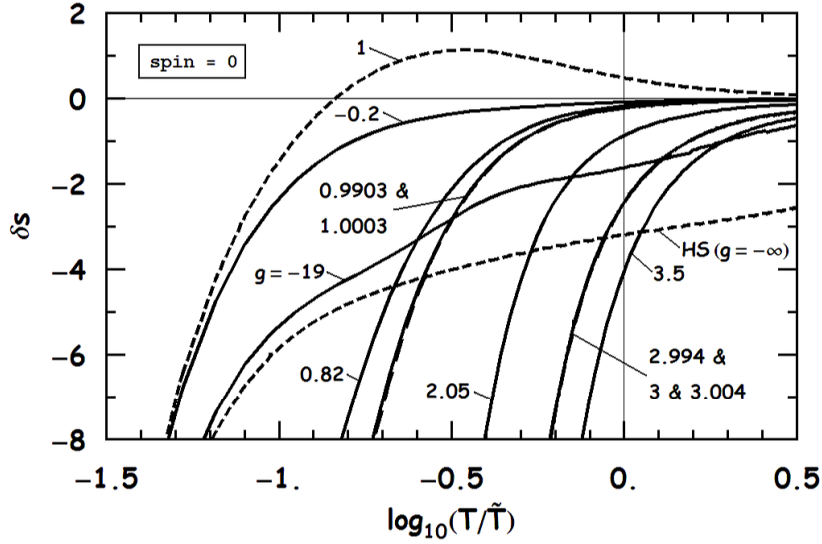}}
\vskip -20pt
\caption{Dimensionless entropy density shift for spin zero particles as a function of temperature for various values of the parameter $g$. Special cases are shown by dashed lines.}
\label{deltasLgTs0}
\end{figure*}

 \textit{Spin-zero particles}: The dimensionless entropy density shift $\delta s$ for spin zero particles is shown in Fig. \ref{deltasLgTs0} as a function of temperature for various values of the parameter $g$. The overall shift is negative, becoming small as the temperature increases. The S-wave resonance stands out and also acquires positive values, but odd $l$ partial waves do not contribute.	

\begin{figure*}[tb]
\centerline{\includegraphics[width=11cm]{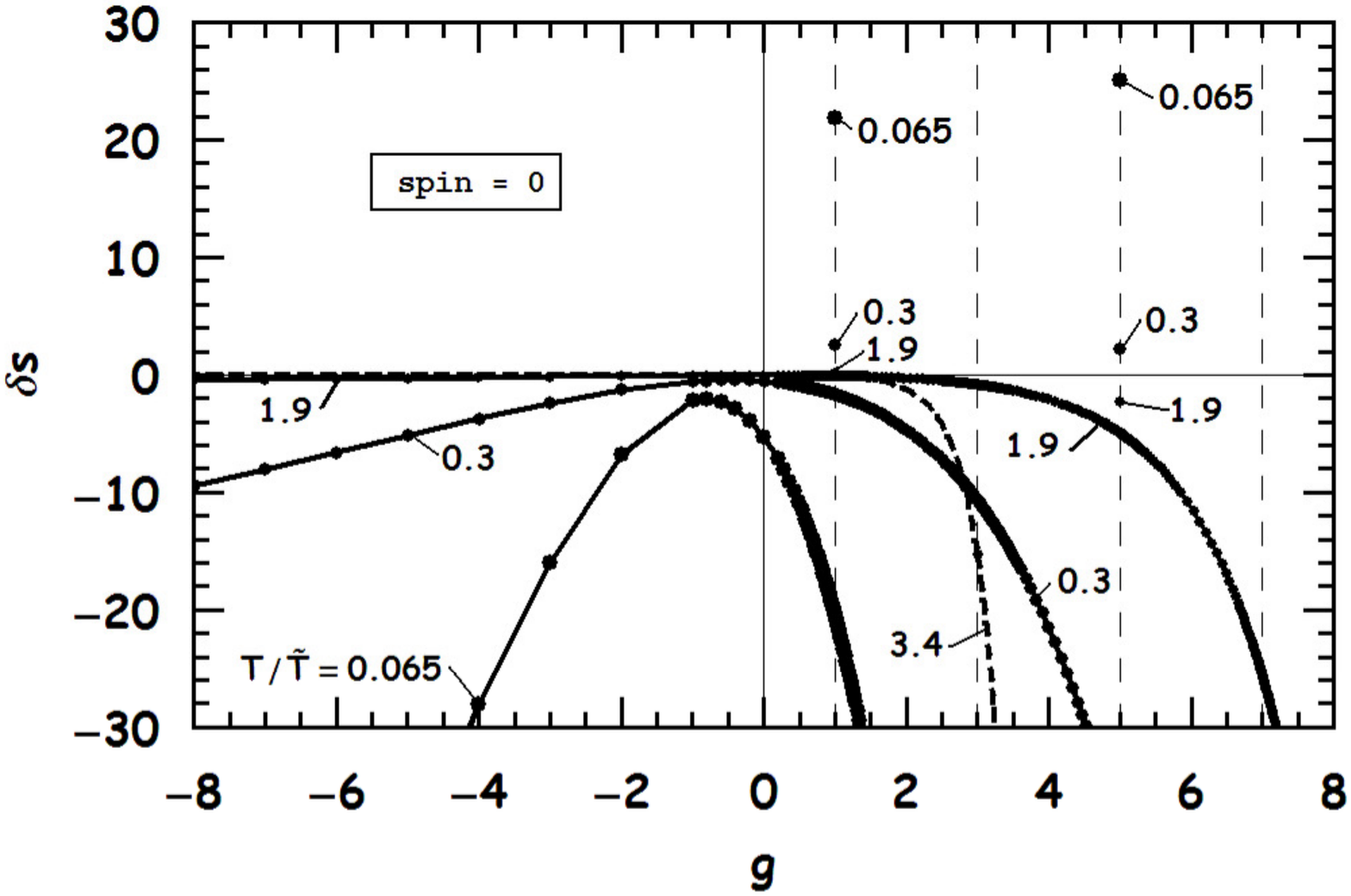}}
\vskip-20pt
\caption{Dimensionless entropy density shift for spin zero particles as a function of the parameter $g$ for various temperatures. Resonance values of $g$ are shown by vertical dashed lines.}
\label{deltasgs0_}
\end{figure*}

Figure \ref{deltasgs0_} shows the same shift as a function of the strength parameter for various values of temperature. The entropy density shift increases with the strength of the interaction and has some positive values at resonances which are separated from the curves.

\par
\begin{figure*}[tb]
\centerline{\includegraphics[width=11cm]{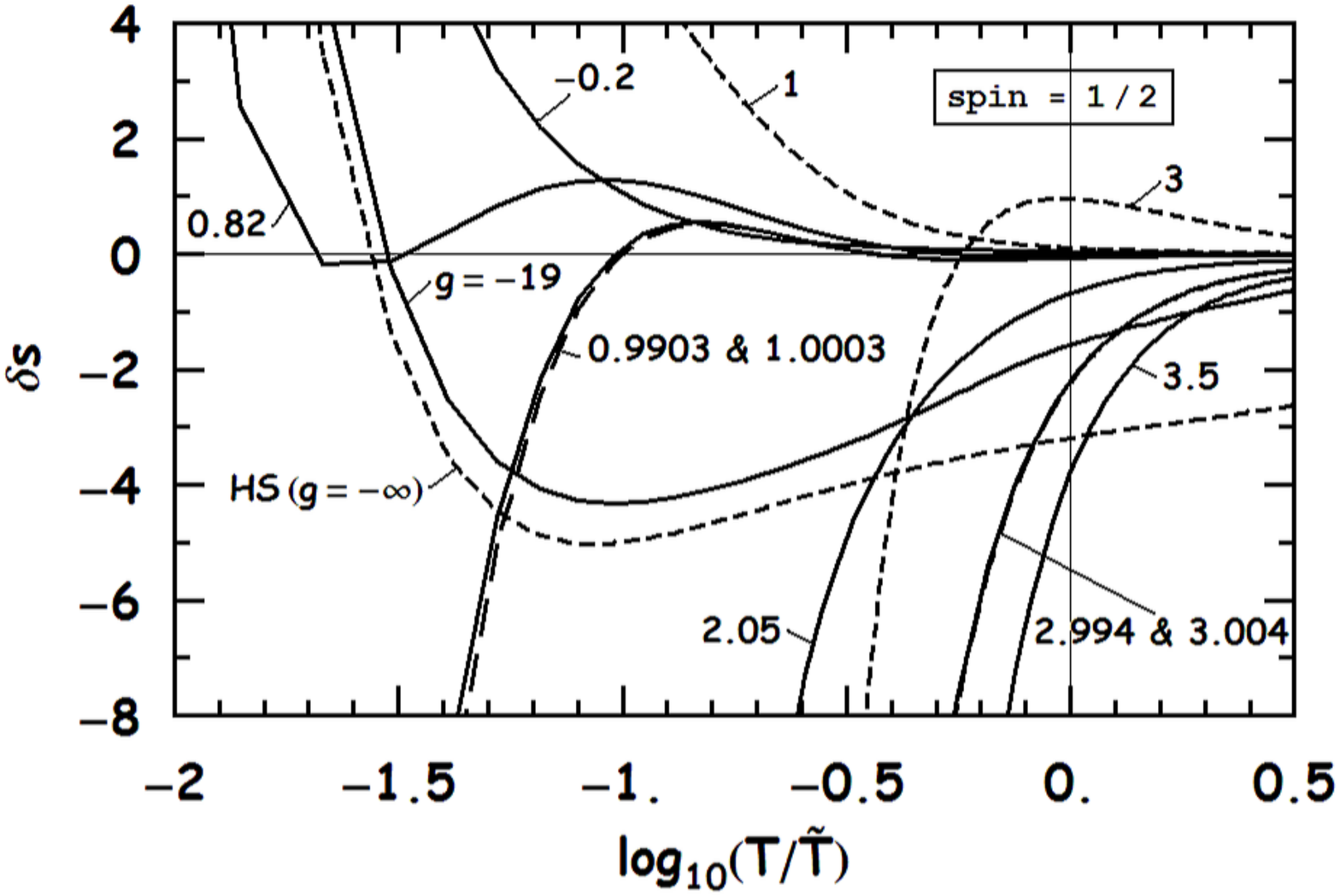}}
\vskip -20pt
\caption{Dimensionless entropy density shift for spin half particles as a function of temperature for various values of the parameter $g$. Special cases are shown by dashed lines.}
\label{deltasLgTs1o2}
\end{figure*}

\begin{figure*}[tb]
\centerline{\includegraphics[width=11cm]{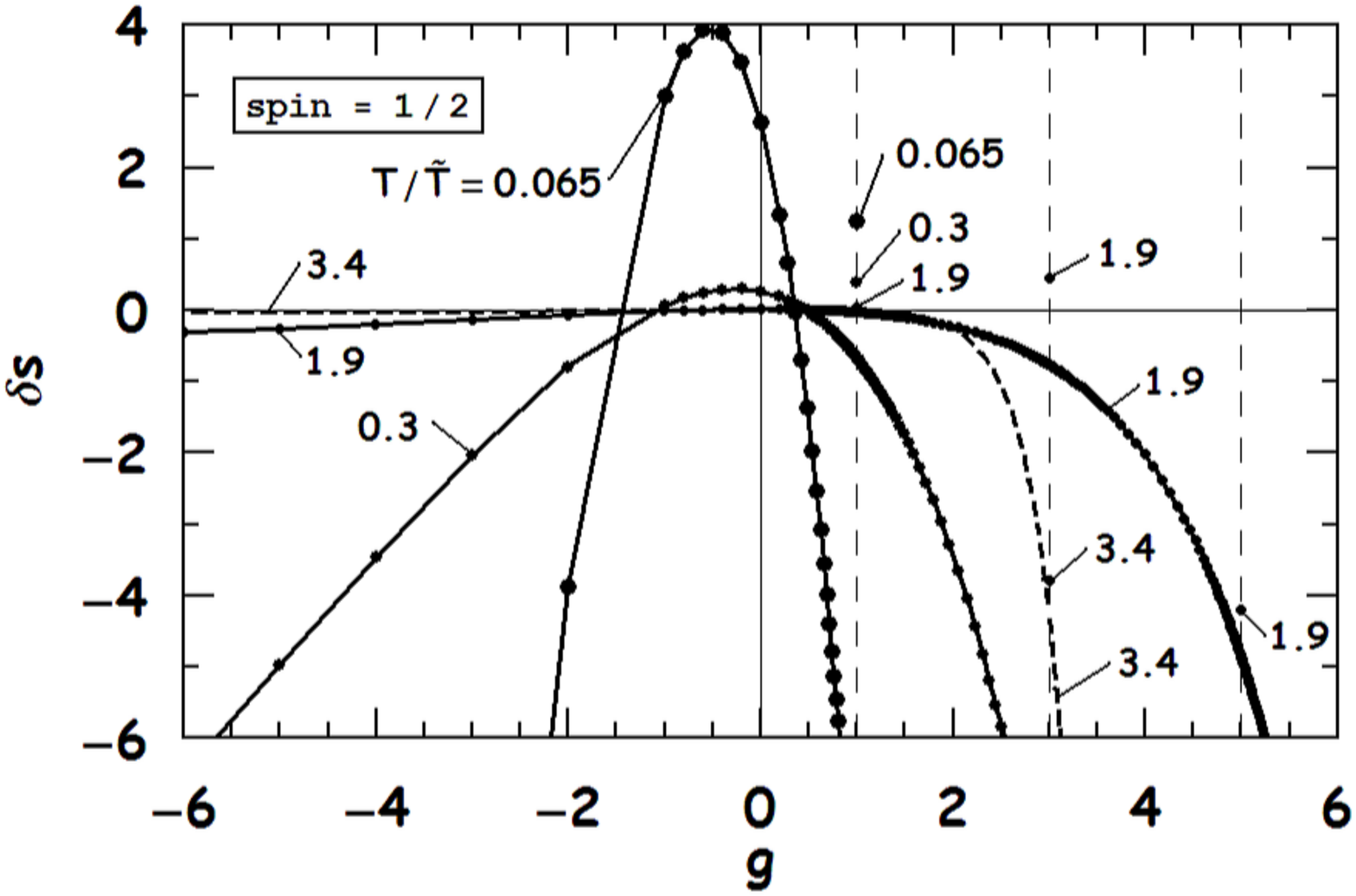}}
\vskip -20pt
\caption{Dimensionless entropy density shift for spin half particles as a function of the parameter $g$ for various temperatures. Resonance values of $g$ are shown by vertical dashed lines.}
\label{deltasgs1o2_}
\end{figure*}

 \textit{Spin-half particles}: Results for spin half particles are shown in Fig. \ref{deltasLgTs1o2} and Fig. \ref{deltasgs1o2_}. In this case all partial waves contribute. The entropy density shift can become positive at lower temperatures and also receives contributions from the P-wave resonance ($g=3$)  which are positive for higher temperatures. Positive shifts occur for the weakly interacting system (small $g$).

\par
\begin{figure*}[tb]
\centerline{\includegraphics[width=11cm]{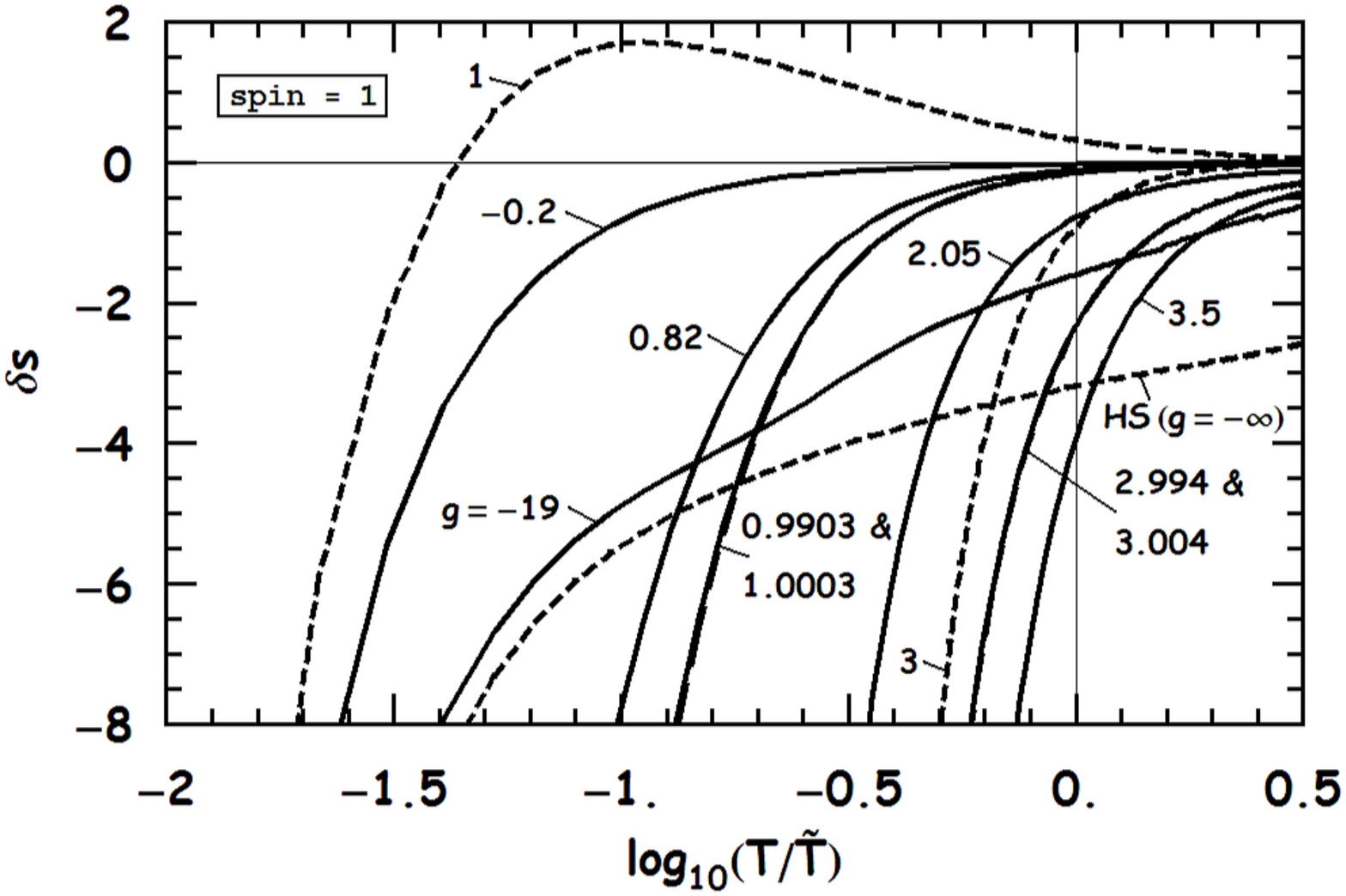}}
\vskip -20pt
\caption{Dimensionless entropy density shift for spin-one particles as a function of temperature for various values of the parameter $g$. Special cases are shown by dashed lines.}
\label{deltasLgTs1}
\end{figure*}

\begin{figure*}[tb]
\centerline{\includegraphics[width=11cm]{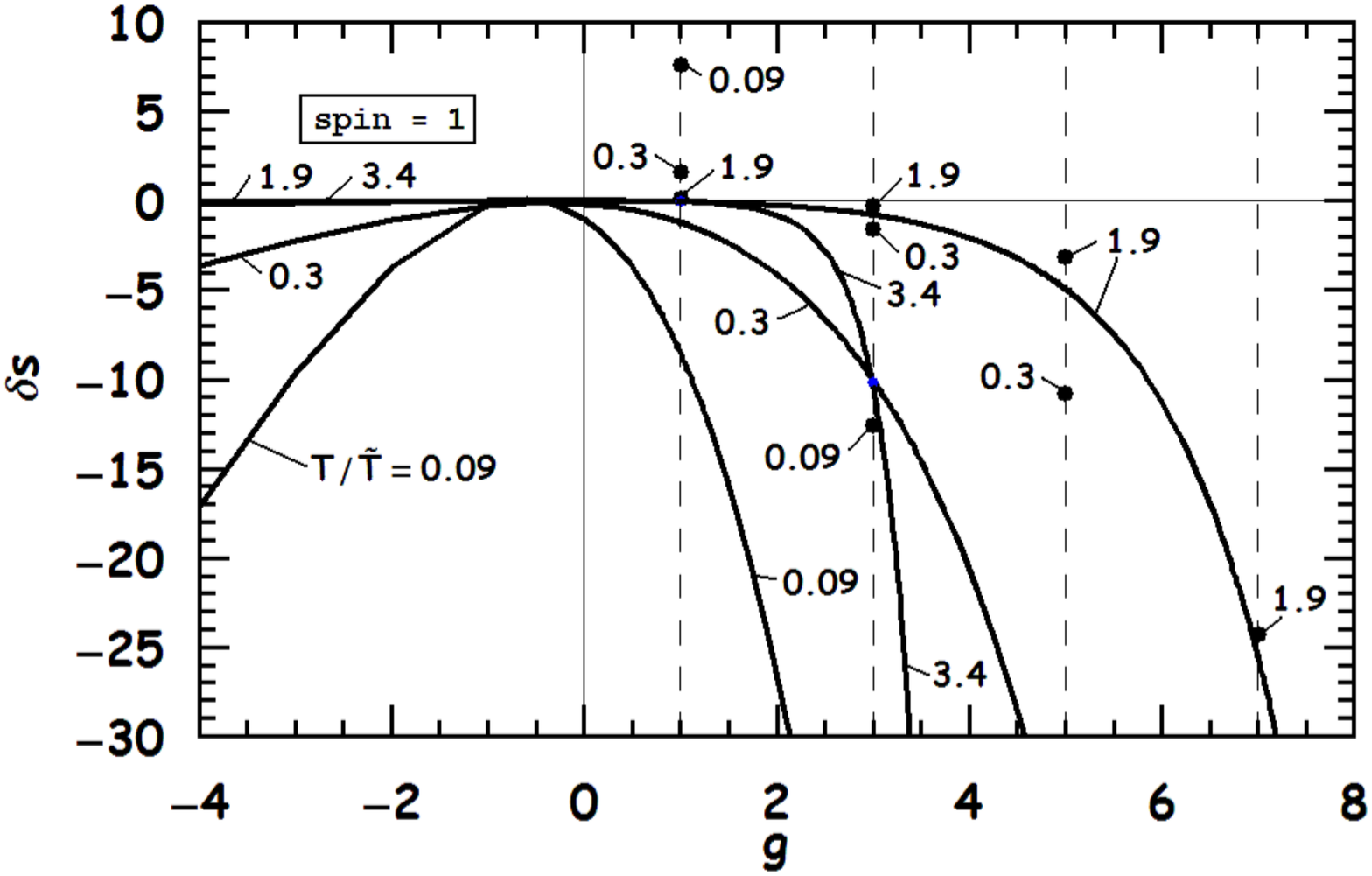}}
\vskip -20pt
\caption{Dimensionless entropy density shift for spin-one particles as a function of the parameter $g$ for various temperatures. Resonance values of $g$ are shown by vertical dashed lines.}
\label{deltasgs1}
\end{figure*}

 \textit{Spin-one particles}: Results for particles with spin unity, shown in Fig. \ref{deltasLgTs1} and Fig. \ref{deltasgs1}, are qualitatively the same as for spin zero particles, but with the addition of the special case of the P-wave resonance ($g=3$).


For Eq. (\ref{ds_a2}) to be a correction, the requirement
\begin{equation}
\label{ds_cond}
    \delta s \ll \frac{s_0}{\tilde{s}}
\end{equation}
must be met.
We can define the dimensionless entropy density shift boundary function
\begin{equation}
\label{DS}
    \Delta s(n,T) \equiv \frac{s_0}{\tilde{s}}=\frac{\frac{5}{2}-\log(n R^3)+\frac{3}{2}\log\left(\frac{R}{\lambda}\right)^2}{n R^3},
\end{equation}
which is important to know once the density $n$ is specified.

The thermodynamic identity 
\begin{equation}
\label{TDI}
    u=-p+T s+\mu n,
\end{equation}
used in conjunction with Eq. (\ref{pa2vir}) and Eq. (\ref{s_a2}) gives the kinetic energy density
\begin{equation}
\label{u_a2}
u-\frac{3}{2}n k_B T = n k_B T\left(\frac{3}{2} a_2(T)-T\frac{d a_2(T)}{T}\right)(n \lambda^3)+O\left((n \lambda^3)^2\right),
\end{equation}
Shifts due to interactions can be gauged by the dimensionless quantity
\begin{equation}
\label{du_a2}
\delta u\equiv\frac{u-u_0}{\tilde{u}}=\left(\frac{3}{2} a_2(T)-T\frac{d a_2(T)}{T}\right)\left(\frac{\lambda}{R}\right)^3
+O\left( \left(\frac{\lambda}{R}\right)^6 (n R^3) \right).
\end{equation}
For an ideal gas
\begin{equation}
\label{u0_a2}
u_0=\frac{3}{2}n k_B T ,
\end{equation}
with the energy density scale
\begin{equation}
\label{ut_a2}
\tilde{u}\equiv n k_B T (n R^3)=T \tilde{s}.
\end{equation}
Similar to Eq. (\ref{DS}), we can define a dimensionless energy density shift boundary function
\begin{equation}
\label{DU}
    \Delta u(n) \equiv \frac{u_0}{\tilde{u}}=\frac{3}{2}\frac{1}{n R^3}.
\end{equation}

The total energy density includes the mass term which dominates as $\epsilon\simeq m c^2 n$
\begin{equation}
\label{eps_a2}
\epsilon-\left(\frac{3}{2}n k_B T + m c^2 n\right)= n k_B T\left(\frac{3}{2} a_2(T)-T\frac{d a_2(T)}{dT}\right)(n \lambda^3)+O\left((n \lambda^3)^2\right),
\end{equation}
or
\begin{equation}
\label{deps_a2}
\delta \epsilon\equiv\frac{\epsilon-\epsilon_0}{\tilde{u}}=\left(\frac{3}{2} a_2(T)-T\frac{d a_2(T)}{T}\right)\left(\frac{\lambda}{R}\right)^3
+O\left( \left(\frac{\lambda}{R}\right)^6 (n R^3) \right),
\end{equation}
\begin{equation}
\label{eps0_a2}
\epsilon_0=\frac{3}{2}n k_B T + m c^2 n.
\end{equation}
Therefore the enthalpy, $h=\epsilon+p$, will be
\begin{equation}
\label{h_a2}
h-\left(\frac{5}{2}n k_B T+ m c^2 n\right)=n k_B T \left(\frac{5}{2}a_2(T)-T\frac{d a_2(T)}{dT}\right)(n \lambda^3)+O\left((n \lambda^3)^2\right),
\end{equation}
or
\begin{equation}
\label{dh_a2}
\delta h\equiv \frac{h-h_0}{\tilde{u}}= \left(\frac{5}{2}a_2(T)-T\frac{d a_2(T)}{dT}\right)\left(\frac{\lambda}{R}\right)^3
+O\left( \left(\frac{\lambda}{R}\right)^6 (n R^3) \right),
\end{equation}
\begin{equation}
\label{h0_a2}
h_0=\frac{5}{2}n k_B T+ m c^2 n,
\end{equation}
whence
\begin{equation}
\label{DEp}
    \Delta \epsilon(n,T) \equiv \frac{\epsilon_0}{\tilde{u}}=\frac{3}{2}\frac{1}{n R^3}+\frac{m c^2}{k \tilde{T}}\left(\frac{\lambda}{R}\right)^2,
\end{equation}
and
\begin{equation}
\label{DH}
    \Delta h(n,T) \equiv \frac{h_0}{\tilde{u}}=\frac{5}{2}\frac{1}{n R^3}+\frac{m c^2}{k \tilde{T}}\left(\frac{\lambda}{R}\right)^2.
\end{equation}
The specific heat per particle at constant volume is given by the integral \cite{Hirschfelder67}
\begin{equation}
\label{cvint}
    c_v-\frac{3}{2}k_B=-T\int_0^n \frac{\partial^2 p(\tilde{n},T)}{\partial T^2} \frac{d \tilde{n}}{\tilde{n}^2},
\end{equation}
where using the virial Eq. (\ref{pa2vir}), we obtain
\begin{equation}
\label{cv_a2}
c_v-\frac{3}{2}k_B =- k_B \left(\frac{3}{4}a_2(T)-T\frac{d a_2(T)}{dT}+T^2\frac{d^2 a_2(T)}{{dT}^2}\right)(n \lambda^3)+O\left((n \lambda^3)^2\right).
\end{equation}
Again, shifts due to interactions can be seen through the dimensionless quantity
\begin{equation}
\label{dcv_a2}
\delta c_v\equiv\frac{c_v-c_{v0}}{\tilde{c}}= \left(\frac{3}{4}a_2(T)-T\frac{d a_2(T)}{dT}+T^2\frac{d^2 a_2(T)}{{dT}^2}\right)\left(\frac{\lambda}{R}\right)^3
+O\left( \left(\frac{\lambda}{R}\right)^6 (n R^3) \right),
\end{equation}
with
\begin{equation}
\label{cv0_a2}
c_{v0}=\frac{3}{2}k_B,
\end{equation}
\begin{equation}
\label{ct_a2}
\tilde{c} \equiv k_B (n R^3),
\end{equation}
whence
\begin{equation}
\label{DCV}
    \Delta c_v(n) \equiv \frac{c_{v0}}{\tilde{c}}=\frac{3}{2}\frac{1}{n R^3}.
\end{equation}

\begin{figure*}[tb]
\centerline{\includegraphics[width=11cm]{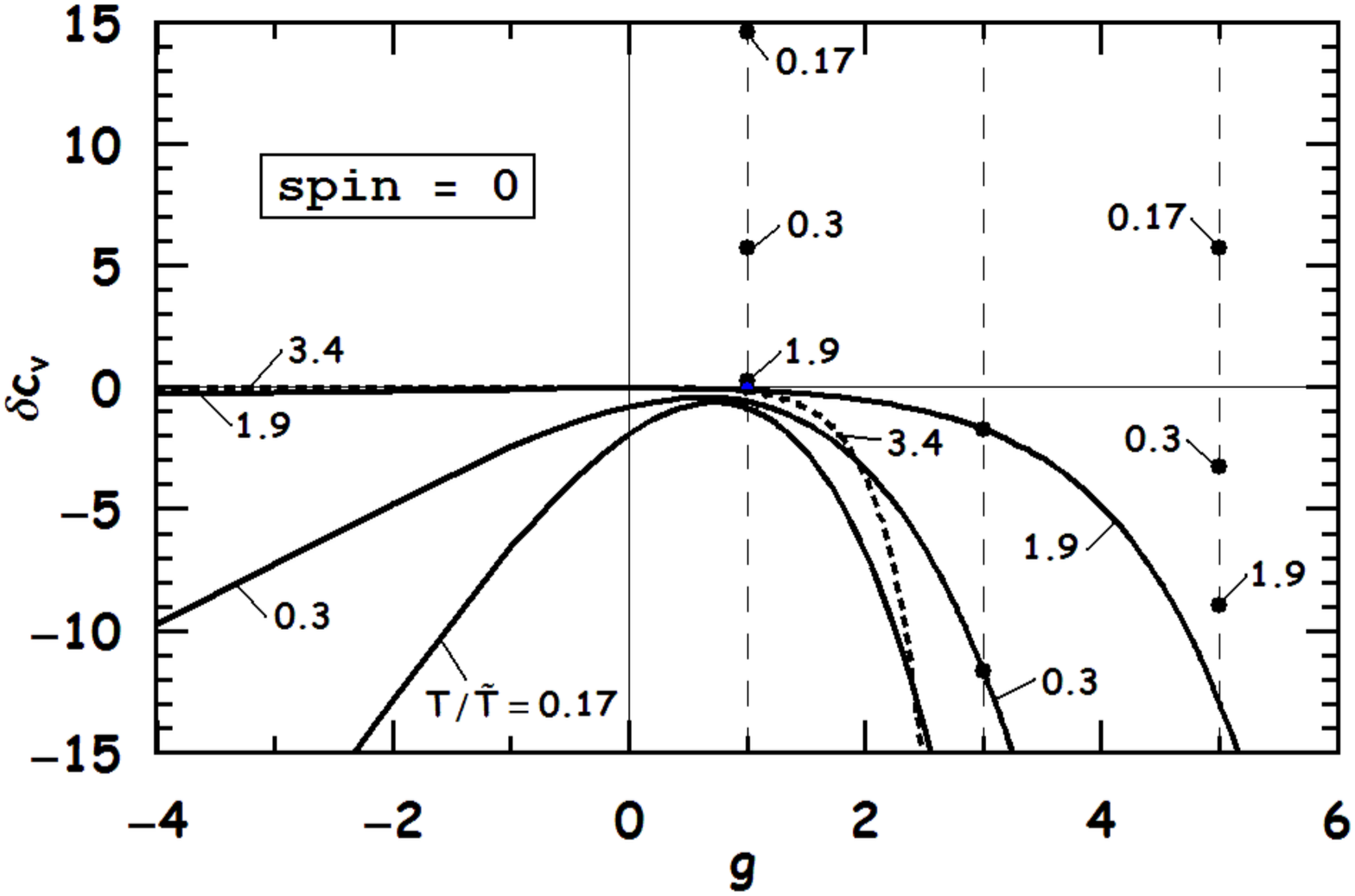}}
\vskip -20pt
\caption{Dimensionless density boundary of entropy for spin zero particles as a function of the parameter $g$ for various temperatures. Resonance values of $g$ are shown by vertical dashed lines.}
\label{deltacvgs0}
\end{figure*}

\begin{figure*}[tb]
\centerline{\includegraphics[width=11cm]{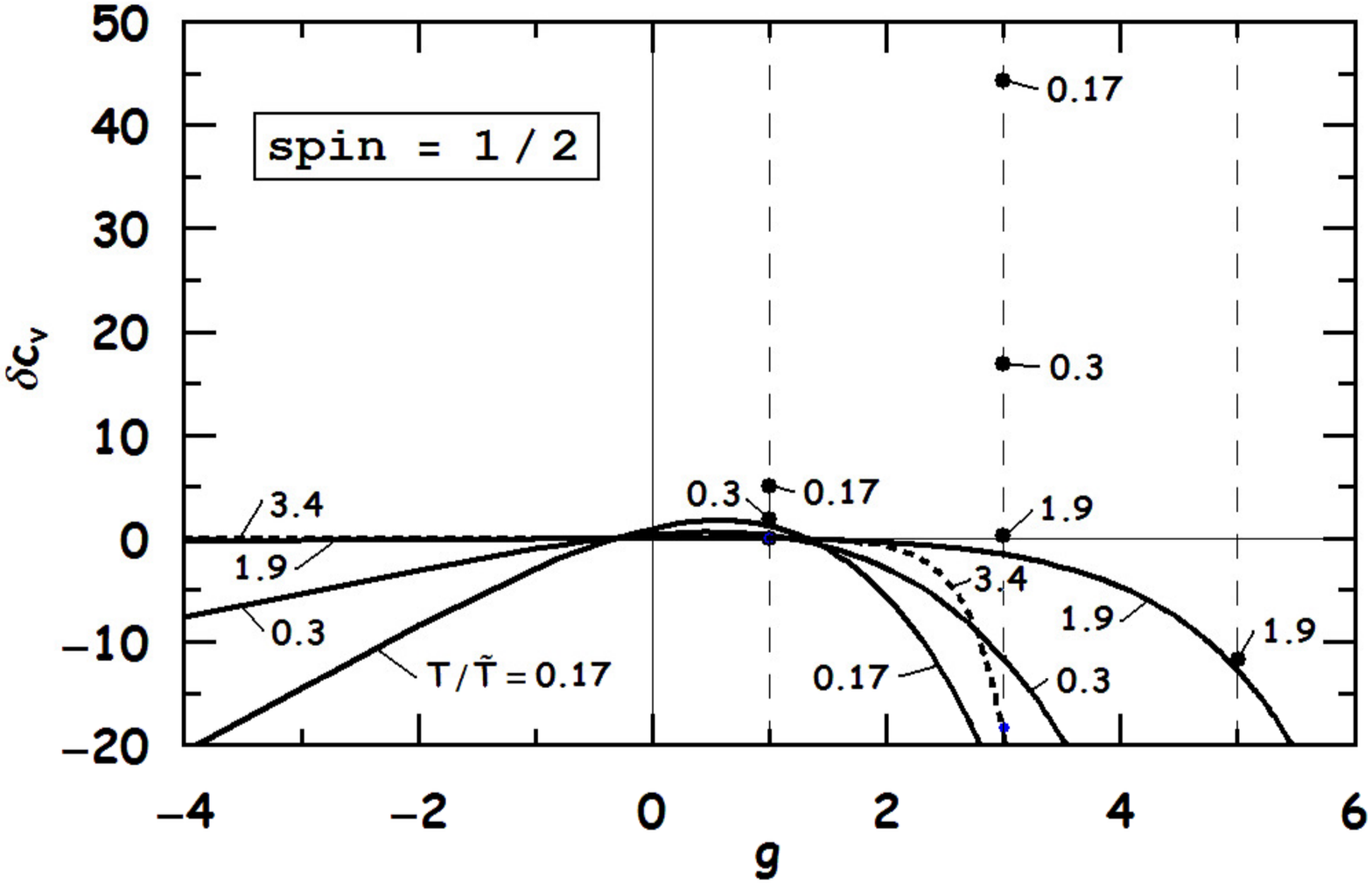}}
\vskip -20pt
\caption{Dimensionless density boundary of entropy for spin half particles as a function of the parameter $g$ for various temperatures. Resonance values of $g$ are shown by vertical dashed lines.}
\label{deltacvgs1o2}
\end{figure*}

\begin{figure*}[tb]
\centerline{\includegraphics[width=11cm]{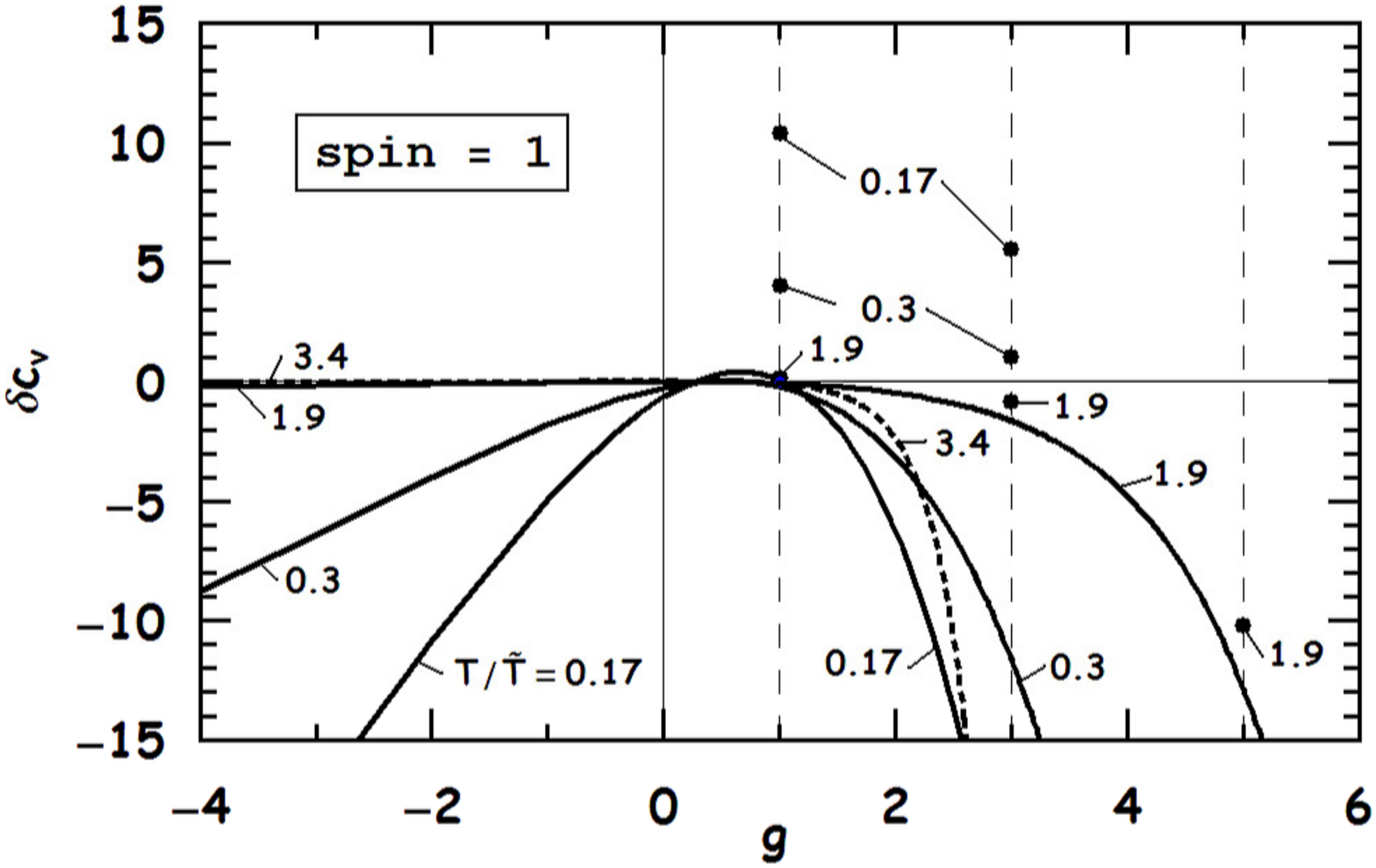}}
\vskip -20pt
\caption{Dimensionless density boundary of entropy for spin-one particles as a function of the parameter $g$ for various temperatures. Resonance values of $g$ are shown by vertical dashed lines.}
\label{deltacvgs1}
\end{figure*}

Fig. \ref{deltacvgs0}, Fig. \ref{deltacvgs1o2} and Fig. \ref{deltacvgs1} show the shifts in specific heat per particle. They have maximum values close to zero near $g=1$ and separated points for values at exact resonances.

The specific heat per particle at constant pressure is given by the ratio \cite{Hirschfelder67}
\begin{equation}
\label{cpder}
    c_p-c_v=\frac{T \left(\frac{\partial p(n,T)}{\partial T} \right)^2}{\left(\frac{\partial p(n,T)}{\partial n}\right) n^2},
\end{equation}
where using the virial Eq. (\ref{pa2vir}) again, we obtain
\begin{equation}
\label{cp_a2}
\begin{split}
c_p-\frac{5}{2}k_B =- k_B \left(\frac{15}{4}a_2(T)-3T\frac{d a_2(T)}{dT}+T^2\frac{d^2 a_2(T)}{{dT}^2}\right)(n \lambda^3)\\
+O\left((n \lambda^3)^2\right).
\end{split}
\end{equation}
or
\begin{equation}
\label{dcp_a2}
\begin{split}
\delta c_p \equiv \frac{c_p-c_{p0}}{\tilde{c}}=\left(\frac{15}{4}a_2(T)-3T\frac{d a_2(T)}{dT}+T^2\frac{d^2 a_2(T)}{{dT}^2}\right)\left(\frac{\lambda}{R}\right)^3\\
+O\left( \left(\frac{\lambda}{R}\right)^6 (n R^3) \right),
\end{split}
\end{equation}
and
\begin{equation}
\label{cp0_a2}
c_{p0}=\frac{5}{2}k_B.
\end{equation}
The dimensionless boundary function for $c_p$ is
\begin{equation}
\label{DCP}
    \Delta c_p(n) \equiv \frac{c_{p0}}{\tilde{c}}=\frac{5}{2}\frac{1}{n R^3}.
\end{equation}
Hence, the adiabatic constant $\gamma_a=\frac{c_p}{c_v}$ has the form
\begin{equation}
\label{gamma_adiam}
\gamma_a - \frac{5}{3}=-\left(\frac{15}{9}a_2(T)-\frac{8}{9}T\frac{d a_2(T)}{dT}-\frac{4}{9}T^2\frac{d^2 a_2(T)}{{dT}^2}\right)(n \lambda^3)+O\left((n \lambda^3)^2\right).
\end{equation}
The adiabatic sound speed is given by the expression \cite{Venugopalan92}
\begin{equation}
\label{c02exp}
    \left(\frac{c_0}{c}\right)^2=\frac{\gamma_a n}{h}\left(\frac{\partial p(n,T)}{\partial n}\right),
\end{equation}
where $c$ is the speed of light in vacuum. In the non-relativistic case, for which $h\simeq m c^2 n$,
\begin{equation}
\label{c02expNR}
    c_0^2=\frac{\gamma_a}{m}\left(\frac{\partial p(n,T)}{\partial n}\right),
\end{equation}
whence
\begin{equation}
\label{c02_a2}
c_0^2 - \frac{5}{3}\frac{k_B T}{m}=\frac{k_B T}{m}\left(\frac{5}{3}a_2(T)+\frac{8}{9}T\frac{d a_2(T)}{dT}+\frac{4}{9}T^2\frac{d^2 a_2(T)}{{dT}^2}\right)(n \lambda^3)+O\left((n \lambda^3)^2\right).
\end{equation}
Making it dimensionless, we obtain
\begin{equation}
\label{dc02_a2}
\begin{split}
\delta c_0^2 \equiv \frac{c_0^2-c_i^2}{\tilde{c}_0^2}=\left(\frac{5}{3}a_2(T)+\frac{8}{9}T\frac{d a_2(T)}{dT}+\frac{4}{9}T^2\frac{d^2 a_2(T)}{{dT}^2}\right)
\left(\frac{\lambda}{R}\right)^3\\
+O\left( \left(\frac{\lambda}{R}\right)^6 (n R^3) \right),
\end{split}
\end{equation}
\begin{equation}
\label{ci2_a2}
c_i^2 = \frac{5}{3}\frac{k_B T}{m},
\end{equation}
\begin{equation}
\label{c02t_a2}
\tilde{c}_0^2 = \frac{k_B T}{m}(n R^3).
\end{equation}
The boundary is provided by
\begin{equation}
\label{DC2}
    \Delta c_0^2(n) \equiv \frac{c_{i}^2}{\tilde{c}_0^2}=\frac{5}{3}\frac{1}{n R^3}.
\end{equation}
Let us emphasize that the corrections above are only valid for the range of temperatures and densities for which
\begin{equation}
\label{corr_val}
    \Delta F \gg \delta F,
\end{equation}
where $F$ is any of the thermodynamical property calculated in this section.
Therefore, we define the density function $n_F(T)$ which is found from the equation
\begin{equation}
\label{corr_val_eqn}
    \left| \Delta F(n_{F}(T),T) \right|= \left| \delta F(T) \right|,
\end{equation}
and the requirement $n \ll n_F(T)$ reflects the fact that the system at fixed temperature $T$ must be dilute enough 
for the virial approximation for $F$ to be valid. The critical quantum density 
\begin{equation}
\label{nQ}
	n_Q(T) \equiv \frac{1}{\lambda^3(T)},
\end{equation}
also signals the degenerate quantum regime as the system becomes dense. 

\begin{figure*}[tb]
\centerline{\includegraphics[width=11cm]{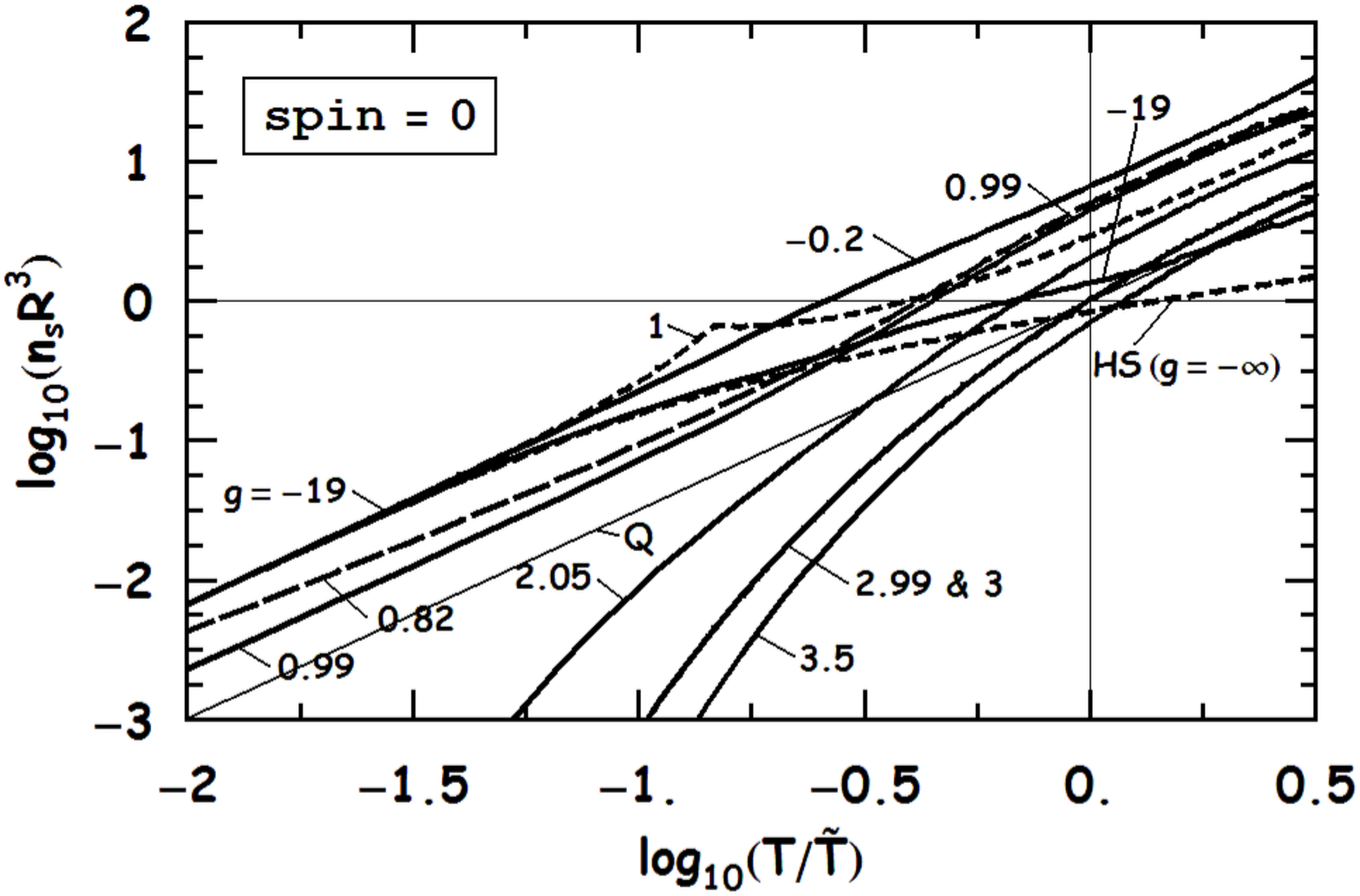}}
\vskip -20pt
\caption{Dimensionless density boundary of entropy for spin zero particles as a function of temperature for various values of the parameter $g$. Special cases are shown by dashed lines and the critical quantum density is labeled by 'Q'.}
\label{nsR3Ts0}
\end{figure*}

\begin{figure*}[tb]
\centerline{\includegraphics[width=11cm]{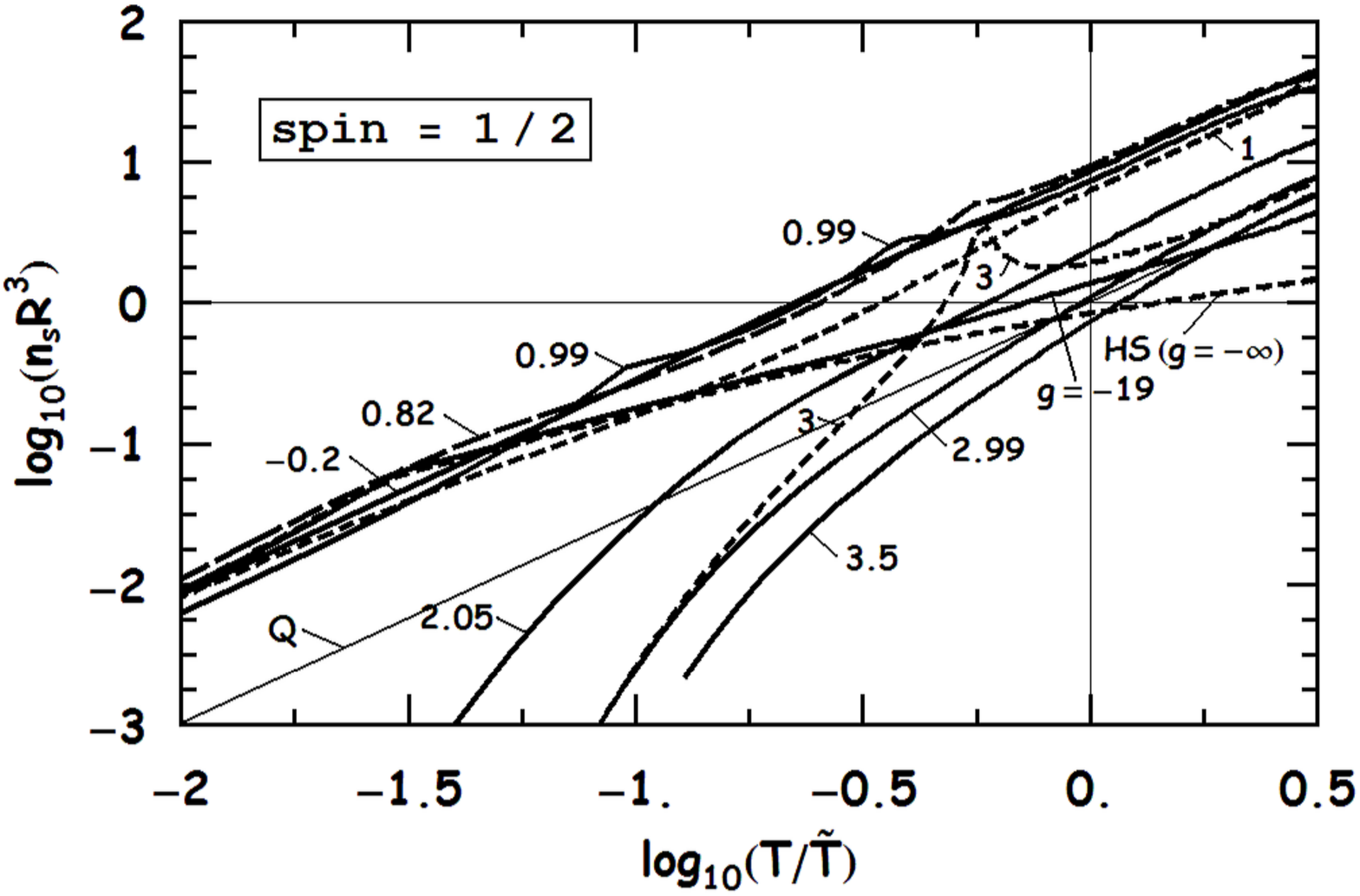}}
\vskip -20pt
\caption{Dimensionless density boundary of entropy for spin half particles as a function of temperature for various values of the parameter $g$. Special cases are shown by dashed lines and critical quantum density is labeled by 'Q'.}
\label{nsR3Ts1o2}
\end{figure*}

\begin{figure*}[tb]
\centerline{\includegraphics[width=11cm]{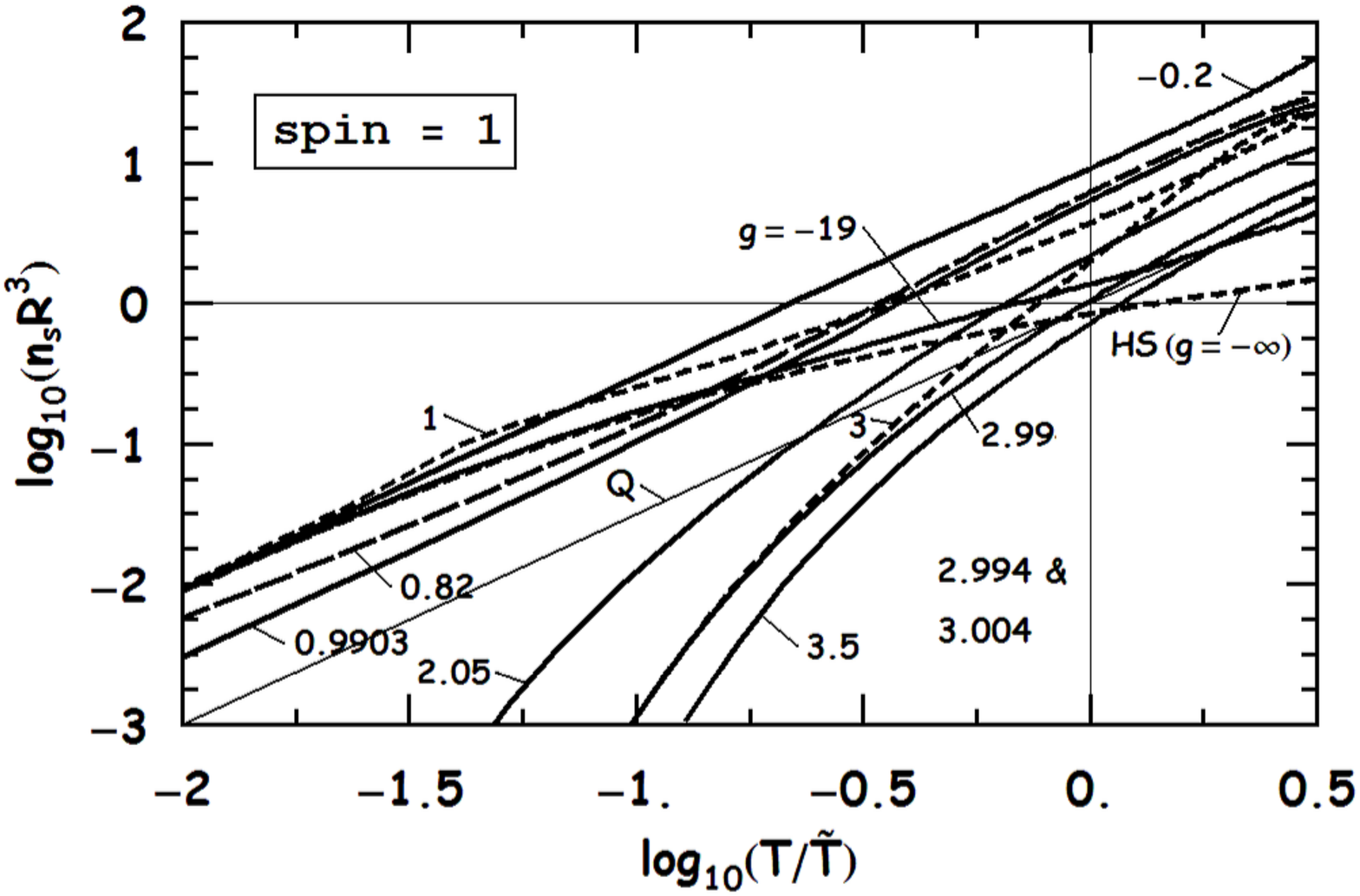}}
\vskip-20pt
\caption{Dimensionless density boundary of entropy for spin-one particles as a function of temperature for various values of the parameter $g$. Special cases are shown by dashed lines and critical quantum density is labeled by 'Q'.}
\label{nsR3Ts1}
\end{figure*}

The dimensionless boundary density $n_s R^3$ for the entropy is shown in Fig. \ref{nsR3Ts0}, Fig. \ref{nsR3Ts1o2} and Fig. \ref{nsR3Ts1} as a function of temperature for various values of $g$. For the system to be dilute, the density must fall below these boundary values of density.

\begin{figure*}[tb]
\centerline{\includegraphics[width=11cm]{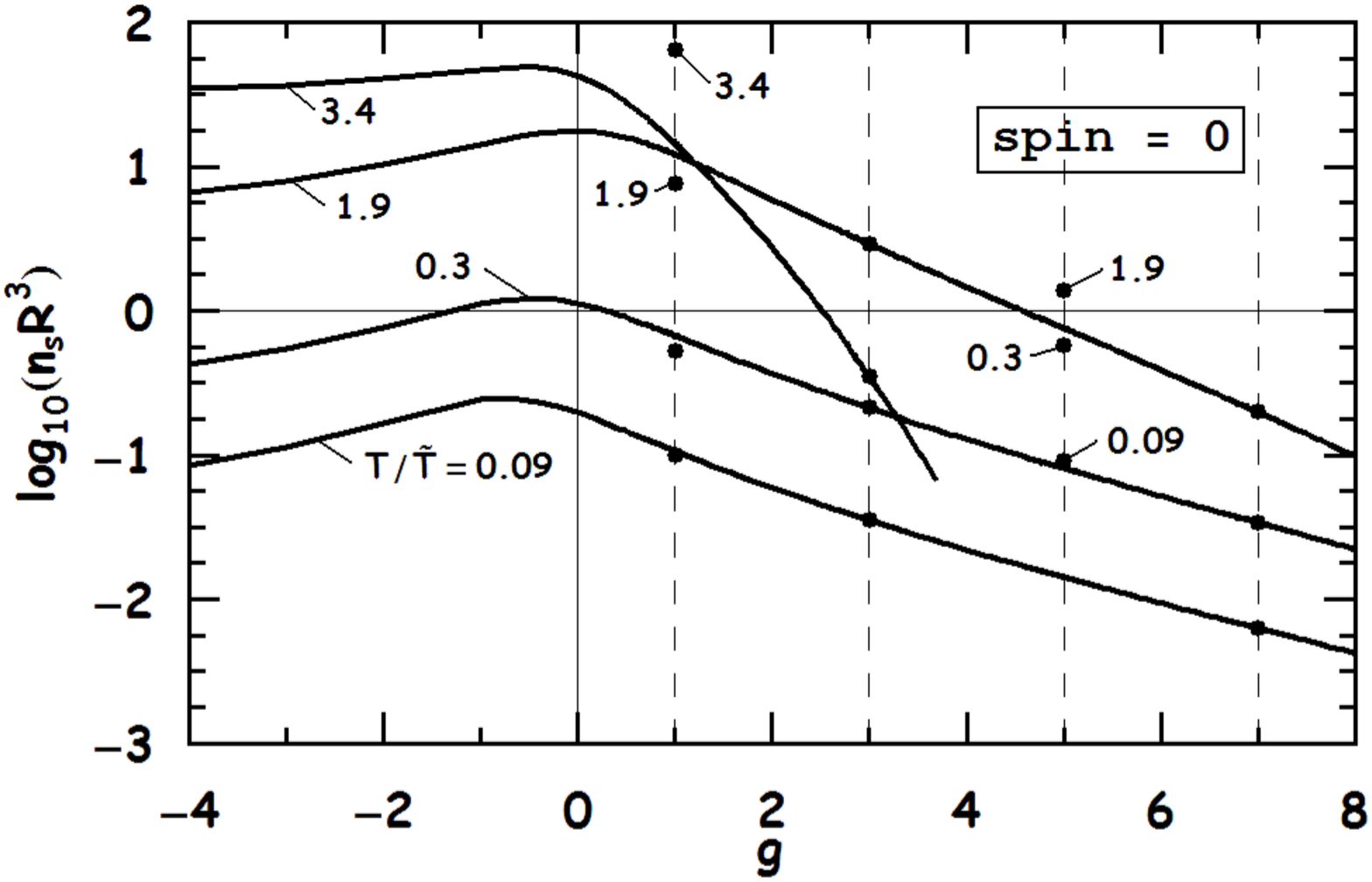}}
\vskip -20pt
\caption{Dimensionless density boundary of entropy for spin zero particles as a function of the parameter $g$ for various temperatures. Resonance values of $g$ are shown by vertical dashed lines.}
\label{nsR3gs0}
\end{figure*}

\begin{figure*}[tb]
\centerline{\includegraphics[width=11cm]{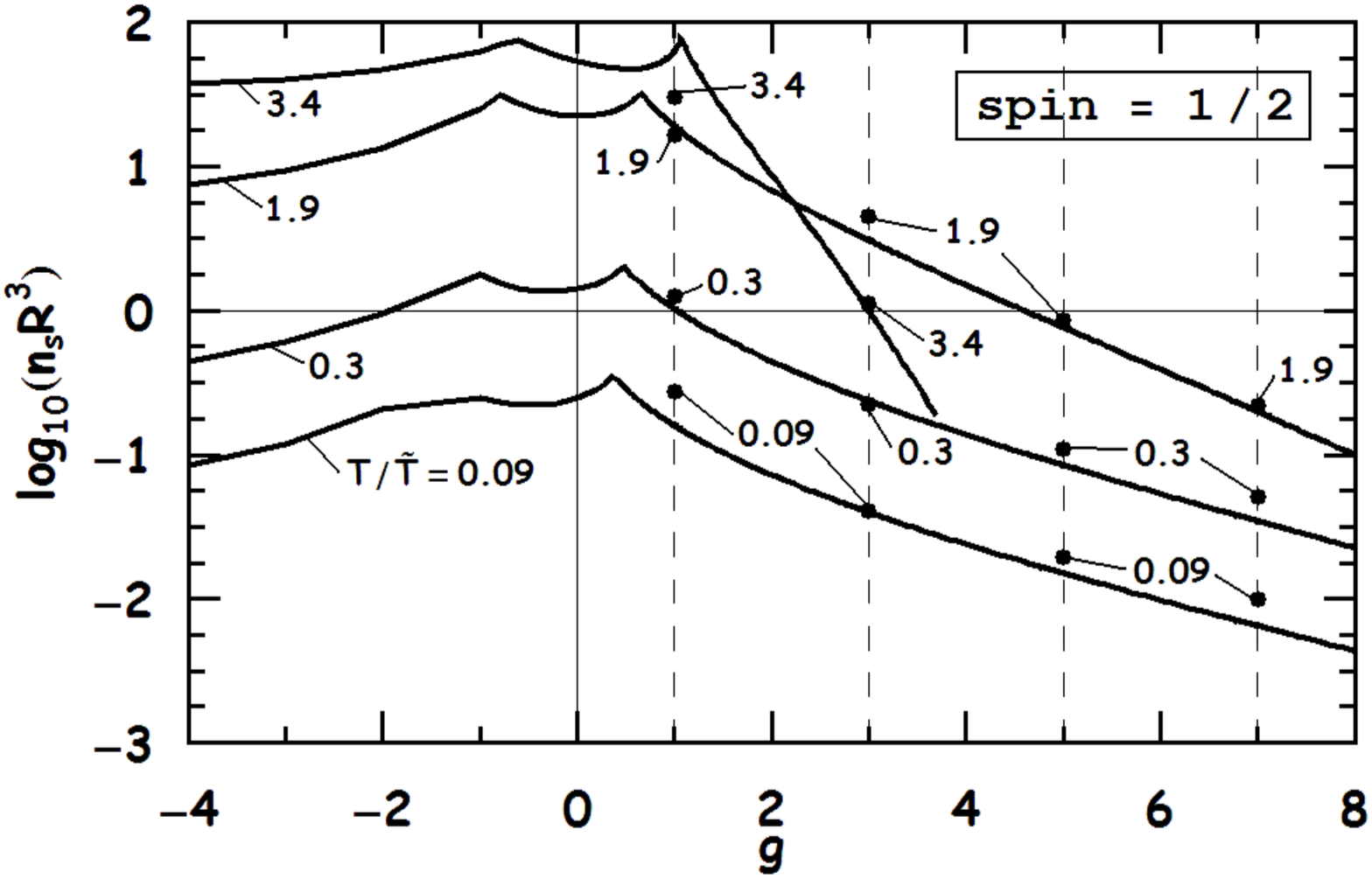}}
\vskip-20pt
\caption{Dimensionless density boundary of entropy for spin half particles as a function of the parameter $g$ for various temperatures. Resonance values of $g$ are shown by vertical dashed lines.}
\label{nsR3gs1o2}
\end{figure*}

\begin{figure*}[tb]
\centerline{\includegraphics[width=11cm]{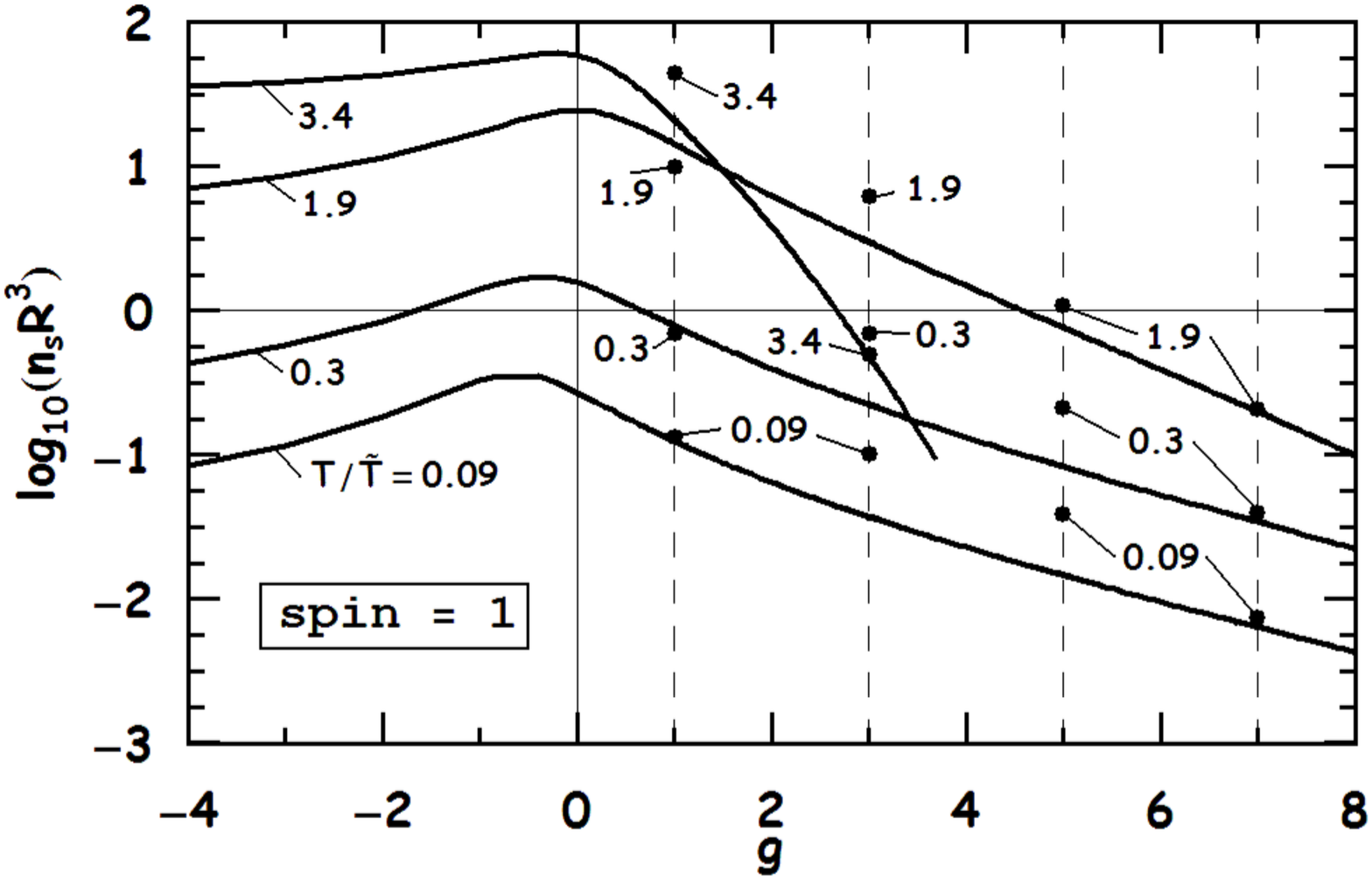}}
\vskip -20pt
\caption{Dimensionless density boundary of entropy for spin-one particles as a function of the parameter $g$ for various temperatures. Resonance values of $g$ are shown by vertical dashed lines.}
\label{nsR3gs1}
\end{figure*}

The same dimensionless boundary density, in Fig. \ref{nsR3gs0}, Fig. \ref{nsR3gs1o2} and Fig. \ref{nsR3gs1}, is shown as a function of the strength parameter $g$ for various temperatures.

\begin{figure*}[tb]
\centerline{\includegraphics[width=11cm]{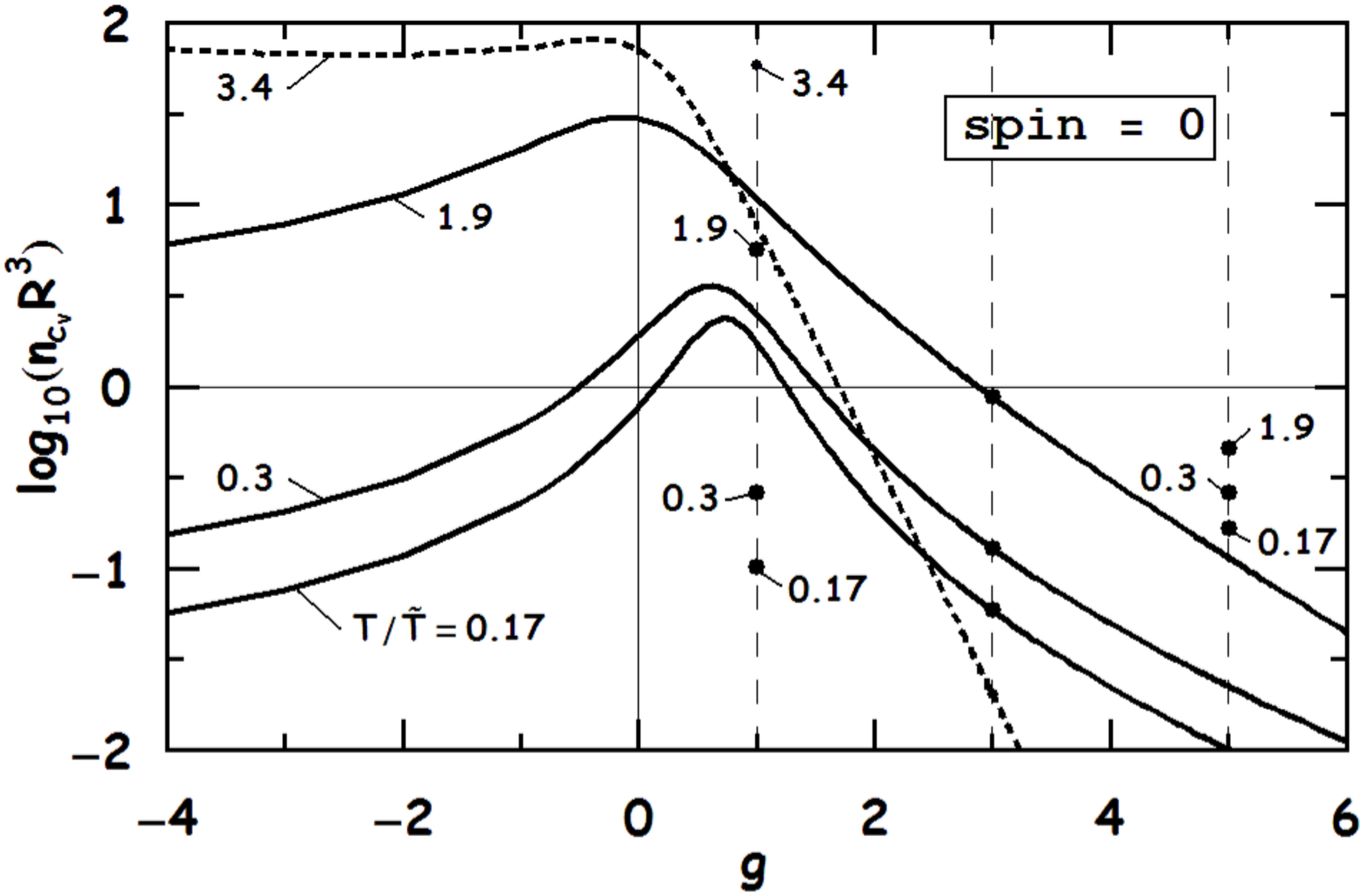}}
\vskip -20pt
\caption{Dimensionless density boundary of entropy for spin zero particles as a function of the parameter $g$ for various temperatures. Resonance values of $g$ are shown by vertical dashed lines.}
\label{nsR3cvgs0}
\end{figure*}

\begin{figure*}[tb]
\centerline{\includegraphics[width=11cm]{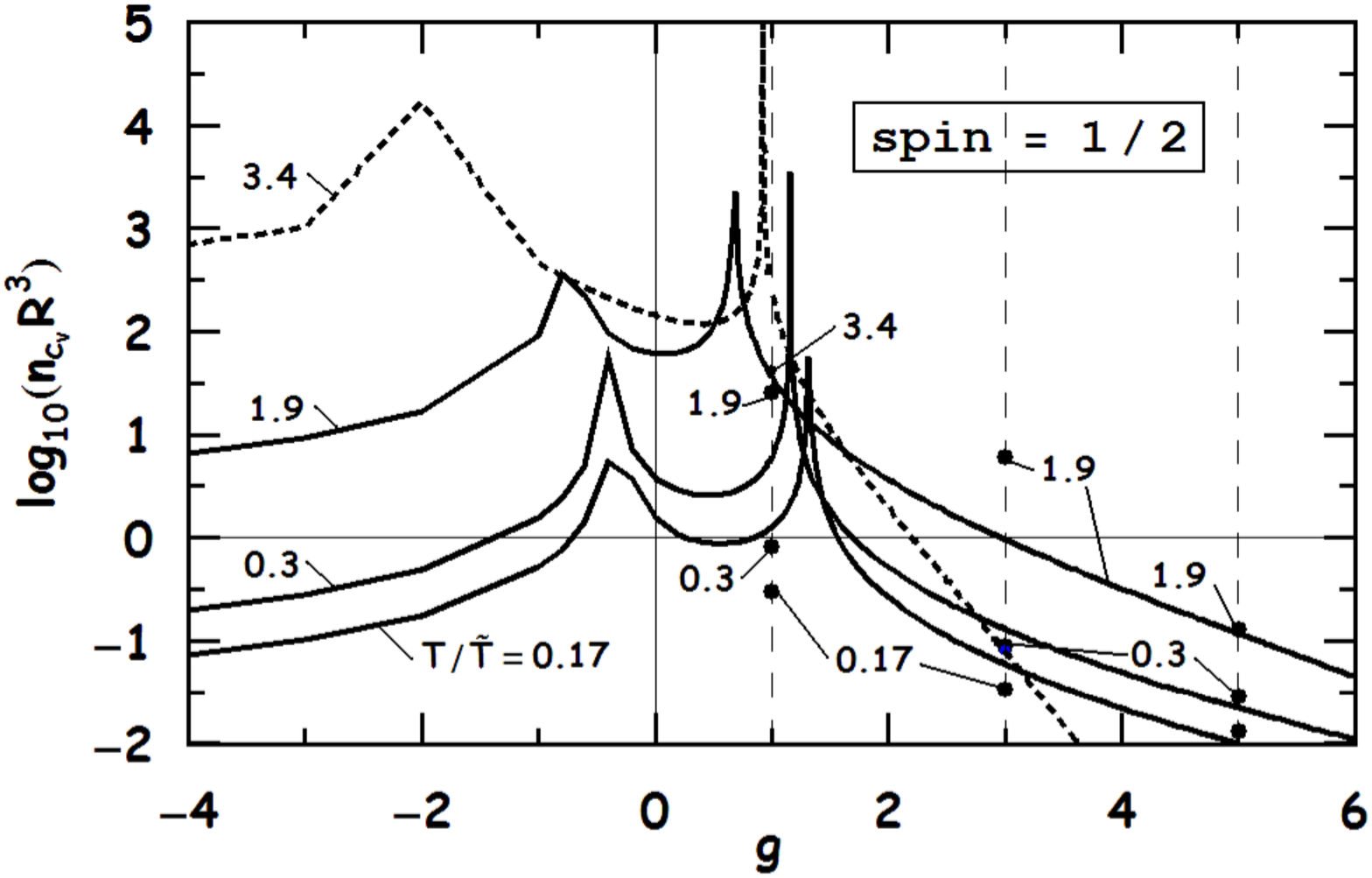}}
\vskip -20pt
\caption{Dimensionless density boundary of entropy for spin half particles as a function of the parameter $g$ for various temperatures. Resonance values of $g$ are shown by vertical dashed lines.}
\label{nsR3cvgs1o2}
\end{figure*}

\begin{figure*}[tb]
\centerline{\includegraphics[width=11cm]{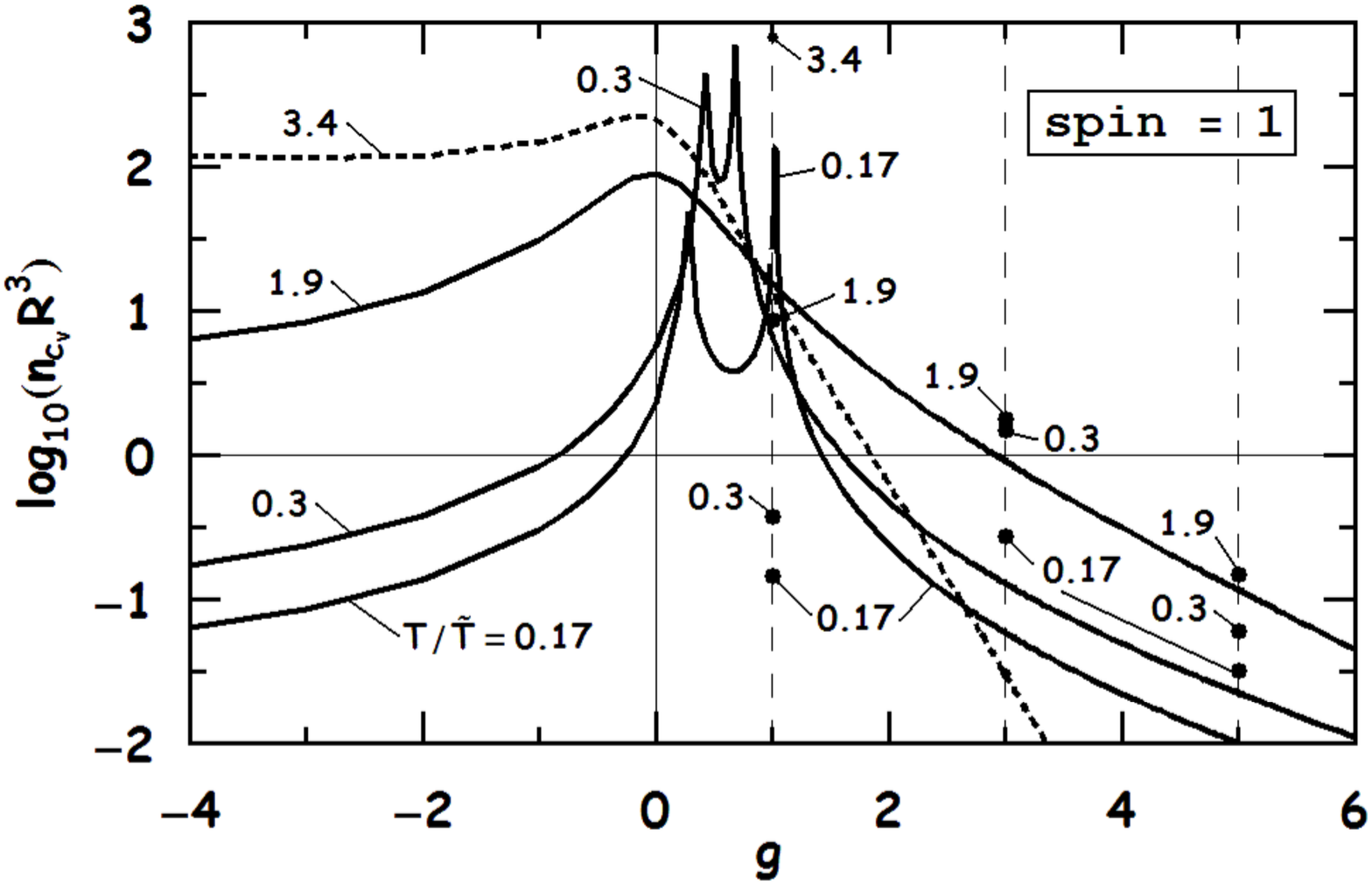}}
\vskip -20pt
\caption{Dimensionless density boundary of entropy for spin-one particles as a function of the parameter $g$ for various temperatures. Resonance values of $g$ are shown by vertical dashed lines.}
\label{nsR3cvgs1}
\end{figure*}

The boundary density for the specific heat is shown in Fig. \ref{nsR3cvgs0}, Fig. \ref{nsR3cvgs1o2} and Fig. \ref{nsR3cvgs1} as a function of the strength parameter $g$ for various temperatures. Values at resonances stand out as pointed out by the solid dots.

The first and second derivatives of the second virial coefficient are found by calculating the following integrals
\begin{equation}
\label{I2l}
    I^{(2)}_l(T)\equiv\frac{T}{\gamma}\frac{d I_l(T)}{dT}=\int_0^\infty dx \left [ -\frac{\partial \delta_l(x)}{\partial x}\right ] x^2 e^{-\gamma(T) \, x^2},
\end{equation}
and
\begin{equation}
\label{I4l}
    I^{(4)}_l(T)\equiv\frac{T}{\gamma}\frac{d I^{(2)}_l(T)}{dT}=\int_0^\infty dx \left [ -\frac{\partial \delta_l(x)}{\partial x}\right ] x^4 e^{-\gamma(T) \, x^2}.
\end{equation}
Hence
\begin{equation}
\label{da2I2}
    T\frac{d a_2(T)}{dT}=-2^{\frac{3}{2}}\left({\sum_{Bound}}'g_E \, \beta E_{Bound} e^{-\beta E_{Bound}}
    -\frac{\gamma}{\pi}{\sum_l}'(2l+1)I^{(2)}_l(T)\right),
\end{equation}
and
\begin{equation}
\label{da2I4}
\begin{split}
    T^2\frac{d^2 a_2(T)}{{dT}^2}=&-2^{\frac{3}{2}}\bigl[{\sum_{Bound}}'g_E \, \beta E_{Bound}(\beta E_{Bound}-2) e^{-\beta E_{Bound}}\\
    &+\frac{\gamma}{\pi}{\sum_l}'(2l+1)\left(2 I^{(2)}_l(T)-\gamma I^{(4)}_l(T)\right) \bigr],
\end{split}
\end{equation}
where $g_E=2l+1$ is the multiplicity of the bound state.


To conclude this section, we note that 
the second virial coefficient corrections to the thermodynamic state variables  for a dilute gas of particles interacting through a delta-shell potential
exhibit distinct features when the interaction is tuned to the unitary limit and when resonances occur for different partial waves. 


\clearpage

\section*{Transport properties of a dilute non-relativistic delta-shell gas}

\section{Transport coefficients}
\label{p3sec1}

If a system is disturbed from equilibrium, net flows of mass,
energy and momentum are generated. In the first approximation, these flows are
described by coefficients of diffusion, thermal conductivity and
viscosity. A detailed description of the theory
based on an approximate solution of the Boltzmann equation is given
in Chapman\cite{Chapman70} and Hirschfelder\cite{Hirschfelder67}.
The formalism described in these references is adopted here to
calculate results for the delta-shell potential.

The transport cross-section of order $n$ is
given by the integral\cite{Chapman70}
\begin{equation}
  \phi^{(n)}=2 \pi
  \int_{-1}^{+1}d\cos\theta(1-\cos^n\theta)\frac{d\sigma(k,\theta)}{d\Omega}\bigr|_{c.m.},
\label{tr_crosssec}
\end{equation}
where the scattering angle $\theta$ and the collisional differential
cross-section $\frac{d\sigma(k,\theta)}{d\Omega}\bigr|_{c.m.}$ are
calculated in the center of mass reference frame of the two colliding
particles. For indistinguishable particles, an expansion of the amplitude in partial
waves $ \, \sum'_{l} (2l+1)(e^{i 2 \delta_l}-1) P_l(\cos\theta)$
and the orthogonality of the Legendre polynomials $P_l$ simplifies the
above integral to the infinite sums
\begin{equation}
q^{(1)}_{Fermi,Bose}\equiv\frac{\phi^{(1)}}{4\pi
R^2}=\frac{2}{x^2}{\sum_{l}}'(2l+1)\sin^2(\delta_l),
 \label{tr_crosssec_sum1}
\end{equation}
\begin{equation}
q^{(2)}_{Fermi,Bose}\equiv\frac{\phi^{(2)}}{4\pi
R^2}=\frac{2}{x^2}{\sum_{l}}'\frac{(l+1)(l+2)}{(2l+3)}\sin^2(\delta_{l+2}-\delta_l),
 \label{tr_crosssec_sum2}
\end{equation}
where the prime on the summation sign indicates the use of even $l$
for Bosons and odd $l$ for Fermions.  In
Eqs.
~(\ref{tr_crosssec_sum2}), the low
energy hard-sphere cross section $4 \pi R^2$
has been used to render the transport cross sections dimensionless. If
the particles possess spin $s$, then the properly symmetrized forms
are:
\begin{equation}
\begin{array}{ll}
q^{(n)}_{(s)} = \frac{s+1}{2s+1}~q^{(n)}_{Bose} + 
\frac{s}{2s+1}~q^{(n)}_{Fermi},& \text{for integer } s\,,\\
q^{(n)}_{(s)} = \frac{s+1}{2s+1}~q^{(n)}_{Fermi} + 
\frac{s}{2s+1}~q^{(n)}_{Bose},& \text{for half-integer } s\,. \nonumber
\end{array}
\end{equation}
The transport coefficients are given in terms of the transport integrals
\begin{equation}
  \omega_\alpha^{(n,t)}(T)\equiv 
\int_0^{\infty}d\gamma \, e^{-\alpha\gamma^2}\gamma^{2t+3}q^{(n)}(x)\,,
\label{tr_ints}
\end{equation}
where $\alpha$ is a pure number and $\gamma=\frac{\hbar k}{\sqrt{2 \mu k_B T}}=\frac{x}{\sqrt{2
\pi}} \left(\frac{\lambda(T)}{R}\right)$ (the quantity $\gamma$ can also be thought of as the ratio of relative velocity to the average thermal velocity) with the thermal de-Broglie
wavelength $\lambda=(2\pi\hbar^2/m \, k_BT)^{1/2}$. We also note the
useful relation
\begin{equation}
  \omega_\alpha^{(n,t)}(T)=\frac{1}{\alpha^{t+2}}\omega_{1}^{(n,t)}(T/\alpha).
\label{tr_ints_rel}
\end{equation}

For numerical calculations it is useful to rewrite the expressions using the definitions
\begin{equation}
  \Omega_{\alpha,l}^{(1,t)}(T)\equiv 
\int_0^{\infty}d\gamma \, e^{-\alpha\gamma^2}\gamma^{2t+3} \sin^2 \left( \delta_l(\gamma) \right)\,,
\label{tr_ints_s1}
\end{equation}
\begin{equation}
  \Omega_{\alpha,l}^{(2,t)}(T)\equiv 
\int_0^{\infty}d\gamma \, e^{-\alpha\gamma^2}\gamma^{2t+3} \sin^2 \left( \delta_{l+2}(\gamma)-\delta_{l}(\gamma) \right)\,.
\label{tr_ints_s2}
\end{equation}
Then, for integer $s$,
\begin{eqnarray}
  \omega_{\alpha,s}^{(1,t)}(T)=\frac{(\lambda/R)^2}{\pi}\left[ \frac{s+1}{2s+1}\sum_{even \,l}(2l+1)\Omega_{\alpha,l}^{(1,t)}(T)\right.\\
  \left.+\frac{s}{2s+1}\sum_{odd \, l}(2l+1)\Omega_{\alpha,l}^{(1,t)}(T)\right],
\label{om1ts}
\end{eqnarray}
\begin{eqnarray}
  \omega_{\alpha,s}^{(2,t)}(T)=\frac{(\lambda/R)^2}{\pi}\left[ \frac{s+1}{2s+1}\sum_{even \, l}\frac{(l+1)(l+2)}{(2l+3)}\Omega_{\alpha,l}^{(2,t)}(T)\right.\\
  \left.+\frac{s}{2s+1}\sum_{odd \, l}\frac{(l+1)(l+2)}{(2l+3)}\Omega_{\alpha,l}^{(2,t)}(T)\right],
\label{om2ts}
\end{eqnarray}
and, for half-integer $s$,
\begin{eqnarray}
  \omega_{\alpha,s}^{(1,t)}(T)=\frac{(\lambda/R)^2}{\pi}\left[ \frac{s}{2s+1}\sum_{even \, l}(2l+1)\Omega_{\alpha,l}^{(1,t)}(T)\right.\\
   \left.+\frac{s+1}{2s+1}\sum_{odd \, l}(2l+1)\Omega_{\alpha,l}^{(1,t)}(T)\right],
\label{om1tsh}
\end{eqnarray}
\begin{eqnarray}
  \omega_{\alpha,s}^{(2,t)}(T)=\frac{(\lambda/R)^2}{\pi}\left[ \frac{s}{2s+1}\sum_{even \, l}\frac{(l+1)(l+2)}{(2l+3)}\Omega_{\alpha,l}^{(2,t)}(T)\right.\\
   \left.+\frac{s+1}{2s+1}\sum_{odd \, l}\frac{(l+1)(l+2)}{(2l+3)}\Omega_{\alpha,l}^{(2,t)}(T)\right].
\label{om2tsh}
\end{eqnarray}

In what follows, the coefficients of self diffusion
$\mathcal{D}$, shear viscosity $\eta$, and thermal conductivity $\kappa$
are normalized to the corresponding hard-sphere-like values 
\begin{equation}
\tilde{\mathcal{D}}=\frac{3\sqrt{2}}{32}\frac{\hbar}{m n R^3 }\,,\quad
\tilde{\eta} =\frac{5 \sqrt {2}}{32}\frac{\hbar}{R^3}\,,~~
\text{and} ~~
\tilde{\kappa}=\frac{75}{64 \sqrt{2}} \frac{\hbar k_B}{m R^3}\,.  
\label{chars}
\end{equation}
In the first order of deviations from the equilibrium distribution
function, the transport coefficients are 
\begin{eqnarray}
\frac{\left[\mathcal{D}\right]_1}{\tilde{\mathcal{D}}} = 
\left(\frac{R}{\lambda(T)}\right)\frac{1}{\omega_{1,s}^{(1,1)}(T)}\,, \nonumber\\
\frac{\left[\eta \right]_1}{\tilde{\eta}} = 
\frac{\left[\kappa \right]_1}{\tilde{\kappa}} = 
\left(\frac{R}{\lambda(T)}\right)\frac{1}{\omega_{1,s}^{(2,2)}(T)}\,,
\label{Diffvisq}
\end{eqnarray}
Eq. (\ref{Diffvisq}) shows clearly that if $\omega_{1,s}^{(2,2)}$ is
$T$-independent (as for hard-spheres with a constant cross section),
the shear viscosity exhibits a $T^{1/2}$ dependence which arises
solely from its inverse dependence with $\lambda (T)$. For
energy-dependent cross sections, however, the temperature dependence
of the viscosity is sensitive also to the temperature dependence of
the omega-integral.

Note that in the first-order approximation, \\
the specific heat capacity for a
monoatomic gas is expressible as
\begin{equation}
c_{vo} = \frac{2m}{5}\frac{\left[\kappa \right]_1}{\left[\eta
\right]_1}=\frac{3 k_B}{2}.
\label{termo}
\end{equation}
The second-order results can be cast as 
\begin{equation}
\frac{\left[\mathcal{C}\right]_2}{\left[\mathcal{C}\right]_1} = 
\left(1+\delta_{\mathcal{C}}(T)\right)
\left(1\pm n \lambda^3 \epsilon_{\mathcal{C}}(T)\right)\,,
\label{Diff2}
\end{equation}
where $\mathcal{C}$ is $\mathcal{D}$ or $\eta$ or $\kappa$, and  the $\pm$
refers to Bose ($+$) and Fermi ($-$) statistics. Explicitly, 
\begin{equation}
\delta_{\eta} \equiv 
\frac{3(7\omega_{1,s}^{(2,2)}-2\omega_{1,s}^{(2,3)})^2}
{2\left(\omega_{1,s}^{(2,2)}\left(77\omega_{1,s}^{(2,2)} + 
6\omega_{1,s}^{(2,4)}\right)-6\left(\omega_{1,s}^{(2,3)}\right)^2\right)}\,,
\label{delta_eta} \\
\end{equation}
\begin{equation}
\delta_{\mathcal{D}}\equiv
\frac{(5\omega_{1,s}^{(1,1)}-2\omega_{1,s}^{(1,2)})^2/\omega_{1,s}^{(1,1)}}
{\left(30\omega_{1,s}^{(1,1)} +
4\omega_{1,s}^{(1,3)} + 
8\omega_{1,s}^{(2,2)}\right) - 
4\frac{\left(\omega_{1,s}^{(1,2)}\right)^2}{\omega_{1,s}^{(1,1)}}},
\label{dDiff2}
\end{equation}
\begin{equation}
\delta_\kappa\equiv\frac{(7\omega_{1,s}^{(2,2)}-2\omega_{1,s}^{(2,3)})^2}
{4\left(\omega_{1,s}^{(2,2)}\left(7\omega_{1,s}^{(2,2)}(T) + 
\omega_{1,s}^{(2,4)}\right)-\left(\omega_{1,s}^{(2,3)}\right)^2\right)},
\label{dtermo2}
\end{equation}
and
\begin{equation}
\epsilon_{\kappa}\equiv 2^{-7/2} 
\left[7-\frac{128}{3^{3/2}}
\frac{\omega_{4/3,s}^{(2,3)}+6\omega_{4/3,s}^{(2,2)}}{9\omega_{1,s}^{(2,2)}}\right],
\label{epskappa}
\end{equation}
\begin{equation}
\epsilon_{\eta} \equiv 
2^{-7/2} \left[4-\frac{128}{3^{3/2}}
\frac{\omega_{4/3,s}^{(2,2)}}{\omega_{1,s}^{(2,2)}}\right]\,.
\label{epseta}
\end{equation}
It is worthwhile to note that at the first order of deviations from the
equilibrium distribution function, all the dimensionless transport coefficients are
independent of density, unless density dependent cut-offs are used to
delimit the transport cross sections. An explicit density dependence
arises only at the second order.

It is useful to define a characteristic temperature
\begin{equation}
\tilde{T}\equiv\frac{2\pi\hbar^2}{k_Bm R^2}~~\text{or}~~
\frac{T}{\tilde{T}}=\left(\frac{R}{\lambda}\right)^2
\label{Ttilde}
\end{equation}
in terms of which limiting forms of the transport coefficients can be
studied.

\section{Delta-shell gas and resonances}
\label{p3sec2}

\begin{figure*}[th]
\centerline{\includegraphics[width=11cm]{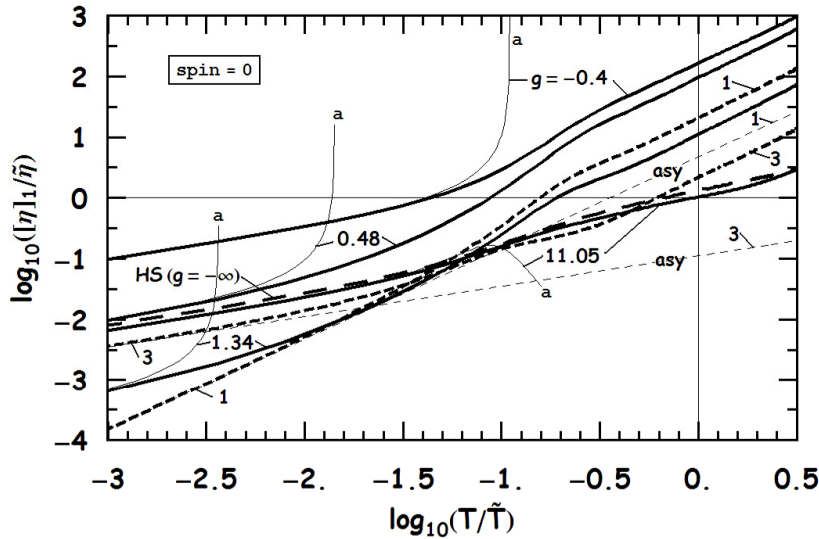}}
\vskip -30pt
  \caption{Normalized (to $\tilde{\eta}$ in Eq.~(\ref{chars}))
diffusion coefficient for spin zero particles as a function of normalized (to $\tilde{T}$ in
Eq.~(\ref{Ttilde})) temperature for various values of the strength
parameter $g$ in Eq.~(\ref{gdef}). Thin curves show the
asymptotic trends (also labeled as 'a' and 'asy') for $T\ll \tilde T$. The dashed curves highlight special cases.}
  \label{le1OetlTs0}
\end{figure*}

\begin{figure*}[th]
\centerline{\includegraphics[width=11cm]{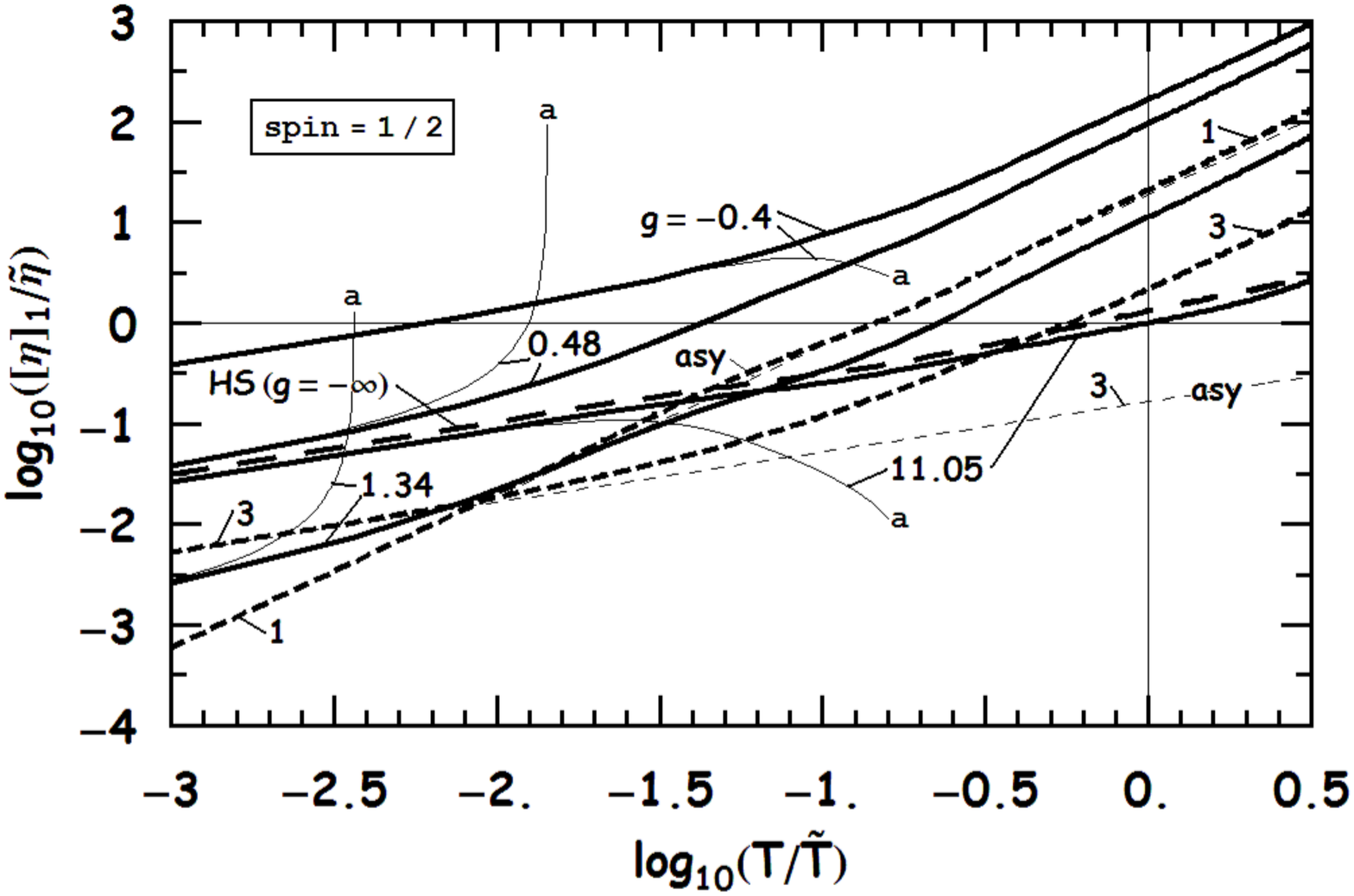}}
\vskip -30pt
  \caption{Normalized (to $\tilde{\eta}$ in Eq.~(\ref{chars}))
diffusion coefficient for spin half particles as a function of normalized (to $\tilde{T}$ in
Eq.~(\ref{Ttilde})) temperature for various values of the strength
parameter $g$ in Eq.~(\ref{gdef}). Thin curves show the
asymptotic trends (also labeled as 'a' and 'asy') for $T\ll \tilde T$ . The dashed curves highlight special cases.}
  \label{le1OetlTs1o2}
\end{figure*}

\begin{figure*}[th]
\centerline{\includegraphics[width=11cm]{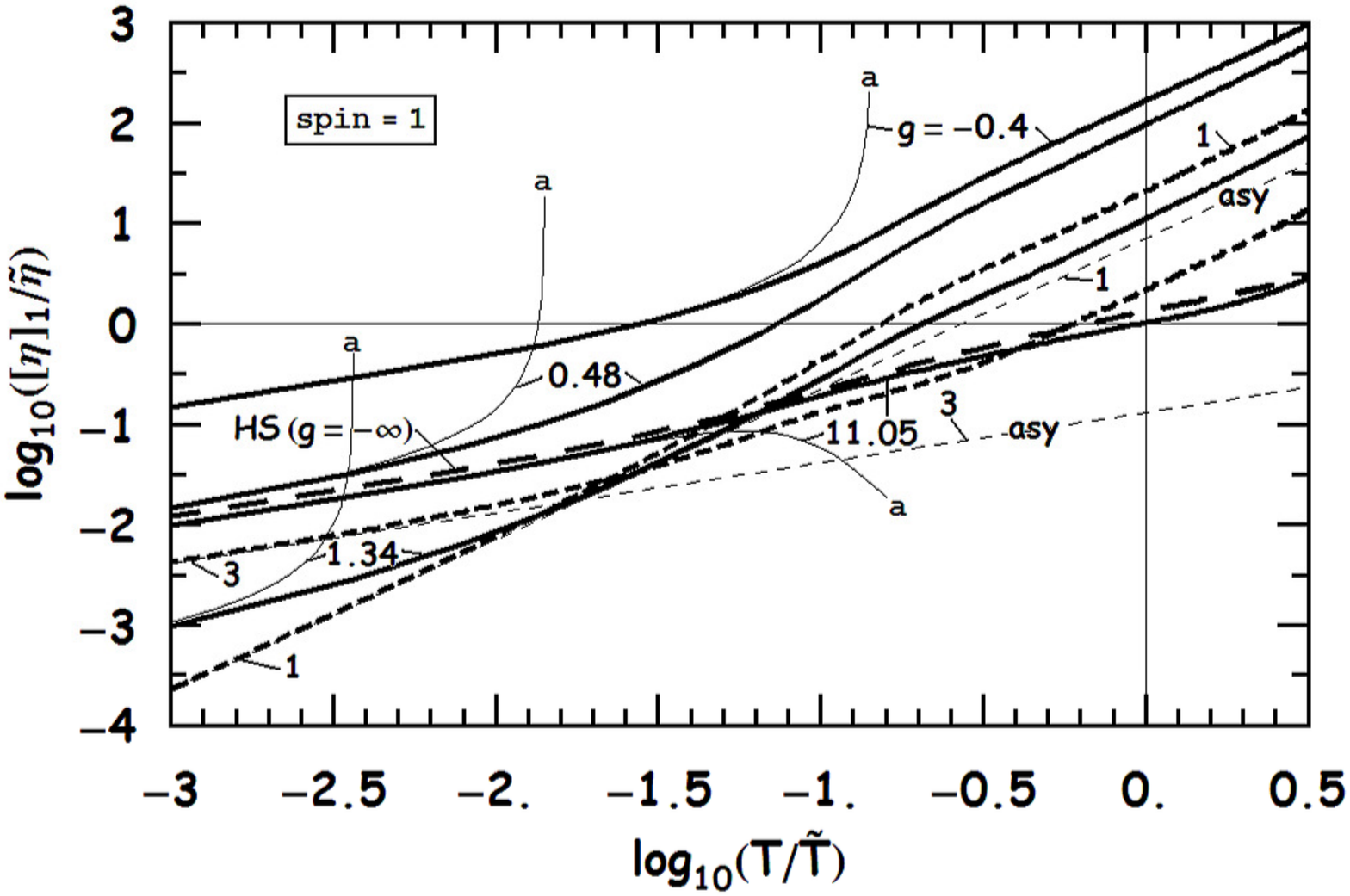}}
\vskip -30pt
  \caption{Normalized (to $\tilde{\eta}$ in Eq.~(\ref{chars}))
diffusion coefficient for spin-one particles as a function of normalized (to $\tilde{T}$ in
Eq.~(\ref{Ttilde})) temperature for various values of the strength
parameter $g$ in Eq.~(\ref{gdef}). Thin curves show the
asymptotic trends (also labeled as 'a' and 'asy') for $T\ll \tilde T$. The dashed curves highlight special cases.}
  \label{le1OetlTs1}
\end{figure*}

Figures \ref{le1OetlTs0}, \ref{le1OetlTs1o2} and \ref{le1OetlTs1}  show
$\left[\eta\right]_1$ from Eq. (\ref{Diffvisq}) normalized to
$\tilde{\eta}$ from Eq. (\ref{chars}).  As expected, the shear viscosity grows steadily with temperature.
At low temperatures the results acquire the asymptotic trends inferred from Chapter \ref{asychap}. Also at low temperatures
the slope of the curves is $1/2$ except for the case of S-wave resonance for which the slope equals $3/2$. For high temperatures the
slope tends to the classical value of $1/2$. The spin of the particles affects only the low temperature result. For high temperatures,
the results become spin independent. Qualitatively the same features are exhibited by the coefficient of diffusion.

\begin{figure*}[tb]
\centerline{\includegraphics[width=11cm]{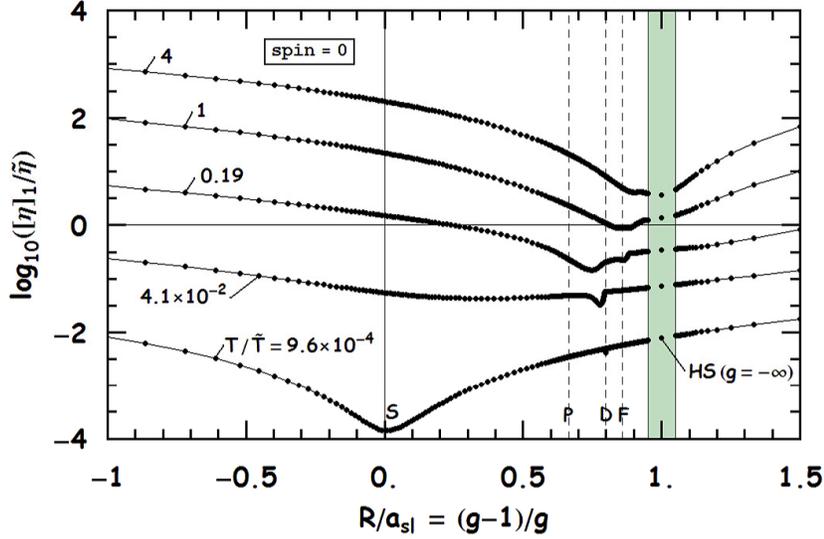}}
\vskip -30pt
  \caption{The normalized viscosity coefficient in
  Eq.~(\ref{Diffvisq}) for spin zero particles  as a function of the inverse scattering length. Effects due to resonances associated with the partial waves
  $l=0,1,2,3$ are indicated by the letters $S,P,D,F$,
  respectively. For states of two spin zero particles odd $l$ partial waves are absent and do not produce any feature. In the vertical shaded region, a large number of
  partial waves are required to obtain convergent results.}
  \label{le1OetISLs0}
\end{figure*}
\begin{figure*}[tb]
\centerline{\includegraphics[width=11cm]{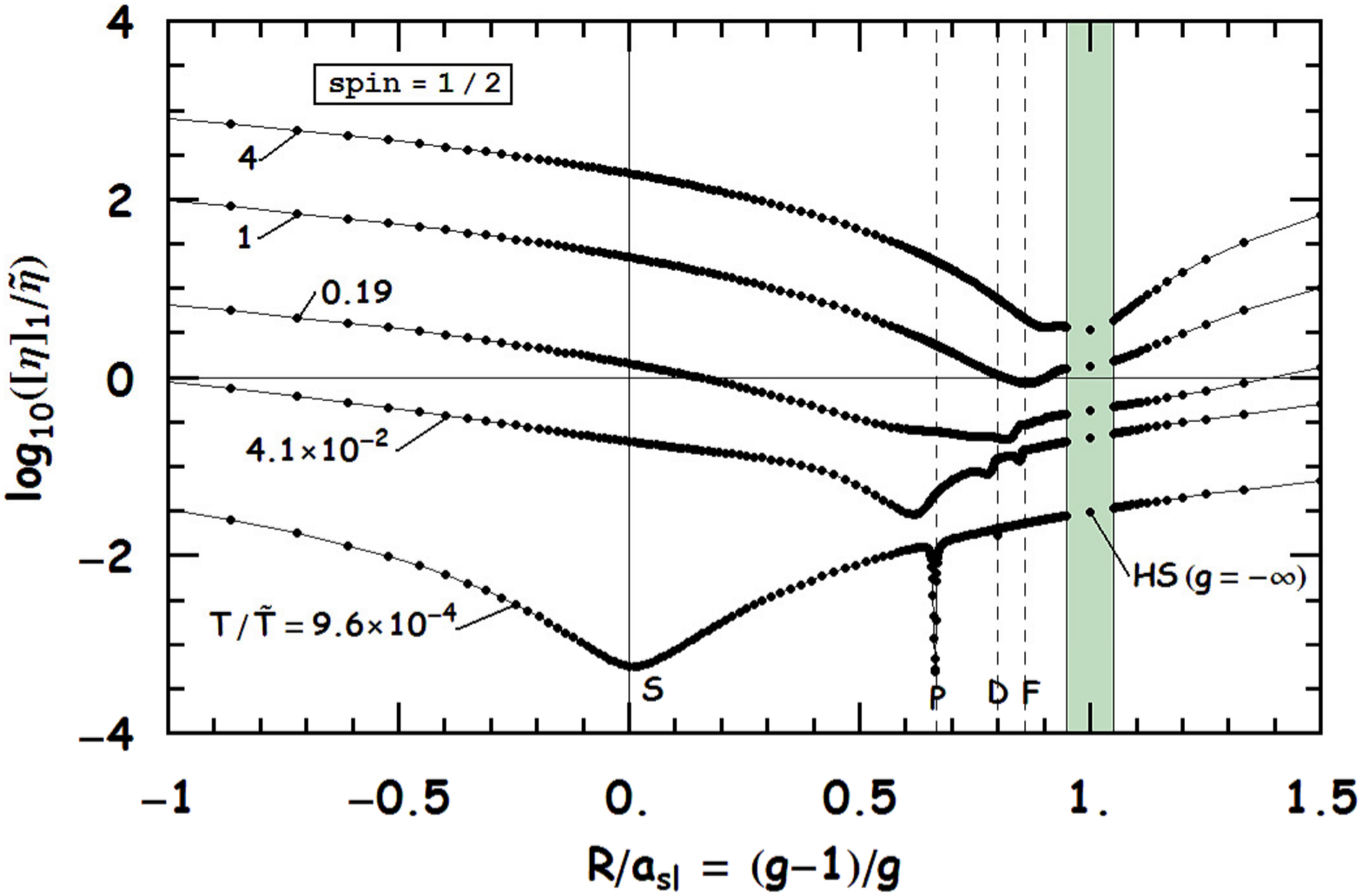}}
\vskip -30pt
  \caption{The normalized viscosity coefficient in Eq.~(\ref{Diffvisq}) for spin half particles  as a function of the inverse scattering length. Effects due to resonances associated with the partial waves
  $l=0,1,2,3$ are indicated by the letters $S,P,D,F$,
  respectively. In the vertical shaded region, a large number of
  partial waves are required to obtain convergent results.}
  \label{le1OetISLs1o2}
\end{figure*}
\begin{figure*}[tb]
\centerline{\includegraphics[width=11cm]{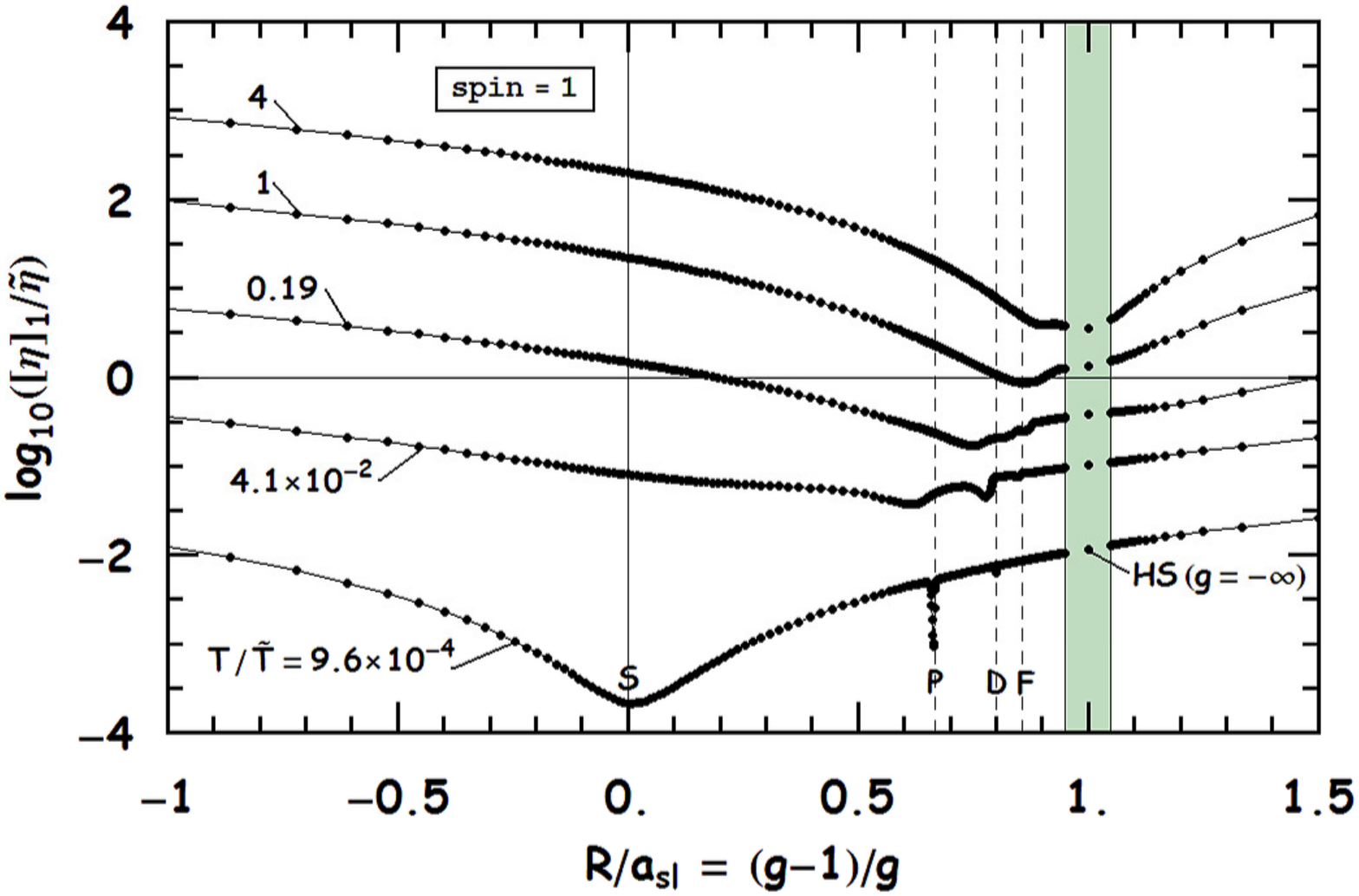}}
\vskip -30pt
  \caption{The normalized viscosity coefficient in
  Eq.~(\ref{Diffvisq}) for spin-one particles  as a function of the inverse scattering length. Effects due to resonances associated with the partial waves
  $l=0,1,2,3$ are indicated by the letters $S,P,D,F$,
  respectively. In the vertical shaded region, a large number of
  partial waves are required to obtain convergent results.}
  \label{le1OetISLs1}
\end{figure*}
%

Figures \ref{le1OetISLs0}, \ref{le1OetISLs1o2} and \ref{le1OetISLs1} show the normalized viscosity coefficient as
a function of $\frac{R}{a_{sl}}=\frac{g-1}{g}$ (inverse scattering length) for $T <
\tilde{T}$. Enhanced cross sections at resonances produce significant
drops in the viscosity as $g\rightarrow 2l+1$. In the case of zero spin, odd $l$ partial waves do not contribute, but for non-zero spins all
$l$-wave resonances are present. The widths of the dips
in viscosity decrease with increasing values of $l$. For $T\leq
\tilde{T}$, the dips become less prominent and disappear for $T \geq
\tilde T$. Spin affects only the low temperature values of the transport coefficient.

\begin{figure*}[tb]
\centerline{\includegraphics[width=11cm]{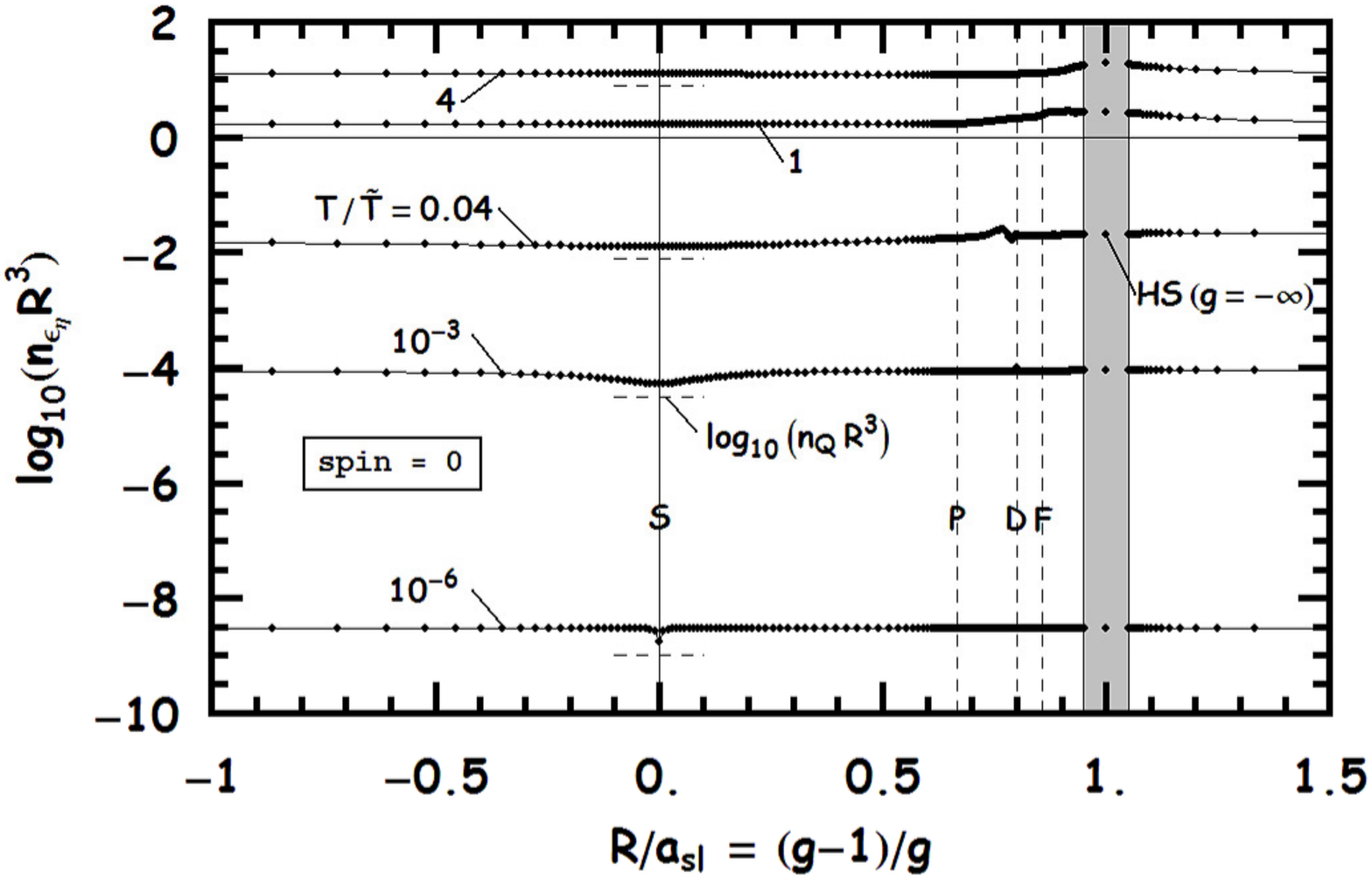}}
\vskip -30pt
  \caption{Characteristic density for spin zero particles  as a function of the inverse scattering length. Resonances associated with the partial waves
  $l=0,1,2,3$ are indicated by the letters $S,P,D,F$,
  respectively. In the vertical shaded region, a large number of
  partial waves are required to obtain convergent results.}
  \label{lgNeeR3ISLs0}
\end{figure*}

\begin{figure*}[tb]
\centerline{\includegraphics[width=11cm]{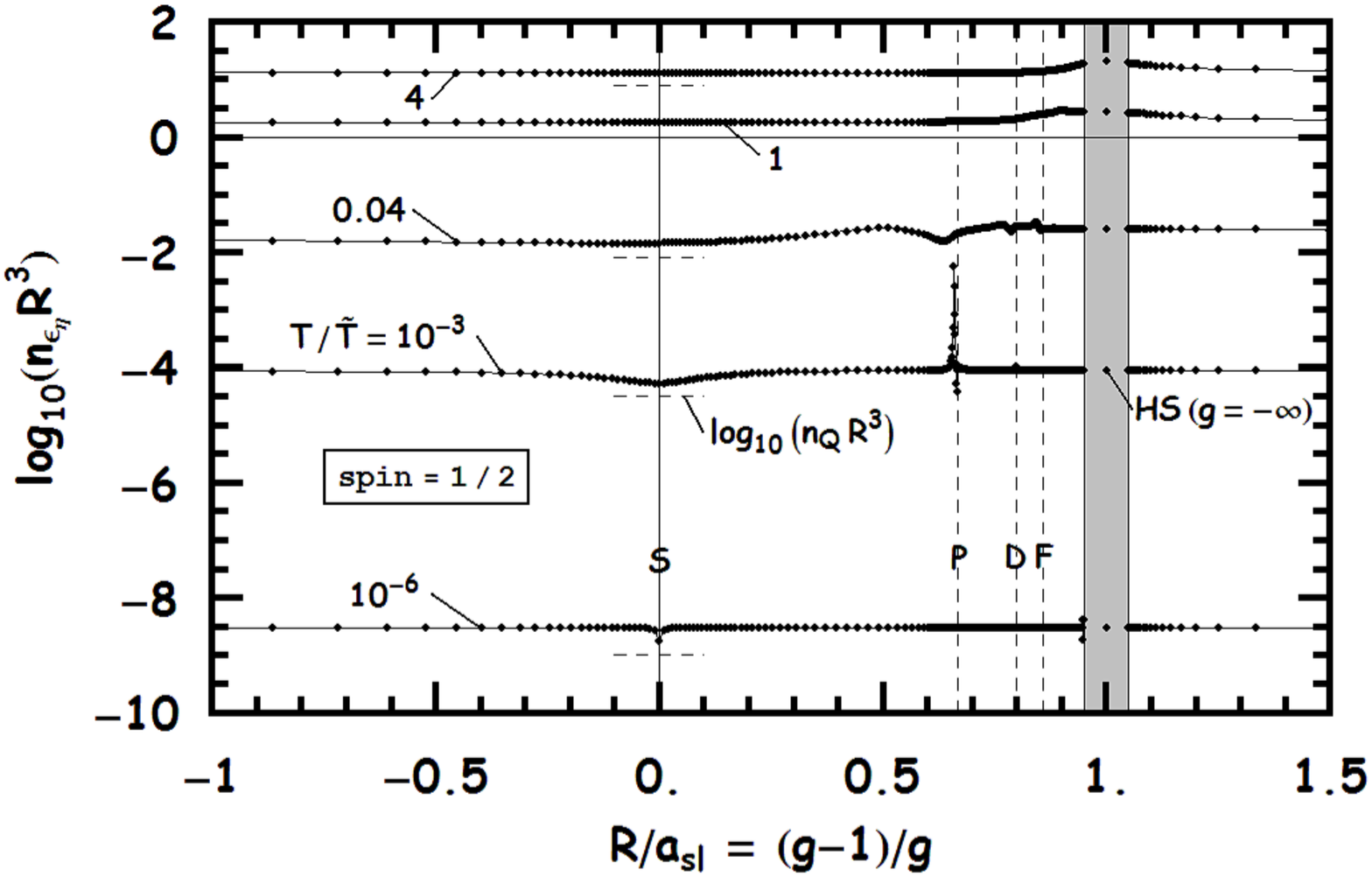}}
\vskip -30pt
  \caption{Characteristic density for spin half particles as a function of the inverse scattering length. Resonances associated with the partial waves
  $l=0,1,2,3$ are indicated by the letters $S,P,D,F$,
  respectively. In the vertical shaded region, a large number of
  partial waves are required to obtain convergent results.}
  \label{lgNeeR3ISLs1o2}
\end{figure*}

For the densities and temperatures considered, the
contribution from the second approximation is given by $\delta_{\mathcal{C}}$ and $\epsilon_{\mathcal{C}}$. 
The density dependent correction is proportional to $\epsilon_{\mathcal{C}} n \lambda^3$ and allows us to define the characteristic density
\begin{equation}
	\label{nR3eps}
	n_{\epsilon_{\mathcal{C}}}R^3 \equiv \frac{(T/\tilde{T})^{3/2}}{\epsilon_{\mathcal{C}}(T)}.
\end{equation}
As shown in Fig. \ref{lgNeeR3ISLs0} and Fig. \ref{lgNeeR3ISLs1o2}, it approximately equals the quantum critical density from Eq. (\ref{nQ}), $n_{\epsilon_{\mathcal{C}}}\approx n_Q$.

\begin{figure*}[tb]
\centerline{\includegraphics[width=11cm]{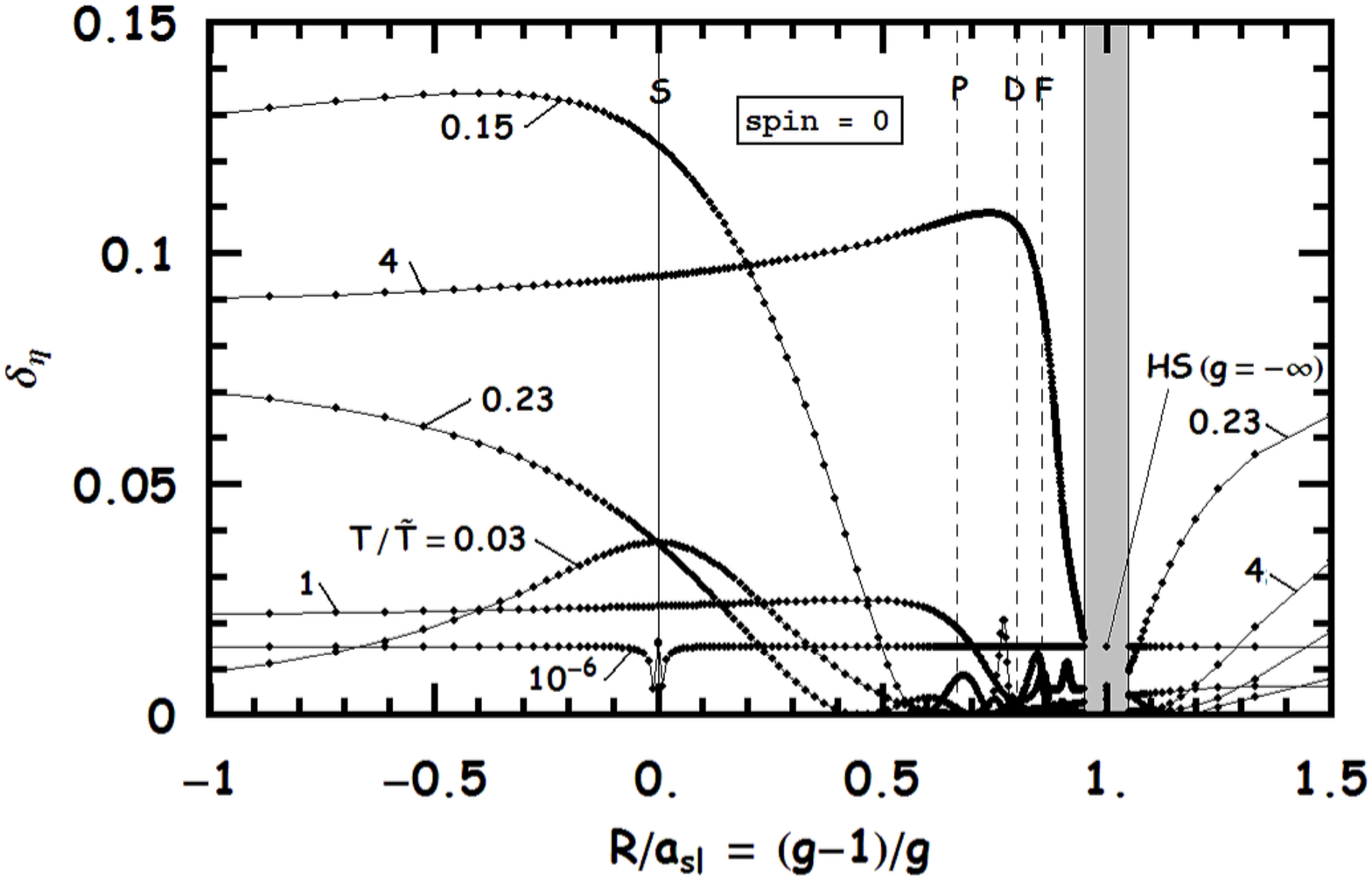}}
\vskip -30pt
  \caption{Second order correction $\delta_{\eta}$ for spin zero particles as a function of the inverse scattering length. Resonances associated with the partial waves  $l=0,1,2,3$ are indicated by the letters $S,P,D,F$,
  respectively. In the vertical shaded region, a large number of
  partial waves are required to obtain convergent results.}
  \label{deltaEtaISLs0}
\end{figure*}
\begin{figure*}[tb]
\centerline{\includegraphics[width=11cm]{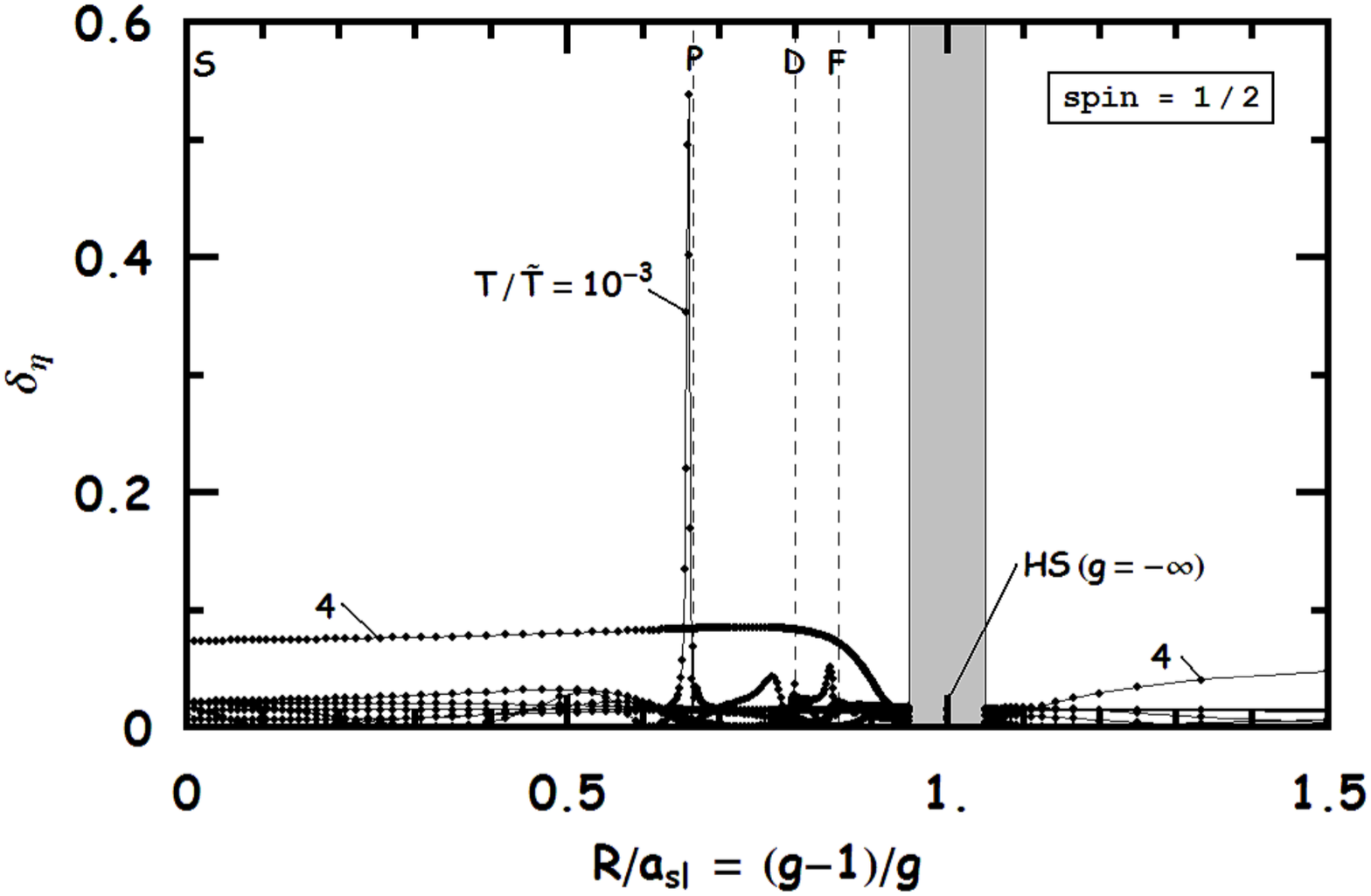}}
\vskip -30pt
  \caption{Second order correction $\delta_{\eta}$ for spin half particles as a function of the inverse scattering length. Resonances associated with the partial waves  $l=0,1,2,3$ are indicated by the letters $S,P,D,F$,
  respectively. In the vertical shaded region, a large number of
  partial waves are required to obtain convergent results.}
  \label{deltaEtaISLs1o2}
\end{figure*}

The density independent part $\delta_{\eta}$ is shown in Fig. \ref{deltaEtaISLs0} for spin=0 and in Fig. \ref{deltaEtaISLs1o2} for spin=12/. The second order corrections  are below $15\%$ for all cases shown.

\begin{figure*}[tb]
\centerline{\includegraphics[width=11cm]{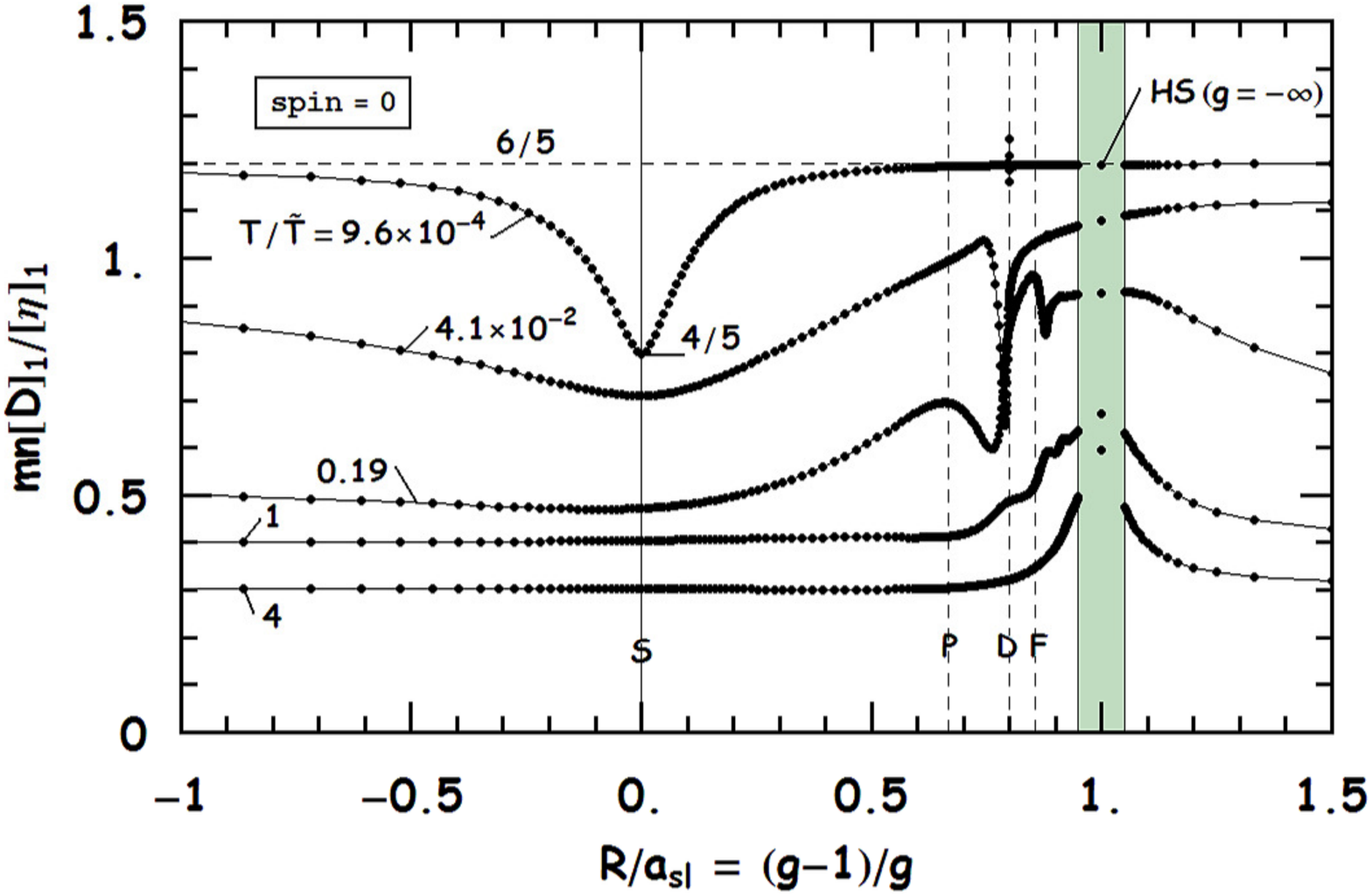}}
\vskip -30pt
  \caption{Ratio of diffusion (times $mn$) to shear
  viscosity versus inverse scattering length for spin zero particles. In the vertical shaded
  region, a large number of partial waves are required to obtain
  convergent results. The hard-sphere results ($g\to-\infty$) are
  shown by solid circles.}
  \label{mnD1oE1isls0}
\end{figure*}
\begin{figure*}[tb]
\centerline{\includegraphics[width=11cm]{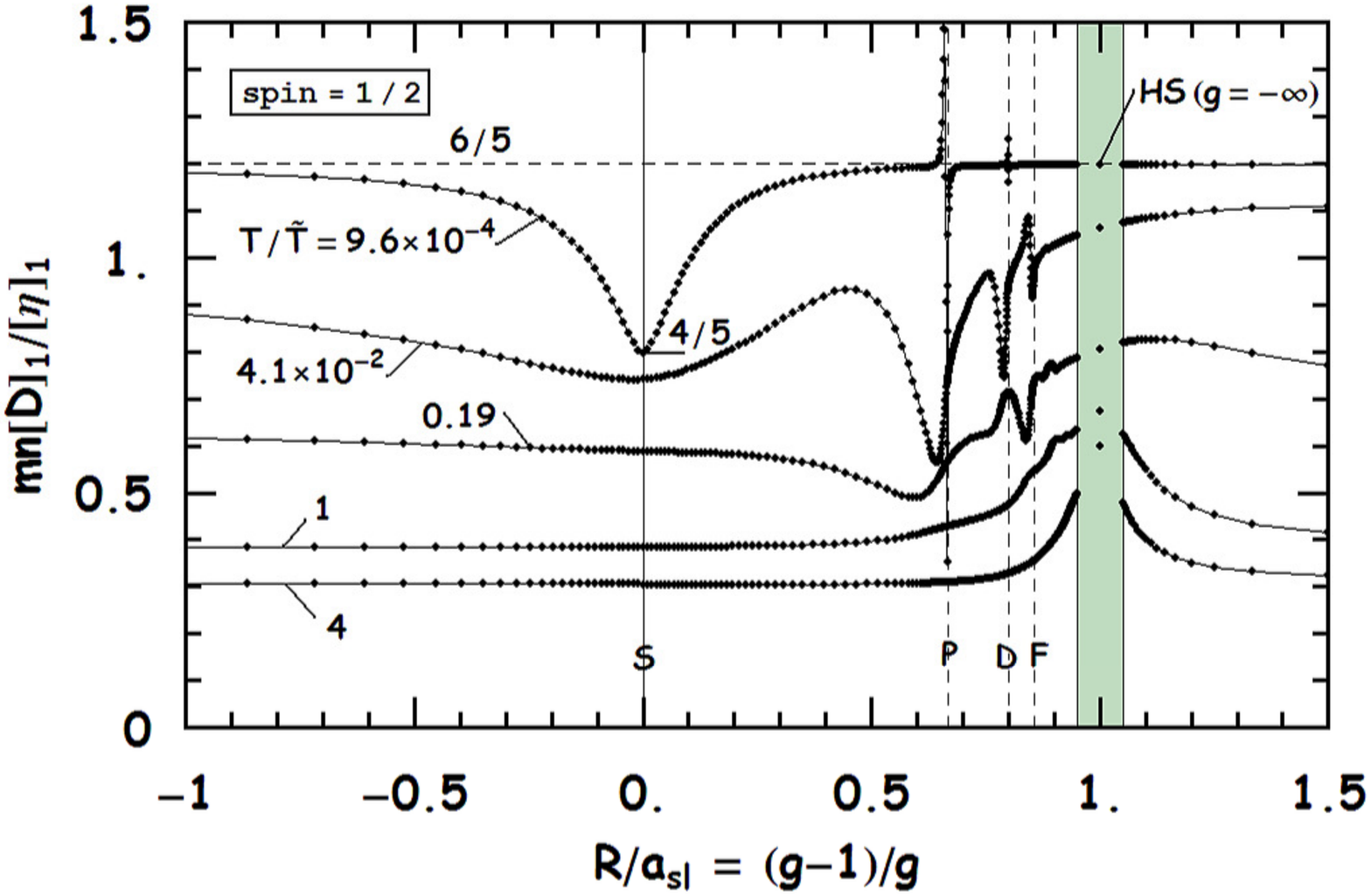}}
\vskip -30pt
  \caption{Ratio of diffusion (times $mn$) to shear
  viscosity versus inverse scattering length for spin half particles. In the vertical shaded
  region, a large number of partial waves are required to obtain
  convergent results. The hard-sphere results ($g\to-\infty$) are
  shown by solid circles.}
  \label{mnD1oE1isls1o2}
\end{figure*}
\begin{figure*}[tb]
\centerline{\includegraphics[width=11cm]{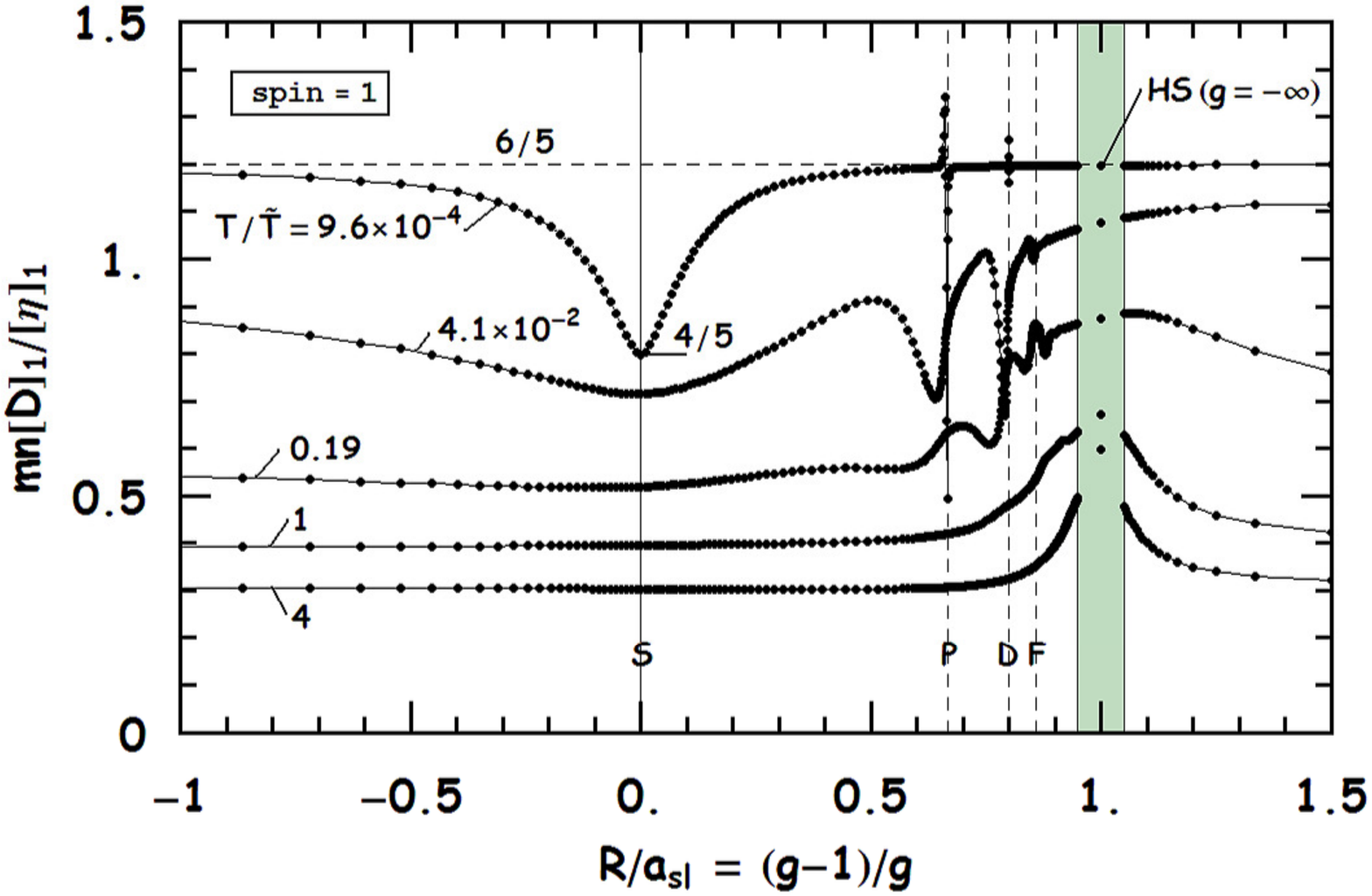}}
\vskip -30pt
  \caption{Ratio of diffusion (times $mn$) to shear
  viscosity versus inverse scattering length for spin-one particles. In the vertical shaded
  region, a large number of partial waves are required to obtain
  convergent results. The hard-sphere results ($g\to-\infty$) are
  shown by solid circles.}
  \label{mnD1oE1isls1}
\end{figure*}
%

In Fig.~\ref{mnD1oE1isls0}, Fig.~\ref{mnD1oE1isls1o2} and Fig.~\ref{mnD1oE1isls1}, the ratio of the coefficients of diffusion
(times $mn$ to make it dimensionless) and viscosity is shown as function of $R/a_{sl}$ for various values of
$T\leq \tilde T$. As expected, the largest variations in this ratio occur
as $g\rightarrow 2l+1$, that is, as resonances are approached.  For
$T\ll\tilde{T}$, the ratio approaches the asymptotic value $6/5$ away
from resonances and $4/5$ for the $S$-wave resonance (a point). With
increasing $T$, resonances become progressively broader with
diminishing strengths; for $T\approx\tilde{T}$ resonances disappear.
Different spin values affect the ratio only at low temperatures.
For comparison, this figure also includes the results for hard spheres (HS).


\section{Asymptotic behavior at low temperatures}
\label{asychap}

At low temperatures, the integrands in Eq. (\ref{tr_ints_s1}) and Eq. (\ref{tr_ints_s2}) can be expanded
in a series involving $\gamma$. The first order terms can be integrated leading to expressions
\begin{equation}
\begin{split}
  \Omega_{\alpha,0}^{(2,t)}(T\to 0) \to \Omega_{\alpha,0}^{(1,t)}(T\to 0) \\
  \to \pi g^2 \alpha^{-(t+3)}\Gamma(t+2)\\
  \times\left[ 
  \frac{\alpha}{(g-1)^2}-\frac{2\pi(2+g^2)(2+t)}{3(g-1)^4}(T/\tilde{T})
  \right](T/\tilde{T}),
\end{split} 
\label{tr_ints_saa}
\end{equation}
for $l=0$ and $g\neq1$. For $l=0$ and $g=1$
\begin{equation}
\begin{split} 
  \Omega_{\alpha,0}^{(2,t)}(T\to 0) \to \Omega_{\alpha,0}^{(1,t)}(T\to 0) \\
  \to (1/18)\alpha^{-(t+2)}\Gamma(t+1)\\
  \times\left[ 
  9 \alpha - 8 \pi (2+t)(T/\tilde{T})
  \right].
\end{split} 
\label{tr_ints_saa1}
\end{equation}
For $l>0$ and $g\neq 2l+1$
\begin{equation}
\begin{split} 
  \Omega_{\alpha,l}^{(2,t)}(T\to 0) \to \Omega_{\alpha,l}^{(1,t)}(T\to 0) \\
  \to 4^{l}\pi^{2l+1} \frac{g^2(2l+1)^2\alpha^{-(2l+2+t)}}{(2l+1-g)^2((2l+1)!!)^4}\\
  \times\Gamma(2l+2+t)(T/\tilde{T})^{2l+1},
\end{split} 
\label{tr_ints_sag}
\end{equation}
and, finally, for $l>0$ and $g=2l+1$
\begin{equation}
\begin{split} 
  \Omega_{\alpha,l}^{(2,t)}(T\to 0) \to \Omega_{\alpha,l}^{(1,t)}(T\to 0) \\
  \to 4^{l-2}\pi^{2l-1} \frac{(2l+3)^2 (2l+1)^2(2l-1)^2\alpha^{-(2l+t)}}{((2l+1)!!)^4}\\
  \times\Gamma(2l+t)(T/\tilde{T})^{2l-1}.
\end{split}
\label{tr_ints_sagr}
\end{equation}
Above $\Gamma(t)$ stands for Euler's gamma function\cite{AbramowitzANDStegun}.

It is seen that the cases $g\rightarrow 1,3$ require special consideration.  In these
cases, the asymptotic behavior for $T\ll \tilde T$ is obtained from a
series expansion in $x=k R$ of the transport cross sections
$q^{(n)}(x)$ in Eqs. 
~(\ref{tr_crosssec_sum2}), and subsequent integrations of
Eq.~(\ref{tr_ints}). The limiting forms we found are listed in Table 1 and Table 2.
\begin{table}[pt]
\tbl{The asymptotic behavior of diffusion coefficient $\left[\mathcal{D}\right]_1 / \tilde{\mathcal{D}}$ at low temperatures.}
{\begin{tabular}{|l|ccc|l|} \toprule
Case & $s=0$ & $s=1/2$ & $s=1$ & Common multiplier\\
\hline
$g=1$ & $2\pi$ & $8\pi$ & $3\pi$ & $\times(T/\tilde{T})^{3/2}$\\
$g=3$ & $2/9$ & $4/17$ & $18/79$ & $\times(T/\tilde{T})^{1/2}$\\
$g\neq1,3$ & $1/2$ & $2$ & $3/4$ & $\times\left(\frac{g-1}{g}\right)^2(T/\tilde{T})^{1/2}$\\
\hline
\end{tabular}}
\label{Tabdiffusion}
\end{table}
The asymptotic behavior (for $T \ll \tilde T$) of the 
shear viscosity $\left[\eta \right]_1$ in Table 2 is similar to that of
the diffusion coefficient in Table 1.
\begin{table}[pt]
\tbl{The asymptotic behavior of shear viscosity coefficient $[\eta]_1 / \tilde{\eta}$ at low temperatures.}
{\begin{tabular}{|l|ccc|l|} \toprule
Case & $s=0$ & $s=1/2$ & $s=1$ & Common multiplier \\
\hline
$g=1$ & $3\pi/2$ & $6\pi$ & $9\pi/4$ & $\times(T/\tilde{T})^{3/2}$\\
$g=3$ & $1/9$ & $1/6$ & $3/23$ & $\times(T/\tilde{T})^{1/2}$\\
$g\neq1,3$ & $1/4$ & $1$ & $3/8$ & $\times\left(\frac{g-1}{g}\right)^2(T/\tilde{T})^{1/2}$\\
\hline
\end{tabular}}
\label{Tabvisco}
\end{table}
It is interesting that even at the two-body level, the coefficients of
diffusion, thermal conductivity and viscosity acquire a significantly
larger temperature dependence as the scattering length 
$a_{sl} \rightarrow \infty~(g \rightarrow 1)$.

The coefficient of viscosity (as also $mn$ times the coefficient of
diffusion) has the dimension of action ($\hbar$) per unit volume.  The
manner in which the effective physical volume $\mathcal{V}$ changes as the
strength parameter $g$ is varied is illuminating in our results for
$T\ll \tilde T$ (see next section) as Table 3 shows. In the unitary limit ($g=1$), the
relevant volume is ${\mathcal{V}}\propto \lambda^3$. For $g=3$, ${\mathcal{V}}
\propto \lambda R^2$ (independent of $a_{sl})$, and for $g\neq 1,3$,
${\mathcal{V}} \propto \lambda a_{sl}^2$ (independent of $R$).

\begin{table}[t]
\tbl{First order coefficients of diffusion (times $mn$), shear
viscosity, and their ratios  for $T\ll \tilde T$ for select strength
parameters $g$ and spins $s$ (c. m. is common multiplier). The unitary limit ($g=1$) result for $\eta$ was
obtained earlier\cite{Massignaan2007}.}
{\begin{tabular}{|l|ccc|ccc|c|ccc|} 
\hline
Case & & $mn{\mathcal{D}}$ & & & $\eta$ & & c. m. & & $mn{\mathcal{D}}/\eta$ & \\
\hline
 $s=$ & 0 & 1/2 & 1 & 0 & 1/2 & 1 & & 0 & 1/2 & 1 \\
\hline
$g=1$ & $\frac{3\sqrt{2}\pi}{16}$ &  $\frac{3\sqrt{2}\pi}{4}$ &  $\frac{9\sqrt{2}\pi}{32}$ &
$\frac{15\pi}{32\sqrt{2}}$ & $\frac{15\pi}{8\sqrt{2}}$ & $\frac{45\pi}{64\sqrt{2}}$ & $\times\frac{\hbar}{\lambda^3}$ & 4/5 & 4/5 & 4/5 \\
$g=3$ & $\frac{\sqrt{2}}{48}$ &  $\frac{3\sqrt{2}}{136}$ &  $\frac{27\sqrt{2}}{1264}$ &
$\frac{5}{144\sqrt{2}}$ & $\frac{5}{96\sqrt{2}}$ & $\frac{15}{368\sqrt{2}}$ & $\times\frac{\hbar}{\lambda R^2}$ & 6/5 & 72/85 & 414/395 \\
$g\neq1,3$ & $\frac{3\sqrt{2}}{64}$ &  $\frac{3\sqrt{2}}{16}$ &  $\frac{9\sqrt{2}}{128}$ &
$\frac{5}{64\sqrt{2}}$ & $\frac{5}{16\sqrt{2}}$ & $\frac{15}{128\sqrt{2}}$ & $\times\frac{\hbar}{\lambda {a_{sl}}^2}$ & 6/5 & 6/5 & 6/5 \\
\hline
\end{tabular}}
\label{Taball}
\end{table}

\section{Ratio of shear viscosity to entropy density}
\label{p3sec4}

A lower limit to the ratio of shear viscosity to
entropy density is being sought \cite{Kats2009} with results even 
lower than $(4\pi)^{-1}(\hbar/k_B)$ first proposed by
Kovtun\cite{Kovtun2005}. Using string theory methods, they show that this ratio is equal to a universal value
for a large class of strongly interacting quantum field theories whose dual description involves
black holes in anti$-$de Sitter space.

We therefore will use our simple model to examine $[\eta]_{1,2}/s$ in the light of the results of this work.
Also we recall the entropy density
\begin{eqnarray}
s &=& (5/2-\ln(n \lambda^3)+\delta s(T)\, n
R^3)n k_B \,,
\label{entropy}
\end{eqnarray}
which includes the second virial correction from Eq. (\ref{ds_a2}) to the
ideal gas entropy density. The validity is checked through Eq. (\ref{corr_val_eqn}).

\begin{figure}[tb]
\centerline{\includegraphics[width=12cm]{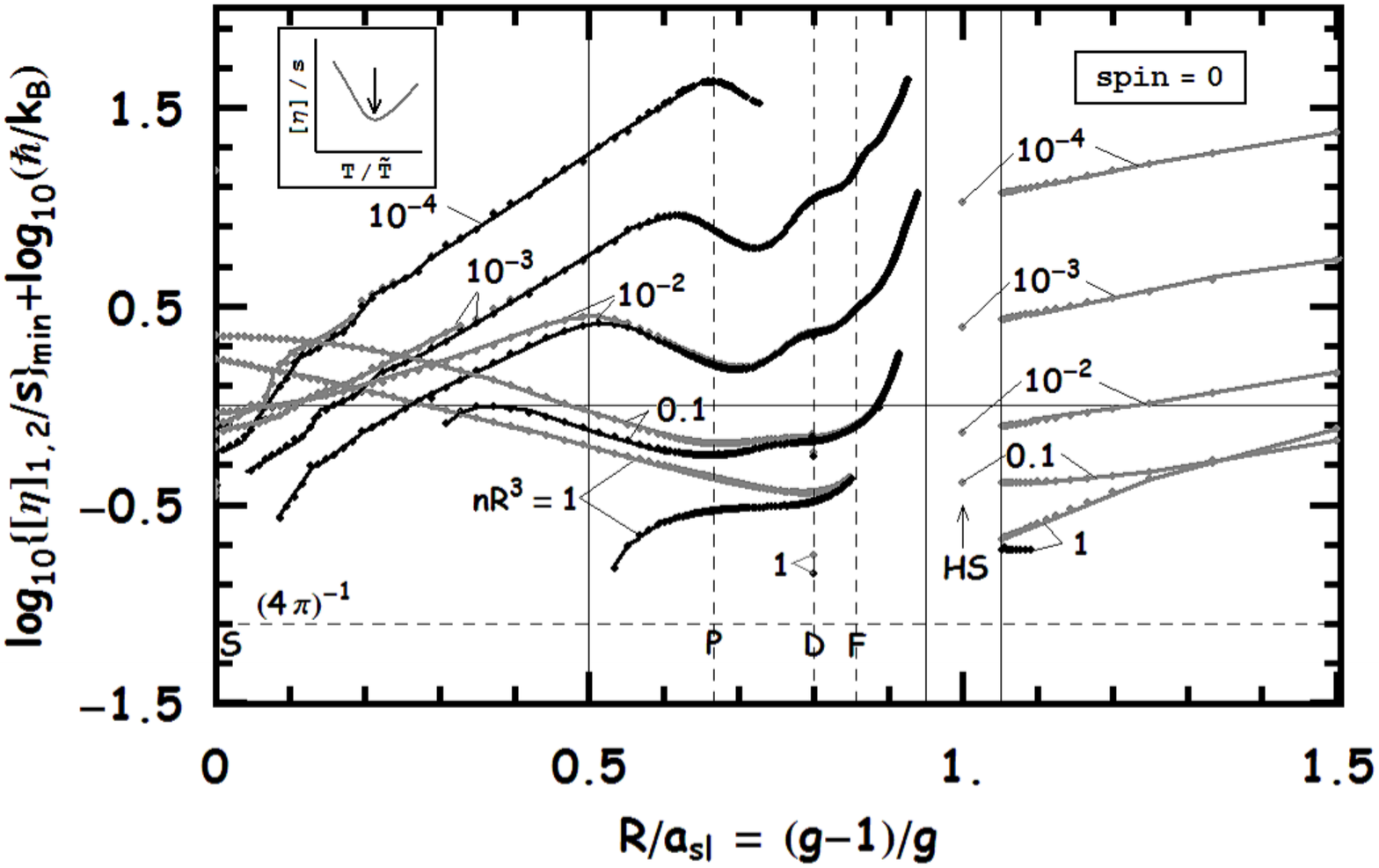}}
\vskip -30pt
   \caption{For each $n R^3$, the minimum value in the first (lower curve) and
  second (upper curve) order calculations of shear viscosity is
  plotted versus $R/a_{sl}$ for the indicated $n R^3$ for spin zero particles.  In the vertical
  blank region, a large number partial waves are needed.  Vertical
  lines with letters $S,P,D$, and $F$, respectively, indicate
  resonances associated with the partial waves $l=0,1,2$ and $3$. The
  symbol HS denotes hard-spheres for which $g \to -
  \infty$. The horizontal dashed line shows the conjectured lower
  bound $1/(4 \pi)$ \cite{Kovtun2005}. The inset shows the occurrence of a
  minimum in the ratio of viscosity to entropy density versus $T$.}
  \label{lgEosMINs0}
\end{figure}

\begin{figure}[tb]
\centerline{\includegraphics[width=12cm]{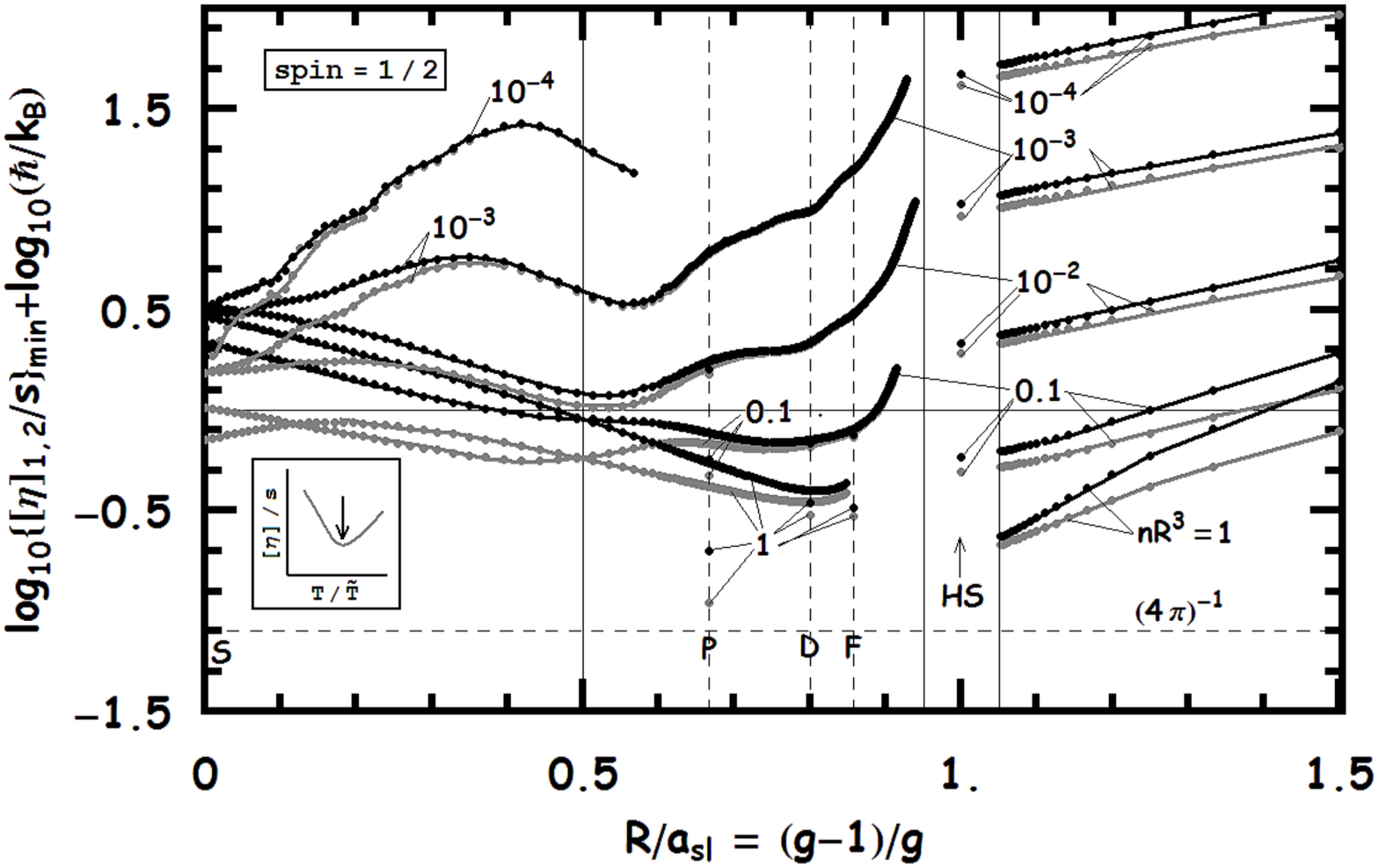}}
\vskip -30pt
   \caption{For each $n R^3$, the minimum value in the first (lower curve) and
  second (upper curve) order calculations of shear viscosity is
  plotted versus $R/a_{sl}$ for the indicated $n R^3$ for spin half particles.  In the vertical
  blank region, a large number partial waves are needed.  Vertical
  lines with letters $S,P,D$, and $F$, respectively, indicate
  resonances associated with the partial waves $l=0,1,2$ and $3$. The
  symbol HS denotes hard-spheres for which $g \to -
  \infty$. The horizontal dashed line shows the conjectured lower
  bound $1/(4 \pi)$ \cite{Kovtun2005}. The inset shows the occurrence of a
  minimum in the ratio of viscosity to entropy density versus $T$.}
  \label{lgEosMINs1o2}
\end{figure}

\begin{figure}[tb]
\centerline{\includegraphics[width=12cm]{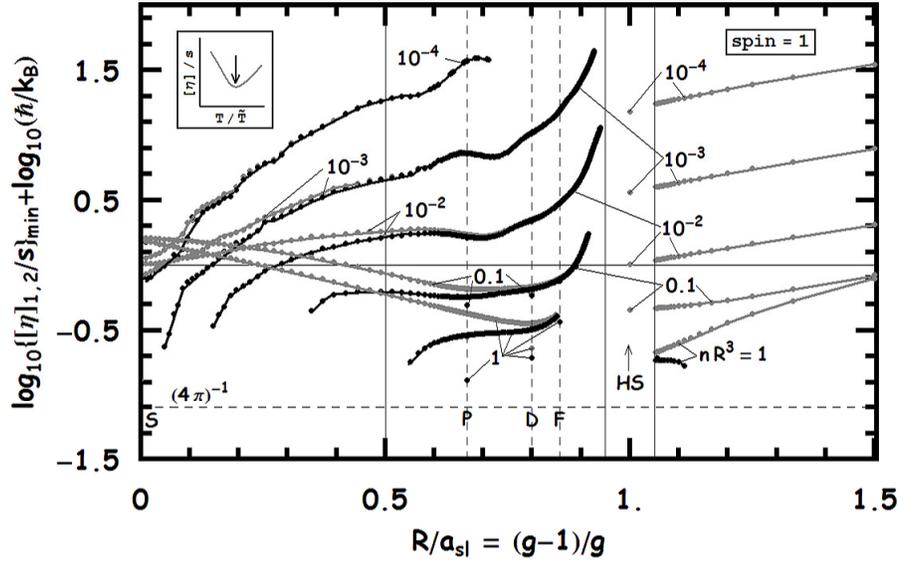}}
\vskip -30pt
   \caption{For each $n R^3$, the minimum value in the first (lower curve) and
  second (upper curve) order calculations of shear viscosity is
  plotted versus $R/a_{sl}$ for the indicated $n R^3$ for spin-one particles.  In the vertical
  blank region, a large number partial waves are needed.  Vertical
  lines with letters $S,P,D$, and $F$, respectively, indicate
  resonances associated with the partial waves $l=0,1,2$ and $3$. The
  symbol HS denotes hard-spheres for which $g \to -
  \infty$. The horizontal dashed line shows the conjectured lower
  bound $1/(4 \pi)$ \cite{Kovtun2005}. The inset shows the occurrence of a
  minimum in the ratio of viscosity to entropy density versus $T$.}
  \label{lgEosMINs1}
\end{figure}

The inserts in Fig. \ref{lgEosMINs0}, Fig. \ref{lgEosMINs1o2} and Fig. \ref{lgEosMINs1}
show a characteristic minimum of
$\eta/s$ versus $T$ for fixed dilution $n R^3$ and strength $g$.
Values of $[\eta]_{1,2}/s$ at the minimum are also shown as functions
of $R/a_{sl}$ (inverse scattering length) for several $n R^3$. The lower (upper)
curve for each $n R^3$ corresponds to the first (second) order
calculations of $\eta$.  The large role of the improved estimates of
$\eta$ on the ratio $\eta/s$ is noticeable. The case of $n R^3=1$
possibly requires an adequate treatment of many-body effects not
considered here.  We can, however, conclude that 
in the dilute gas limit $\eta/s$ for the delta-shell gas
remains above $(1/4\pi)\hbar/k_B$.

Our analysis here of the transport coefficients of particles subject
to a delta-shell potential has been devoted to the dilute gas
(non-degenerate) limit, in which two-particle interactions dominate,
but with scattering lengths that can take various values including
infinity. Even at the two-body level considered, a rich structure in
the temperature dependence and the effective physical volume
responsible for the overall behavior of the transport coefficients are
evident.  The role of resonances in reducing the transport
coefficients are amply delineated. The cases of intermediate and extreme degeneracies cases are considered in Refs. \cite{Massignaan2007} \cite{Bruun2005} \cite{Bruun2007} and \cite{Rupak2007} which highlight the additional roles of many-body effects. Their result in the limiting case $T\gg \frac{\hbar^2/a_{sl}^2}{m}$ (and also well above the Fermi temperature) is
\begin{equation}
	\eta=\frac{15 (m k_B T)^{3/2}}{32 \sqrt{\pi} \hbar^2},
	\label{refetares1}
\end{equation}
and for $T\ll \frac{\hbar^2/a_{sl}^2}{m}$ the result is
\begin{equation}
	\eta=\frac{5 \sqrt{m k_B T}}{32 \sqrt{\pi} a_{sl}^2},
	\label{refetares2}
\end{equation}
which coincides with the corresponding ($g=1$ and $g\neq1,3$) results in Table 2.

\clearpage

\section{Summary and conclusions}
\label{concl}

In this work, we have extended the seminal work of Gottfried~\cite{Gottfried66,Gottfried04} on the two-body quantum mechanics of the delta-shell potential to the many-body context by studying the thermal and transport properties of a dilute gas. 
We start with the two-body physics of the delta-shell potential $V(r)=-v \, \delta(r-R)$. The potential has two parameters: strength $v$ and range $R$. The Schr\"odinger equation for this potential has analytical solutions and enables the analysis of bound, resonant and scattering states. The physics of interaction depends only on the dimensionless strength parameter $g \propto v R$, where $R$ sets the length scale. Low energy scattering is effectively described by the scattering length and the effective range, which, for this particular case, are $a_{sl}/R=g/(g-1)$ and $r_0/R=(2/3)(1+1/g)$. Therefore, for values of $g \approx 1$ the scattering length $a_{sl}$ becomes large and diverges at $g=1$ (unitary limit). This provides a simple model to study properties of the system near the unitary limit. Also $g$ controls how many bound states the system has (integer part of $(g+1)/2$), so that contributions from bound and scattering states can be investigated.

Scattering states are described by the partial wave phase shifts $\delta_l$ ($l$ is the angular momentum quantum number). The derivative of the phase shifts with respect to the wave number $k$, $d \delta_l / dk$, is proportional to the density of states. Additionally, higher-$l$ scattering lengths are derived and diverge at values of $g=2l+1$. When the corresponding partial-wave scattering length is large, it produces loosely bound $l$-states with energy close to zero. It appears as a delta-function-like feature in the $d \delta_l / dk$ vs. $k$ plot and represents a resonance. Near resonance values, $g\approx1$ for S-wave ($l=0$) and $g \approx 3$ for P-wave ($l=1$), approximate analytical expressions for the phase shift derivatives were obtained, and they confirm the $\pi/2$ and $\pi$ values for zero-energy phase shifts from Levinson's theorem. Using the scattering wave differential equation, the S-wave scattering length for the delta-shell potential is alternatively obtained.

The bound state wave-function normalization was found with the help of the Feynman-Hellmann theorem. The delta-shell model of the deuteron was examined by adjusting parameters to reproduce the binding energy of $2.22$ MeV and the scattering length of $5.42$ fm (in the triplet configuration, the deuteron is dominated by the S-state). This model gives an effective range $r_0=1.79$ fm, which is close to the experimental value of $1.76$ fm. The Feynman-Hellmann theorem when applied to the delta-shell potential also allowed us to confirm the virial theorem for bound-states and obtain Kramers-Pasternak-like relations for moments of the bound states.

With the two-body physics inputs in hand, the statistical physics of a dilute delta-shell gas allows to study its thermal and transport properties, especially near resonances. Two-body interactions can be tuned by the parameter $g$ and they contribute to the thermal properties through the second virial coefficient, $a_2$. This coefficient is the first correction to the ideal-gas behavior and incorporates scattering and bound states. Near S- and P-resonances, the scattering part of the virial integral can be well approximated by loosely bound discrete states, signaling the formation of an admixture of long-living dimers. Limits for the virial approximation are discussed and require the density to be small ($n R^3<1$, that is, dilute) and the thermal de-Broglie wavelength to be large ($\lambda(T)>R$, that is, low temperature $T$). Also dimers have to be the minority to avoid significant contribution from three-body interactions and the bound state energy $E_b$ should not be much larger than the temperature (since it contributes in $a_2$ as $e^{-E_b/(k_B T)}$). Nevertheless, the effective range expansion for the second virial coefficient can be obtained, including the unitary limit and resonances.

A knowledge of the second virial coefficient and its derivatives with respect to temperatures enables us to calculate virial corrections to the ideal-gas thermodynamical properties, such as the entropy density. When varying $g$, all state variables experience a sudden change in value only at exact resonance $g=2l+1$.

The Chapman-Enskog method for solving the Boltzmann equation  involves small perturbations of the distribution function from its equilibrium state. For dilute systems only two-body collisions affect the particle's distribution function, thus the collisional integral can be expressed in terms of thermally weighted differential cross-sections. Hence, the transport coefficients are given by transport integrals ($\omega_1^{(1,1)}(T)$ and $\omega_1^{(2,2)}(T)$). Use of classical or quantum cross sections will produce corresponding results for the transport coefficients. The second-order calculations include higher-order transport integrals ($\omega_1^{(1,2)}(T)$, $\omega_1^{(2,3)}(T)$ and etc.) and corrections due to the quantum higher-density effects ($n \lambda^3 \approx 1$ and $\omega_{4/3}^{(2,2)}(T)$,...).

It is found that when interactions are tuned close to resonance values it results in a reduction of the transport coefficients. That is, a cold ($\lambda(T)>R$) unitary gas and a gas with P-wave zero-energy dimers will have several orders of magnitude lower (dips) shear viscosity $[\eta]_1$ than in other regimes. Also, it is shown that the delta-shell result coincides with that for hard-spheres when $g$ is set to $-\infty$. These dips also show up in the ratio of self-diffusion to shear viscosity. Investigation of the asymptotic behavior of transport coefficients, when $\lambda(T)\gg R$, reveals that the coefficient $\propto T^{3/2}$ in the unitary limit and $\propto T^{1/2}$ otherwise. The asymptotic value of shear viscosity also uncovers the relevant volumes: $\lambda^3$ for the unitary limit ($g=1$), $\lambda R^2$ for $g=3$ and $\lambda a_{sl}^2$ for the rest.

Taking the ratio of the coefficient of shear viscosity and entropy density and plotting it versus temperature allows us to find its minimum value.
Calculation of this minimum for various interactions (and therefore scattering lengths) at several densities puts it well above the suggested universal value of $(4\pi)^{-1} \hbar/k_B$ \citep{Kovtun2005}. Although the system has to be dilute, significant insights into the physical properties of a unitary and resonant gas are gained.

It would be interesting to study three-body physics with delta-shell interactions. But the difficulty is that it requires the introduction of three-body forces and the summation of many terms (three-body clusters) arising from non-commuting operators. Surmounting this difficulty is a task for future work. 



\section*{Acknowledgments}

Through the course of this work, SP's research was supported in part by the U.S. DOE grants DE-FG02-93ER-40756 (Ohio University),  DE-FG02-87ER40365 (Indiana University) and DE-SC0008808 (NUCLEI  SciDAC Collaboration), and by a grant from Conacyt (CB-2009-01, \#132400). He also acknowledges a postdoctoral fellowship from DGAPA at the Universidad Nacional Autonoma de Mexico. 
MP thanks  research support from the U.S. DOE grant
DE-FG02-93ER-40756.


\clearpage

\bibliography{references5}{}
\bibliographystyle{apsrev}
\end{document}